\documentclass[]{aa} 

\usepackage{epsfig}

\renewcommand{\d}{{\rm d}}
\newcommand{\pl}{\partial}
 
\newcommand{\beq}{\begin{equation}} 
\newcommand{\eeq}{\end{equation}} 
\newcommand{\beqa}{\begin{eqnarray}} 
\newcommand{\eeqa}{\end{eqnarray}} 
\newcommand{\bea}{\begin{array}} 
\newcommand{\ea}{\end{array}} 
\newcommand{\cH}{{\cal H}} 
\newcommand{\rhob}{\overline{\rho}} 
\newcommand{\lag}{\langle} 
\newcommand{\rag}{\rangle}
\newcommand{\bx}{{\bf x}}
\newcommand{\bv}{{\bf v}}
\newcommand{\bk}{{\bf k}}
\newcommand{\Om}{\Omega_{\rm m}}

\newcommand{\cO}{{\cal O}}

\newcommand{\tKs}{{\tilde{K}_s}}

\newcommand{\cC}{{\cal C}}
\newcommand{\bq}{{\bf q}}

\newcommand{\bs}{{\bf s}}

\begin{document} 
 
\title{Expansion schemes for gravitational clustering: computing two-point and
three-point functions}    
\author{P. Valageas}   
\institute{Institut de Physique Th\'eorique, CEA Saclay, 91191 Gif-sur-Yvette, 
France}  
\date{Received / Accepted } 
 
\abstract
{}
{We describe various expansion schemes that can be used to study 
gravitational clustering. Obtained from the equations of motion or their 
path-integral formulation, they provide several perturbative expansions that 
are organized in a different fashion or involve different partial resummations.
We focus on the two-point and three-point correlation functions, but these
methods also apply to all higher-order correlation and response functions.}
{We present the general formalism, which holds for the gravitational dynamics
and for any similar models, such as the Zeldovich dynamics, that obey
similar hydrodynamical equations of motion with a quadratic nonlinearity.
We give our explicit analytical results up to one-loop order for the 
simpler Zeldovich dynamics. For the gravitational dynamics, we compare our
one-loop numerical results with numerical simulations.}
{We check that the standard perturbation theory is recovered from the path
integral by expanding over Feynman's diagrams. However, the latter expansion
is organized
in a different fashion, and it contains some UV divergences that cancel out as we
sum all diagrams of a given order. Resummation schemes modify the scaling of
tree and one-loop diagrams, which exhibit the same scaling over the linear
power spectrum (contrary to the
standard expansion). However, they do not significantly improve over standard
perturbation theory for the bispectrum, unless one uses accurate two-point
functions (e.g. a fit to the nonlinear power spectrum from simulations).
Extending the range of validity to smaller scales, to reach the range described
by phenomenological models, seems to require at least two-loop diagrams.}
{}


\maketitle

\section{Introduction} 
\label{Introduction}

According to standard cosmological scenarios, the large-scale structures
of the present Universe arise from the amplification by gravitational 
instability of small primordial fluctuations (Peebles 1980). Then, from
observations of the recent Universe at redshifts $z<5$, through galaxy surveys
(Tegmark et al. 2006; Cole et al. 2005), weak-lensing studies 
(Massey et al. 2007; Munshi et al. 2007), or studies of baryonic oscillations
(Eisenstein et al. 1998, 2005), we can derive constraints on the cosmological
parameters and on the properties of those primordial fluctuations.
This requires accurate theoretical predictions for the evolution of 
gravitational 
clustering to compare cosmological models with observations. On large scales,
where density fluctuations are small, linear theory is sufficient. On small 
scales, deep into the nonlinear regime associated with virialized structures
such as galaxy clusters, one can use numerical simulations or phenomenological
models, such as the halo model (Cooray \& Sheth 2002), which allow for an 
accuracy
near $10\%$. However, many observational probes (such as weak lensing
or baryonic oscillations) are mostly sensitive to intermediate scales
(wavenumbers $k \sim 0.1 h$ Mpc$^{-1}$ at $z \la 1$) that might be studied
through analytical methods, such as perturbative expansions. 

Thus, there has recently been a renewed interest in building analytical methods 
that could improve over the standard perturbation theory and provide reliable
predictions in this weakly nonlinear regime. Crocce \& Scoccimarro (2006a,b)
present a partial resummation of the diagrammatic series associated with
the response function, which recovers the expected decay into the nonlinear 
regime, in agreement with numerical simulations. This allows one to obtain
a more accurate prediction for the matter power spectrum than with the usual
perturbation theory (Crocce \& Scoccimarro 2007). On the other hand, 
Valageas (2004, 2007a) described the path-integral formalism associated with
the cosmological gravitational dynamics. This allows one to apply the usual
tools of field theory, such as large-$N$ and semi-classical expansions.
This provides several methods to perform partial resummations of the usual 
perturbation theory. Next, Matarrese \& Pietroni (2007a,b) proposed to
use the dependence of the system on a high-$k$ cutoff to derive some
resummation schemes through the path-integral formalism. An alternative
approach developed by McDonald (2007), based on a renormalization group
technique, also permits going beyond the standard perturbative expansion.

These works have focused on two-point functions, such as the matter power
spectrum or two-point correlation, since the latter are the main quantities
obtained from observations. However, the additional information included
in higher-order statistics, such as the three-point correlation, can prove
very useful in breaking degeneracies (Sefusatti et al. 2006; Munshi et al. 2007).
Therefore, we describe in this paper how such methods can be used to
compute higher-order functions. We focus on the bispectrum, which is the quantity
of most practical interest, since higher-order statistics obtained from
observational surveys are increasingly noisy. We consider both the exact
gravitational dynamics and the simpler Zeldovich dynamics. Indeed,
the equations of motion associated with both systems are similar and these
expansion schemes apply in the same manner to both (Valageas 2007b). Then,
the simpler Zeldovich dynamics can be used as a convenient benchmark to track
possible errors in the derivation of the equations associated with various
schemes.

This article is organized as follows. First, in Sect.~\ref{Equations-of-motion}
we recall the equations of motion associated with cosmological gravitational 
clustering. We focus on weakly nonlinear scales hence we restrict ourselves to
the hydrodynamical description. We write the path-integral formulation of this
system and we briefly present two large-$N$ methods which can be derived within
this formalism. We also recall the main properties of the simpler Zeldovich
dynamics. Next, we present in a general manner various perturbative expansions
that can be applied to both cosmological dynamics (or any equation of motion 
of the same form, i.e. with a quadratic nonlinearity). Thus, we first recall
the standard perturbative expansion over powers of the linear density field
in Sect.~\ref{Standard-perturbation-theory}, we describe the equivalent expansion
that can be derived from the path-integral formalism by expanding over the
interaction term of the action in Sect.~\ref{Expansion-over-cubic}, we present
the resummations associated with the large-$N$ methods in Sect.~\ref{LargeN},
and we briefly discuss in Sect.~\ref{large-k-limit} the resummation
associated with the random advection of small-scale structures by the 
larger-scale coherent flow (involving a high-$k$ limit). Then, we apply
these methods to the Zeldovich dynamics in Sect.~\ref{Zeldovich}, up to one-loop 
order. Indeed, the expressions obtained in this case are simpler than
for the gravitational dynamics. This allows us to show explicitly the connection
between the different methods and to point out the possible divergences which
may appear in intermediate steps as diagrams are gathered in different manners. 
Finally, we apply these methods to the actual gravitational dynamics in 
Sect.~\ref{Gravitational-dynamics}, again up to one-loop order, and we compare 
their predictions with numerical simulations. 
We conclude in Sect.~\ref{Conclusion}.

The reader who is not directly interested in the theoretical aspects of 
gravitational clustering (e.g. the discussion of the various formalisms and 
perturbative methods that may be built, and the problems, such as UV divergences, 
that one may encounter) might first read Sect.~\ref{nonlinear-power-spectrum}
and Figs.~\ref{figBksdNL_z0}-\ref{figBtsdNL}. There, we compare the standard
perturbation theory with the most efficient resummation scheme considered in
this paper (at one-loop order), as well as a recent phenomenological model 
and numerical simulations.

\section{Equations of motion}
\label{Equations-of-motion}

\subsection{Fluid dynamics}
\label{Fluid-dynamics}

In this section, we recall the equations that describe the dynamics
of gravitational clustering in a cosmological framework. As in Valageas (2007a),
we focus on weakly nonlinear scales where we can use the single-stream 
approximation. Then, both cold dark matter and baryonic matter are described 
as a pressureless fluid governed by the continuity and Euler equations, 
supplemented by the Poisson equation for the gravitational potential $\phi$ 
(Peebles 1980),
\beq
\frac{\pl\delta}{\pl\tau} + \nabla.[(1+\delta) \bv] = 0,
\label{continuity1}
\eeq
\beq
\frac{\pl\bv}{\pl\tau} + \cH \bv + (\bv .\nabla) \bv = - \nabla \phi,
\label{Euler1}
\eeq
\beq
\Delta \phi = \frac{3}{2} \Om \cH^2 \delta , 
\label{Poisson1}
\eeq
where $\tau=\int \d t/a$ is the conformal time (and $a$ the scale factor), 
$\cH=\d\ln a/\d\tau$ the conformal expansion rate, and $\Om$ the matter 
density cosmological parameter.
Here, $\delta$ is the matter density contrast and $\bv$ the peculiar velocity.
Since the vorticity field decays within linear theory (Peebles 1980), we take
the velocity to be a potential field so that $\bv$ is fully specified by its 
divergence $\theta=\nabla.\bv$. It is convenient to work in Fourier space, 
which we normalize as
\beq
\delta(\bk) = \int\frac{\d\bx}{(2\pi)^3} e^{-i\bk.\bx} \delta(\bx) ,
\label{deltak}
\eeq
and to introduce the time coordinate $\eta$ defined from the linear growing
mode $D_+$ as
\beq
\eta=\ln D_+(\tau) \;\;\; \mbox{with} \;\;\; D_+(z=0)=1 .
\label{eta}
\eeq
Then, as in Crocce \& Scoccimarro (2006a), we define the two-component vector
$\psi$ as
\beq
\psi(\bk,\eta) = \left(\bea{c} \psi_1(\bk,\eta) \\ \psi_2(\bk,\eta)\ea \right)
= \left( \bea{c} \delta(\bk,\eta) \\ -\theta(\bk,\eta)/\cH f \ea \right) ,
\label{psidef}
\eeq
where $f(\tau)$ is defined as
\beq
f(\tau)= \frac{\d\ln D_+}{\d\ln a} = \frac{1}{\cH} \frac{\d\ln D_+}{\d\tau} .
\label{ftau}
\eeq
Then, the equations of motion (\ref{continuity1})-(\ref{Poisson1}) read
(Crocce \& Scoccimarro 2006a; Valageas 2007a):
\beq
\cO(x,x').\psi(x') = K_s(x;x_1,x_2) . \psi(x_1) \psi(x_2) ,
\label{OKsdef}
\eeq
where we have introduced the coordinate $x=(\bk,\eta,i)$ where $i=1,2$ is the 
discrete index of the two-component vectors. In Eq.(\ref{OKsdef}) and in the 
following, we use the convention that repeated coordinates are integrated over
as
\beq
\cO(x,x').\psi(x') \! = \!\! \int\! \d\bk' \! \d\eta'\sum_{i'=1}^2 
\! \cO_{i,i'}(\bk,\eta;\bk',\eta') \psi_{i'}(\bk',\eta') .
\label{defproduct}
\eeq
The matrix $\cO$ reads
\beq
\!\! \cO(x,x') \! = \! \left( \bea{cc} \frac{\pl}{\pl\eta} & -1 \\ 
\! -\frac{3\Om}{2f^2} \! \; & 
\, \frac{\pl}{\pl\eta} \! + \! \frac{3\Om}{2f^2} \! - \! 1 \! \ea \right) 
\! \delta_D(\bk-\bk') \delta_D(\eta-\eta') 
\label{Odef}
\eeq
whereas the symmetric vertex $K_s(x;x_1,x_2)=K_s(x;x_2,x_1)$ writes as
\beqa
K_s(x;x_1,x_2) & = & \delta_D(\bk_1+\bk_2-\bk) \delta_D(\eta_1-\eta) 
\delta_D(\eta_2-\eta) \nonumber \\ && \times \gamma^s_{i;i_1,i_2}(\bk_1,\bk_2)
\label{Ksdef}
\eeqa
with
\beq
\gamma^s_{1;1,2}(\bk_1,\bk_2)= \frac{(\bk_1+\bk_2).\bk_2}{2k_2^2} ,
\label{gamma1}
\eeq
\beq
\gamma^s_{1;2,1}(\bk_1,\bk_2)= \frac{(\bk_1+\bk_2).\bk_1}{2k_1^2} ,
\label{gamma2}
\eeq
\beq
\gamma^s_{2;2,2}(\bk_1,\bk_2)= \frac{|\bk_1+\bk_2|^2 (\bk_1.\bk_2)}
{2 k_1^2 k_2^2} ,
\label{gamma3}
\eeq 
and zero otherwise (Crocce \& Scoccimarro 2006a; Valageas 2007a).
In the following, we use the common approximation
\beq
\frac{\Om}{f^2} \simeq 1 ,
\label{f1}
\eeq
which simplifies the analysis as the equation of motion (\ref{OKsdef}) no longer
shows any explicit dependence on time $\eta$.

\subsection{Linear regime}
\label{Linear-regime}

On large scales or at early times where the density and velocity fluctuations
are small, one can linearize the equation of motion (\ref{OKsdef}) which yields
$\cO.\psi_L=0$. Then, the linear growing mode is
\beq
\psi_L(x) = e^{\eta} \delta_{L0}(\bk) \left(\bea{c} 1 \\ 1 \ea\right) ,
\label{psiL}
\eeq
where $\delta_{L0}(\bk)$ is the linear density contrast today at redshift
$z=0$. As usual, we define the initial conditions by the linear growing mode
(\ref{psiL}) and we assume Gaussian homogeneous and isotropic initial conditions 
defined by the linear power spectrum $P_{L0}(k)$:
\beq
\lag \delta_{L0}(\bk_1) \delta_{L0}(\bk_2) \rag =  P_{L0}(k_1) 
\delta_D(\bk_1+\bk_2) .
\label{PL0}
\eeq
Then, the linear two-point correlation function $C_L(x_1,x_2)$ reads as
\beqa
C_L(x_1,x_2) & = & \lag \psi_L(x_1) \psi_L(x_2) \rag \nonumber \\
& = & e^{\eta_1+\eta_2} P_{L0}(k_1) \delta_D(\bk_1+\bk_2) 
\left(\bea{cc} 1 & 1 \\ 1 & 1 \ea\right) .
\label{CL}
\eeqa
In previous papers (Valageas 2007a,b) we noted the two-point correlation as
``$G$'', but since a restriction of the response function introduced below 
is also labeled ``$G$'' in the literature, we shall note correlation functions 
as ``$C$'' in this article. 
As in Valageas (2004), it is convenient to introduce the response function
$R(x_1,x_2)$. Using the approximation (\ref{f1}) it reads in the linear regime 
(Crocce \& Scoccimarro 2006a; Valageas 2007a)
\beqa
\lefteqn{R_L(x_1,x_2) = \delta_D(\bk_1-\bk_2) \, \theta(\eta_1-\eta_2)} 
\nonumber \\
&& \times \left\{ \frac{e^{\eta_1-\eta_2}}{5} \left(\bea{cc} 3 & 2 \\ 3 & 2 
\ea\right) + \frac{e^{-3(\eta_1-\eta_2)/2}}{5} \left(\bea{cc} 2 & -2 \\ -3 & 3 
\ea\right) \right\} ,
\label{RL}
\eeqa
where the Heaviside factor $\theta(\eta_1-\eta_2)$ ensures that causality is
satisfied.
Thanks to statistical homogeneity and isotropy the correlation 
functions $C$ (both the nonlinear $C$ as well as $C_L$) are symmetric and of 
the form
\beq
C(x_1,x_2) = \delta_D(\bk_1+\bk_2) C_{i_1,i_2}(k_1;\eta_1,\eta_2) ,
\label{Ck}
\eeq
with
\beq
C_{i_1,i_2}(k;\eta_1,\eta_2) = C_{i_2,i_1}(k;\eta_2,\eta_1) ,
\label{Csym}
\eeq
whereas the response functions $R$ (again both the nonlinear 
$R$ as well as $R_L$) are of the form
\beq
R(x_1,x_2) = \delta_D(\bk_1-\bk_2) \theta(\eta_1-\eta_2) 
R_{i_1,i_2}(k_1;\eta_1,\eta_2) .
\label{Rk}
\eeq
Moreover, the linear two-point functions obey (Valageas 2007a,b):
\beq
\cO(x,z).C_L(z,y) = 0 , \;\;\; \cO(x,z).R_L(z,y) = \delta_D(x-y).  
\label{CLRL} 
\eeq
This can also be checked from expressions (\ref{CL}), (\ref{RL}).

\subsection{Path-integral formalism and large-$N$ expansions}
\label{Path-integral}

As described in Valageas (2007a,b), the statistical properties of the
density and velocity fields can be obtained from the generating functional 
\beq
Z[j] = \lag e^{j.\psi} \rag = \int [\d\psi] [\d\lambda] \; 
e^{j.\psi - S[\psi,\lambda]} ,
\label{Zjpsilambda}
\eeq
where we have introduced an imaginary auxiliary field $\lambda(x)$, and the 
action $S[\psi,\lambda]$ reads as\footnote{Note that the action $S$ only takes
such a simple form because the Jacobian, associated with the change of variables
from the initial condition $\psi_L$ to the nonlinear field $\psi$, can be shown
to be an irrelevant constant.}
\beq
S[\psi,\lambda] = \lambda.(\cO.\psi-K_s.\psi\psi) 
- \frac{1}{2}\lambda.\Delta_I.\lambda ,
\label{Spsilambda}
\eeq
where $\Delta_I$ is the two-point correlation of the initial conditions taken
at time $\eta_I$. This matrix disappears in the final equations when we take
the limit $\eta_I\rightarrow-\infty$.
Moreover, the first and second moments of the field $\lambda$ obey
\beq
\lag\lambda\rag=0 , \;\;\; \lag\lambda\lambda\rag=0 , \;\;\;
\lag \psi(x_1) \lambda(x_2) \rag = R(x_1,x_2) .
\label{Rpsilambda}
\eeq
The action (\ref{Spsilambda}) can serve as a basis for several approaches
borrowed from field theory. For instance, it was used in Matarrese \& Pietroni
(2007a,b) to derive a resummation scheme based on the dependence of the
system on a high-$k$ cutoff (see also Valageas 2007b for a discussion
of this approach in the case of the Zeldovich dynamics).
Another approach relies on standard ``large-$N$'' expansion schemes
(Valageas 2007a,b).
Thus, generalizing the problem to $N$ fields or simply multiplying the
exponent in expression (\ref{Zjpsilambda}) by a factor $N$, one can derive
various expansions over powers of $1/N$ (which agree up to the order $1/N^q$
of the expansion but differ by higher-order terms as they involve different
partial resummations).
This path-integral formalism can also be written for the integral form of the
equation of motion, as discussed in Valageas (2007b).

\subsection{Direct steepest-descent method}
\label{def-Direct-steepest-descent-method}

A first large-$N$ expansion is provided by a direct steepest-descent method,
which expands the path integral (\ref{Zjpsilambda}) around a saddle-point
as described in Valageas (2004, 2007a). It introduces auxiliary correlation and 
response functions $C_0$ and $R_0$ defined by evolution equations that happen to
be identical to the linearized equations (\ref{CLRL}). Thus, we have $C_0= C_L$
and $R_0 = R_L$. (This is specific to the hydrodynamic description, where 
$\lag\psi\rag=0$, and does not hold if we use the Vlasov equation of motion
to describe the system.) 
Then, the nonlinear correlation and response functions $C$ and $R$
can be obtained from the coupled linear equations
\beqa
\cO(x,z).C(z,y) &=& \Sigma(x,z).C(z,y) + \Pi(x,z).R^T(z,y) \label{Ceq} \\
\cO(x,z).R(z,y) &=& \delta_D(x-y) + \Sigma(x,z).R(z,y) \label{Rforward}
\eeqa
where we have introduced the self-energy matrices $\Sigma$ and $\Pi$ that
can be expressed in terms of $C_0$ and $R_0$ (hence in terms of $C_L$ and 
$R_L$). Equations (\ref{Ceq})-(\ref{Rforward}) are actually exact (they may 
be seen as the definition of the self-energy matrices) and the order $1/N^q$ of 
the large-$N$ expansion only enters the expression of the self-energy, which 
can be obtained from a series of diagrams. At one-loop order, there is only 
one contribution which reads as
\beqa
\!\!\Sigma(x,y) \!\!\! & = & \!\!\! 4 K_{\!s}(x;x_1,x_2) K_{\!s}(z;y,z_2) 
R_{\!L}\!(x_1,z) C_{\!L}\!(x_2,z_2) \label{S0eq} \\
\!\!\Pi(x,y) \!\!\! & = & \!\!\! 2 K_{\!s}(x;x_1,x_2) K_{\!s}(y;y_1,y_2) 
C_{\!L}\!(x_1,y_1) C_{\!L}\!(x_2,y_2) \label{P0eq} 
\eeqa

\subsection{The 2PI effective action method}
\label{def-2PI-effective-action-method}

A second large-$N$ expansion is provided by a 2PI effective action approach 
that involves a diagrammatic expansion of the double Legendre transform 
$\Gamma[\psi,C]$ of the generating functional of connected correlation functions 
(Valageas 2007a). 
This is also the generating functional of two-particle irreducible (2PI) diagrams, 
that is those that cannot be disconnected by cutting two lines.
This approach yields the same Schwinger-Dyson equations 
(\ref{Ceq})-(\ref{Rforward}), 
but the self-energy is now given in terms of the nonlinear two-point functions
$C$ and $R$. Therefore, this method does not introduce the auxiliary functions 
$C_0$ and $R_0$, but the system (\ref{Ceq})-(\ref{Rforward}) is now a pair of
nonlinear coupled equations ($\Sigma \propto R C$ and $\Pi\propto CC$ at 
one-loop order). As shown in Valageas (2007a), this nonlinearity is the key
to the small-scale damping of the nonlinear response $R$ (at one-loop order). 
At one-loop order, the self-energy has the same structure as for the
direct steepest-descent method and it reads as
\beqa
\Sigma(x,y) \!\!\! & = & \!\!\! 4 K_s(x;x_1,x_2) K_s(z;y,z_2) R(x_1,z) 
C(x_2,z_2) \label{Seq} \\
\Pi(x,y) \!\!\! & = & \!\!\! 2 K_s(x;x_1,x_2) K_s(y;y_1,y_2) C(x_1,y_1) 
C(x_2,y_2) \label{Peq} 
\eeqa
Thus, Eqs.(\ref{Seq})-(\ref{Peq}) are equal to Eqs.(\ref{S0eq})-(\ref{P0eq})
where the linear two-point functions are replaced by the nonlinear ones.
This simple correspondence would not hold at higher orders as the self-energies
associated with the steepest-descent approach would involve additional terms
as compared with the 2PI effective action approach.

Approximation schemes based on equations of the form (\ref{Ceq})-(\ref{Rforward}),
(\ref{Seq})-(\ref{Peq}), have been applied to many fields of theoretical physics,
such as quantum field theory (e.g. Zinn-Justin 1989; Berges 2002).
They can be derived in different manners and are also called 
``mode-coupling approximations'', in condensed matter and studies of glassy systems
(Bouchaud et al. 1996), or ``direct interaction approximation'' in studies of 
turbulent flows (Kraichnan 1961). A derivation inspired from works on turbulence
is presented in Taruya \& Hiramatsu (2007). An advantage of the derivation from the
path integral (\ref{Zjpsilambda}) is its relative simplicity and the well-known
diagrammatics associated with the expansion up to high orders. Moreover, it 
satisfies an exact variational principle (the equations above are associated with 
the saddle-point of the 2PI effective action $\Gamma$ and it is this action which 
is approximated by truncating an expansion over $1/N$, see Valageas 2004). Although 
we do not use this property here, it might prove useful at some point. On the other 
hand, the path integral (\ref{Zjpsilambda}) may serve as a basis for other 
approximation schemes.

\subsection{Comparison between various approaches}
\label{Comparison}

The standard Schwinger-Dyson equations (\ref{Ceq})-(\ref{Rforward}) can also
be obtained in a diagrammatic fashion from the 1PI effective action 
$\Gamma[\psi]$, also called the generating functional of proper vertices
(see any textbook on quantum field theory).
It is the generating functional of one-particle irreducible (1PI) diagrams, 
that is those that cannot be disconnected by cutting one line, and it is also
the Legendre transform of $W=\ln Z$, where $Z$ is given by 
Eq.(\ref{Zjpsilambda}).
Thus, Scoccimarro \& Crocce (2006a) describe a diagrammatic derivation of 
Eqs.(\ref{Ceq})-(\ref{Rforward}), starting from the equation of motion 
(\ref{OKsdef}) written in integral form (as in Eq.(\ref{psiseries}) below).

As noticed above, Eqs.(\ref{Ceq})-(\ref{Rforward}) alone are mostly a rewriting of 
the problem, and one still needs to specify a method to compute the self-energies 
$\Sigma$ and $\Pi$. We have described above two such methods, the direct 
steepest-descent approach, where the self-energy is written in terms of the 
linear two-point functions $C_L$ and $R_L$, and the 2PI effective action approach, 
where the self-energy is written in terms of the nonlinear two-point functions 
$C$ and $R$. Of course, both methods can also be obtained by diagrammatic means. 
For instance, Scoccimarro \& Crocce (2006a) present a loopwise expansion in terms 
of the nonlinear 
two-point functions that coincides with the 2PI effective action approach
(see also L'vov \& Procaccia 1995 for a detailed description of such a diagrammatic
procedure). 

Next, it is possible to reexpress the self-energy in
a more compact fashion by introducing ``dressed vertices'' that go beyond the
tree-order vertex $K_s$. That is, one can recognize three-point diagrams such as 
those of Fig.~\ref{figC3sd} (but written in terms of nonlinear two-point functions)
in parts of larger diagrams, and identify them with the expansion of properly 
defined dressed vertices (L'vov \& Procaccia 1995). However, this leads to 
another expansion scheme than the strict 2PI scheme, that would be more closely 
related to the 1PI effective action approach.

Indeed, looking for an expansion of the self-energies $\Sigma$ and $\Pi$ in terms of
the nonlinear $R$ and $C$ does not uniquely specify the practical procedure used to
compute correlation functions. One can always reorganize the diagrammatic series
in different fashions that lead to different truncations. 
For instance, as discussed in Sect.5 of Valageas (2004), and recalled above, it is
possible to derive the Schwinger-Dyson equations (\ref{Ceq})-(\ref{Rforward})
from the 1PI effective action $\Gamma[\psi]$. Then, the self-energy is given by an
exact expression that involves the proper vertex $\Gamma_3$ which ``renormalizes''
the bare vertex $K_s$ (its first two terms are given in 
Eqs.(\ref{Gamma3tree})-(\ref{Gamma3loop_sd}) below). Next, the proper vertices 
$\Gamma_q$
obey a hierarchy of equations, that must be truncated at some order $q$ (just like
the familiar BBGKY hierarchy obeyed by the correlations $C_q$ can be truncated
at some order to build an approximation scheme in the weakly nonlinear regime).
This approach provides another expansion scheme, written in terms of 
nonlinear two-point functions, that also contains infinite partial resummations
as compared with the standard perturbative expansion recalled in 
Sect.~\ref{Standard-perturbation-theory} below.

For completeness, let us point out that we always consider the limit 
$\eta_I\rightarrow -\infty$ (i.e. the initial conditions are set up at a redshift 
$z_I\rightarrow \infty$) and the response function $R(x_1,x_2)$ is defined as the
deviation from the mean of the fields at time $\eta_1$ with respect to a 
perturbation at an arbitrary earlier time $\eta_2$. By contrast, the calculations 
presented in Scoccimarro \& Crocce (2006a,b) consider the system defined by 
initial conditions set at a finite time $\eta_I$ (which defines the origin of time 
as $\eta_I=0$) and focus on the propagator $G$ defined from the 
restriction of $R(x_1,x_2)$ to $\eta_2=\eta_I$. However, the formalism of 
Scoccimarro \& Crocce (2006a,b) can also be applied to the case 
$\eta_I\rightarrow -\infty$ (e.g. Sect.~\ref{large-k-limit} below).

\subsection{Zeldovich dynamics}
\label{Zeldynamics}

The gravitational dynamics described in the previous sections is rather complex
and analytical computations often lead to complicated expressions. Therefore,
it can be interesting to investigate a closely related dynamics that obeys a
similar equation of motion but has a simpler structure. The Zeldovich dynamics
offers such an opportunity (Zeldovich 1970; Gurbatov et al. 1989). 
As described in Valageas (2007b), this dynamics can also be defined 
from the hydrodynamical equations (\ref{continuity1})-(\ref{Euler1}), where
the velocity potential is used in place of the gravitational potential.
Then, the equation of motion again has the form (\ref{OKsdef}). The quadratic
vertex $K_s$ is unchanged whereas the linear operator $\cO$ is changed to
\beq
\cO(x,x') = \left( \bea{cc} \frac{\pl}{\pl\eta} & -1 \\ 0
& \; \frac{\pl}{\pl\eta} -1 \ea \right) \delta_D(\bk-\bk') \,
\delta_D(\eta-\eta') .
\label{ZelOdef}
\eeq
This yields the same linear growing mode (\ref{psiL}) hence the same linear 
two-point correlation (\ref{CL}). However, the linear decaying mode is modified
by the change of $\cO$; this also modifies the linear response as
\beqa
\lefteqn{R_L(x_1,x_2) = \delta_D(\bk_1-\bk_2) \, \theta(\eta_1-\eta_2) } 
\nonumber \\
&& \times \left\{ e^{\eta_1-\eta_2} \left(\bea{cc} 0 & 1 \\ 0 & 1 \ea\right)
+ \left(\bea{cc} 1 & -1 \\ 0 & 0 \ea\right) \right\} .
\label{ZelRL}
\eeqa
Then, the path-integral formalism of Sect.~\ref{Path-integral} can be applied in
exactly the same manner and we obtain the same action (\ref{Spsilambda}),
where we must only substitute the new linear operator $\cO$ (Valageas 2007b).
Therefore, large-$N$ expansions again lead to Eqs.(\ref{Ceq})-(\ref{Peq}).
Thus, the Eulerian equations of motion have the same structure and any expansion
scheme can be applied in identical manner to both dynamics. However, in
Lagrangian space the Zeldovich dynamics is much simpler as its solution is
given by
\beq
\bx= \bq+D_+(\eta)\bs_{L0}(\bq) ,
\label{xq}
\eeq
where $\bq$ is the Lagrangian coordinate of the particle. That is, the 
trajectories of particles are given by the linear displacement field 
$\bs_L=D_+\bs_{L0}$. Going back to Eulerian space, one can derive from the
solution (\ref{xq}) an explicit expression for the matter density contrast.
Starting from the uniform matter density $\rhob$ at $t\rightarrow 0$, 
the conservation of matter gives, before orbit-crossing,
\beq
\rho(\bx)\d\bx = \rhob \d\bq \;\;\; \mbox{whence}  \;\;\; 
1+\delta(\bx) = \left|\det\left(\frac{\pl\bx}{\pl\bq}\right)\right|^{-1} .
\label{rhox}
\eeq
This also reads from Eq.(\ref{xq}) (Schneider \& Bartelmann 1995; 
Taylor \& Hamilton 1996) as
\beq
\delta(\bx,\eta) = \int\d\bq \; \delta_{D}[\bx-\bq-\bs_L(\bq,\eta)] - 1 .
\label{deltax}
\eeq
In Fourier space we obtain
\beq
\delta(\bk,\eta) = \int\frac{\d\bq}{(2\pi)^3} \; e^{-i\bk.\bq} 
\left[ e^{-i\bk.\bs_L(\bq,\eta)} - 1\right]  ,
\label{deltakq}
\eeq
whereas the linear displacement field is related to the linear density contrast 
by
\beq
\nabla_{\bq}.\bs_L = - \delta_L , \;\;\;  
\bs_L(\bk,\eta) = i \frac{\bk}{k^2} \delta_L(\bk,\eta) .
\label{sLdeltaL}
\eeq
From Eq.(\ref{deltakq}) one can also derive the exact expressions of the
nonlinear correlation $C$ and response $R$ (Schneider \& Bartelmann 1995; 
Taylor \& Hamilton 1996; Valageas 2007b).

\section{Standard perturbation theory}
\label{Standard-perturbation-theory}

In this section, we recall the standard perturbation theory applied to the
cosmological gravitational dynamics (Fry 1984; Goroff et al. 1986; 
Bernardeau et al. 2002; Scoccimarro 1997).
As explained in Sect.~\ref{Zeldynamics}, note that all expressions derived 
below apply as well to the Zeldovich dynamics.
In this approach, one expands the density and velocity fields over powers
of the linear growing mode (\ref{psiL}) as ($\psi_1=\delta$)
\beqa
\delta(\bk,D) & = & \sum_{n=1}^{\infty} D^n \int\d\bq_1 ..\bq_n 
\delta_D(\bq_1+..+\bq_n-\bk) \nonumber \\
&& \times F_n(\bq_1,..,\bq_n) \delta_{L0}(\bq_1) .. \delta_{L0}(\bq_n) ,
\label{Fn}
\eeqa
and ($\psi_2=-\nabla.\bv/\cH f$)
\beqa
\psi_2(\bk,D) & = & \sum_{n=1}^{\infty} D^n \int\d\bq_1 ..\bq_n 
\delta_D(\bq_1+..+\bq_n-\bk) \nonumber \\ 
&& \times E_n(\bq_1,..,\bq_n) \delta_{L0}(\bq_1) .. \delta_{L0}(\bq_n) ,
\label{En}
\eeqa
with $F_1=E_1=1$ (in Eqs.~(\ref{Fn})-(\ref{En}) and in the following $\bq_j$ 
is not a Lagrangian coordinate but a wavenumber associated with a linear 
growing mode). 
Here (and in most cases below) we use the linear growth factor
$D=e^{\eta}$ of Eq.(\ref{eta}) as the time coordinate. The factorization of the 
time dependence as $D^n$ is due to the approximation $\Om/f^2 \simeq 1$ (and 
it is exact for the Einstein-de Sitter cosmology). 
Then, by substituting the expansions (\ref{Fn})-(\ref{En}) into the
equation of motion (\ref{OKsdef}) one obtains
a recursion relation between $(F_n,E_n)$ and 
$(F_{n+1},E_{n+1})$. This enables one to compute the kernels $F_n$
and the density and velocity fields up to the required order.
Moreover, one can easily check that the kernels $F_n$ and $E_n$ obey the 
symmetry $F_n(-\bq_1,..,-\bq_n)=F_n(\bq_1,..,\bq_n)$ since $\delta(\bx)$ and 
$\theta(\bx)$
are real (whence $\delta(-\bk)=\delta(\bk)^*$). They also satisfy 
$F_n(\bq_1,..,\bq_n)=0$ for $\bq_1+..+\bq_n=0$ as $\delta(\bk=0)=0$ at all times
because of the conservation of matter.
It is convenient to introduce the symmetric vertices $F_n^s$ obtained by 
summing over all permutations of $(\bq_1,..,\bq_n)$ as
\beq
F_n^s(\bq_1,..,\bq_n)= \frac{1}{n!} \sum_{\rm perm.} F_n(\bq_1,..,\bq_n) .
\label{Fns}
\eeq

\begin{figure}[htb]
\begin{center}
\epsfxsize=8.8 cm \epsfysize=2.6 cm {\epsfbox{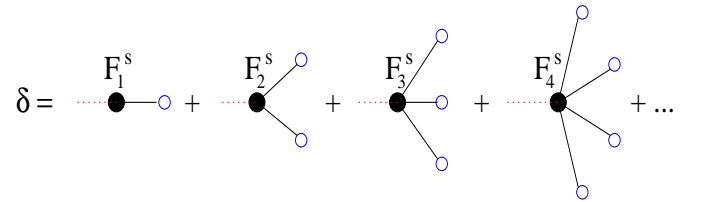}}
\end{center}
\caption{The diagrams associated with the standard perturbative expansion 
(\ref{Fn}) of the density field $\delta(\bk,\eta)$ over powers of the linear 
growing mode $\delta_L$. The filled circles are the vertices $F_n^s$
attached to $n$ linear modes $\delta_L$ shown by the open circles. The left
dotted legs are attached to the external wavenumber $\bk$ of the
nonlinear density $\delta(\bk,\eta)$.}
\label{figdeltastandard}
\end{figure}

The diagrammatic expansion associated with (\ref{Fn}) is shown in 
Fig.~\ref{figdeltastandard}.
The diagrams are very simple to draw but the complexity is hidden in the
vertices $F_n^s$ that are obtained by recursion from lower-order vertices.
Note that one needs to compute a new vertex at each order of the expansion.

\begin{figure}[htb]
\begin{center}
\epsfxsize=8.5 cm \epsfysize=1.25 cm {\epsfbox{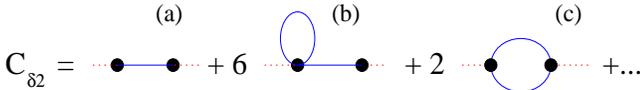}}
\end{center}
\caption{The diagrams associated with the standard perturbative expansion 
(\ref{C2standard}) of the two-point density correlation $C_{\delta2}$ over 
powers of the linear power spectrum $P_L$. The big dots are the 
vertices $F_n^s$ and the solid lines are the linear power spectrum $P_L$. 
We also give the overall multiplicity factor of each diagram, up to one-loop
order (i.e. $P_L^2$ for $C_{\delta2}$).}
\label{figC2standard}
\end{figure}

Then, from the expansion (\ref{Fn}), one obtains for any correlation function
a perturbative expansion over powers of the linear power spectrum $P_L$ 
by using Wick's theorem. 
In particular, the two-point density correlation $C_{\delta2}$ reads (up to
order $\delta_L^4$ whence up to order $P_L^2$) as
\beqa
C_{\delta2} = \lag\delta\delta\rag_c & = & \lag\delta^{(1)}\delta^{(1)}\rag + 
\lag\delta^{(3)}\delta^{(1)}\rag + \lag\delta^{(1)}\delta^{(3)}\rag \nonumber \\
&& + \lag\delta^{(2)}\delta^{(2)}\rag +  ... \, ,
\label{C2standard}
\eeqa
where $\delta^{(n)} \propto F_n^s \delta_L^n$ is the term of order $n$
in the expansion (\ref{Fn}). The diagrammatic expansion associated with 
Eq.(\ref{C2standard}) is displayed in Fig.~\ref{figC2standard}. The
order of the diagrams over $P_L$ scales with the number of loops $\ell$ as
$P_L^{\ell+1}$. Up to order $P_L^2$, we can write the equal-time nonlinear
power spectrum $P(k)$ as
\beq
P(k) = P^{\rm tree}(k) + P^{\rm 1loop}(k) ,
\label{Ptree1loop}
\eeq
with
\beq
P^{\rm tree} = P^{(a)}= P_L(k) ,  \;\;\; 
P^{\rm 1loop} = P^{(b)}+P^{(c)} .
\label{Ptree}
\eeq
Thus, the tree diagram (a) of Fig.~\ref{figC2standard} is merely equal to the 
linear power spectrum $P_L$ whereas the one-loop contributions $P^{(b)}$ and 
$P^{(c)}$ of diagrams (b) and (c) are given by
\beqa
P^{(b)}(k) & = & 6 P_L(k) \int\d\bq P_L(q) F_3^s(\bq,-\bq,\bk) ,
\label{P1loopb}\\
P^{(c)}(k) & = & 2 \int\d\bq P_L(q) P_L(|\bk-\bq|) F_2^s(\bq,\bk-\bq)^2 .
\label{P1loopc}
\eeqa
We give in Fig.~\ref{figC2standard} the overall multiplicity factor of each 
diagram. Thus, the factor 6 for diagram (b) in Fig.~\ref{figC2standard} 
reads as $6=2\times 3$, where the factor 2 comes from Eq.(\ref{C2standard}) 
(associated with the two contributions 
$\lag\delta^{(3)}\delta^{(1)}\rag+\lag\delta^{(1)}\delta^{(3)}\rag$)
whereas the factor 3 comes from the three possible Gaussian pairings which 
can be built for each term $\lag\delta^{(3)}\delta^{(1)}\rag$ (Wick's theorem).
In this article, we focus on equal-time statistics, such as the usual
matter power spectrum $P(k)$, and each factor of $P_L$ gives rise to a 
time-dependent factor $D^2$, where $D$ is the linear growth factor
at the time of interest, but the analysis applies as well to different-time
correlations.

\begin{figure}[htb]
\begin{center}
\epsfxsize=8.6 cm \epsfysize=5.3 cm {\epsfbox{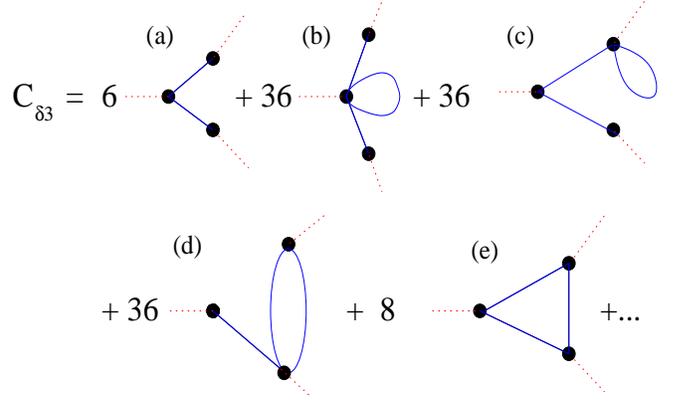}}
\end{center}
\caption{The diagrams associated with the standard perturbative expansion 
(\ref{C3standard}) of the three-point density correlation $C_{\delta3}$ over 
powers of the linear power spectrum $P_L$ (shown up to one-loop order, 
i.e. $P_L^3$ for $C_{\delta3}$). The symbols are as in 
Fig.~\ref{figC2standard}.}
\label{figC3standard}
\end{figure}

On the other hand, the three-point density correlation $C_{\delta3}$ reads (up to
order $P_L^3$) as
\beqa
C_{\delta3} & = & \lag\delta^{(2)}\delta^{(1)}\delta^{(1)}\rag + 2 \, \rm{perm.}
+ \lag\delta^{(4)}\delta^{(1)}\delta^{(1)}\rag + 2 \, \rm{perm.} \nonumber \\
&& + \lag\delta^{(3)}\delta^{(2)}\delta^{(1)}\rag + 5 \, \rm{perm.}
+ \lag\delta^{(2)}\delta^{(2)}\delta^{(2)}\rag + ... \, ,
\label{C3standard}
\eeqa
where each term ``$n$ perm.'' stands for $n$ terms obtained by permutations
over the previous term, such as
\beqa
\lag\delta^{(2)}\delta^{(1)}\delta^{(1)}\rag + 2 \, \rm{perm.} & = &
\lag\delta^{(2)}\delta^{(1)}\delta^{(1)}\rag 
+ \lag\delta^{(1)}\delta^{(2)}\delta^{(1)}\rag \nonumber \\
&& + \lag\delta^{(1)}\delta^{(1)}\delta^{(2)}\rag .
\eeqa
The associated diagrams are shown in Fig.~\ref{figC3standard}. For the 
equal-time bispectrum $B$ defined by
\beq
C_{\delta3}(\bk_1,\bk_2,\bk_3;\eta)= \delta_D(\bk_1+\bk_2+\bk_2) 
B(k_1,k_2,k_3;\eta) ,
\label{Bdef}
\eeq
we obtain at tree-order the well-known result
\beq
B^{\rm tree}(k_1,k_2,k_3) \! = 2 P_L(k_2) P_L(k_3) F_2^s(\bk_2,\bk_3) 
+ 2 \, \rm{perm.}
\label{Btreestandard}
\eeq
and at one-loop order:
\beq
B^{\rm 1loop}(k_1,k_2,k_3) = B^{(b)}+B^{(c)}+B^{(d)}+B^{(e)} ,
\eeq
with (Scoccimarro 1997):
\beqa
B^{(b)} & = & 12 P_L(k_2) P_L(k_3) \int\! \d\bq P_L(q) 
F_4^s(-\bq,\bq,\bk_2,\bk_3) \nonumber \\ 
&& + 2 \, \rm{perm.} ,
\eeqa
\beqa
B^{(c)} & = & 6 P_L(k_2) P_L(k_3) F_2^s(\bk_2,\bk_3) \int\!\d\bq P_L(q)
\nonumber \\ 
&& \times F_3^s(\bq,-\bq,\bk_2) + 5 \, \rm{perm.} ,
\eeqa
\beqa
\lefteqn{ B^{(d)} = 6 P_L(k_1) \int\!\d\bq_1\d\bq_2 \; 
\delta_D(\bq_1+\bq_2-\bk_2) P_L(q_1) } \nonumber \\ 
&& \times P_L(q_2) F_2^s(\bq_1,\bq_2) F_3^s(\bq_1,\bq_2,\bk_1) 
+ 5 \, \rm{perm.} ,
\eeqa
\beqa
B^{(e)} & = & 8 \int \! \d\bq_1\d\bq_2\d\bq_3 \; \delta_D(\bq_1+\bq_2-\bk_1)  
\nonumber \\ 
&& \times \delta_D(\bq_1+\bq_3+\bk_2) P_L(q_1) P_L(q_2) P_L(q_3)
\nonumber \\ 
&& \times F_2^s(\bq_1,\bq_2) F_2^s(\bq_1,\bq_3) F_2^s(-\bq_2,\bq_3) .
\label{Bestandard}
\eeqa

The derivation of higher-order correlations proceeds in exactly the same manner,
by writing $C_{\delta p}=\lag \delta_1 .. \delta_p \rag$, expanding each factor
$\delta_j$ through Eq.(\ref{Fn}), and performing the Gaussian average with
Wick's theorem.

\section{Expansion over cubic interaction term}
\label{Expansion-over-cubic}

In this section, we recall how to recover the results of the standard 
perturbation theory from the path integral (\ref{Zjpsilambda}). As 
described in Valageas (2004, 2007a), the usual perturbative expansion presented
in Sect.~\ref{Standard-perturbation-theory} over powers of the linear growing 
mode $\delta_L$ can also be seen as an expansion over powers of the 
interaction vertex $K_s$ of the equation of motion (\ref{OKsdef}). (Formally 
one may add a coupling constant $g$ as $K_s\rightarrow g K_s$ and expand over 
powers of $g$.) Indeed, by substituting the expansions 
(\ref{Fn})-(\ref{En}) into the equation of motion (\ref{OKsdef}) to derive the
recursion relation between $(F_n,E_n)$ and $(F_{n+1},E_{n+1})$, we can see
that $F_n \propto K_s^{n-1}$ whence $\delta^{(n)} \propto K_s^{n-1}$ (by which we
actually mean $\delta^{(n)} \propto g^{n-1}$). Therefore, the expansion over
powers of $\delta_L$, or over powers of $P_L$ for averaged quantities such
as correlation functions, is identical to the expansion over powers of $K_s$.
Then, this latter expansion can be directly obtained from the path integral 
(\ref{Zjpsilambda}) by expanding over the cubic part $\lambda.K_s.\psi\psi$ of
the action $S[\psi,\lambda]$ of Eq.(\ref{Spsilambda}). 
This equivalence applies to all higher-order correlation functions.

\begin{figure}[htb]
\begin{center}
\epsfxsize=7.6 cm \epsfysize=0.45 cm {\epsfbox{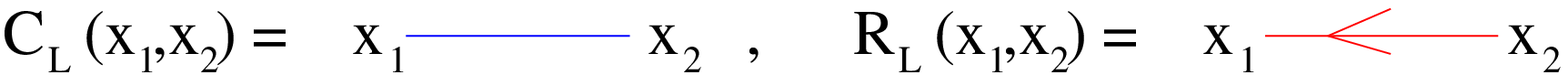}}
\end{center}
\caption{The diagrammatic symbols used for the two-point functions 
$C_L=\lag\psi\psi\rag_0$ and $R_L=\lag\psi\lambda\rag_0$. The third possible 
two-point function vanishes: $\lag\lambda\lambda\rag_0=0$. The propagator 
$R_L=\lag\psi(x_1)\lambda(x_2)\rag_0$ is distinguished from $C_L$ by the 
addition of an arrow which marks the direction of propagation, from the
response-field $\lambda(x_2)$ towards the physical field $\psi(x_1)$, 
that is, from time $D_2$ to time $D_1>D_2$ as enforced by the Heaviside factor 
of Eq.(\ref{RL}). (This is due to causality.)}
\label{figCLRLdef}
\end{figure}

\begin{figure}[htb]
\begin{center}
\epsfxsize=8.5 cm \epsfysize=3.1 cm {\epsfbox{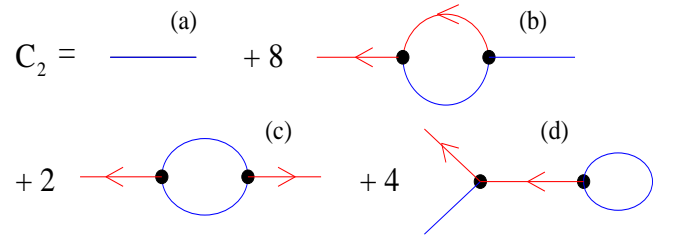}}
\end{center}
\caption{The diagrams associated with the expansion (\ref{C2Z}) over powers 
of $K_s$ for the two-point correlation $C_2$. The big dots are the
three-leg vertex $K_s$. The solid lines are the
linear propagators $C_L$ and $R_L$ as in Fig.~\ref{figCLRLdef}. Note that the
symbols are different from those used in 
Figs.~\ref{figdeltastandard}-\ref{figC3standard} for the standard perturbative
expansions. Again we only display the diagrams obtained up to one-loop order
(i.e. $P_L^2$ for $C_2$).}
\label{figC21loop}
\end{figure}

Thus, we obtain for the
two-point function
\beqa
\lefteqn{ C_2(x_1,x_2) = \lag\psi(x_1)\psi(x_2)\rag \! = \!\! 
\int \! [\d\psi] [\d\lambda] \psi(x_1) \psi(x_2) e^{- S[\psi,\lambda]} } 
\nonumber \\
&& \!\! = \! \int \! [\d\psi] [\d\lambda] \psi(x_1) \psi(x_2) 
\left( \! 1 \! + \! \lambda K_s \psi^2 \! + \! \frac{(\lambda K_s \psi^2)^2}{2} 
\! + \! .. \! \right) e^{-S_0} \nonumber \\
&& \!\! = \! \lag \psi(x_1)\psi(x_2) \left( 1 + \lambda K_s \psi^2 + 
\frac{(\lambda K_s \psi^2)^2}{2} + .. \right) \rag_0
\label{C2Z}
\eeqa
where $S_0=\lambda.\cO.\psi-\frac{1}{2} \lambda.\Delta_I.\lambda$ 
is the quadratic part 
of the action $S[\psi,\lambda]$ and $\lag .. \rag_0$ is the average with respect 
to this Gaussian action $S_0$. Then, one can use Wick's theorem to compute the
Gaussian path integral defined by $S_0$. Since the action $S_0$ is equal to the
action $S$ with $K_s=0$, it actually corresponds to the linear dynamics. Thus
we have (as can also be explicitly checked):
\beqa
\lag\psi\rag_0=\lag\lambda\rag_0=0, \;\;\;\;
\lag\psi(x_1)\psi(x_2)\rag_0 = C_L(x_1,x_2) , \label{WickC0}\\
\lag\psi(x_1)\lambda(x_2)\rag_0 = R_L(x_1,x_2) , \;\;\;\;
\lag\lambda(x_1)\lambda(x_2)\rag_0 = 0 . \label{WickR0}
\eeqa
In terms of diagrams, we note these linear propagators as in 
Fig.~\ref{figCLRLdef}. We add an arrow to the lines associated with the 
propagator $R_L=\lag\psi(x_1)\lambda(x_2)\rag_0$ to mark the arrow of time 
due to causality ($D_1>D_2$), which is enforced by the Heaviside factor of 
Eq.(\ref{RL}). Thus, the
response $R_L$ runs from the response-field $\lambda$ towards the physical field
$\psi$. We could have added an out-going arrow at both ends of $C_L$ 
(both ends are associated with fields $\psi$) but for simplicity we put no 
arrow at all. Since each vertex $K_s$ is associated with one response-field 
$\lambda$ and two physical fields $\psi$, through $\lambda K_s \psi\psi$ as in 
the action (\ref{Spsilambda}), each vertex $K_s$ shown by a big dot in 
Fig.~\ref{figC21loop} and hereafter must be connected to one outgoing line 
(hence a response $R_L$) and to two incoming lines (which can be $R_L$ or $C_L$ 
with the understanding that $C_L$ can be read with two outgoing lines, that is, 
one at each end).

Then, from Eq.(\ref{C2Z}) the two-point correlation
reads up to one-loop order as
\beq
C_2=C_2^{\rm tree} + C_2^{\rm 1loop} + ...  , \;\;\; \mbox{with} \;\;\;
C_2^{\rm tree}= C_L ,
\label{C2tree}
\eeq
and
\beqa
\lefteqn{ C_2^{\rm 1loop} = \frac{1}{2} \lag \psi(x_1)\psi(x_2) 
(\lambda K_s \psi\psi)^2 \rag_0 } \nonumber \\
& = &  8 R_L (K_s R_L C_L K_s) C_L + 2 R_L (K_s C_L C_L K_s) R_L \nonumber \\
&& + 4 R_L C_L K_s (R_L K_s C_L) .
\label{C21loop}
\eeqa
The diagrams associated with Eqs.(\ref{C2tree})-(\ref{C21loop}) are
shown in Fig.~\ref{figC21loop}. In Eq.(\ref{C21loop}) and in 
Fig.~\ref{figC21loop} we have not shown five additional terms and diagrams
that vanish because they contain a two-point function $\lag\lambda\lambda\rag_0$
or a closed loop over response functions $R_L$. Indeed, the former vanishes
because of Eq.(\ref{WickR0}) (and more generally Eq.(\ref{Rpsilambda}))
whereas closed loops over $R_L$ vanish because of the Heaviside factors
$\theta(\eta_1-\eta_2)$ associated with $R_L$ as in Eq.(\ref{RL}).
Moreover, it happens that diagram (d) of Fig.~\ref{figC21loop}
also vanishes because the right vertex $K_s$ attached to the closed loop
over $C_L$ is of the form $K_s(0;\bq,-\bq)$ (where we only write the dependence
on wavenumbers), which is equal to zero from the explicit expressions 
(\ref{gamma1})-(\ref{gamma3}) of the vertices $\gamma^s$.
Then, one can check that the diagram (b) gives back diagram (b) of 
Fig.~\ref{figC2standard} and Eq.(\ref{P1loopb}) whereas diagram (c) gives back 
diagram (c) of Fig.~\ref{figC2standard} and Eq.(\ref{P1loopc}).
Thus, we indeed recover the standard perturbative results.
Nevertheless, these two equivalent expansions have different 
diagrammatic structures. Thus, the expansion of Fig.~\ref{figC21loop},
derived from the path integral (\ref{Zjpsilambda}), only involves the
cubic vertex $K_s$ but two propagators $C_L$ and $R_L$ (with four indices) 
because of the two two-component fields $\psi$ and $\lambda$. By contrast,
the standard expansion of Fig.~\ref{figC2standard} involves new vertices
$F_n^s$ of increasing order as we include higher-order terms but only one
two-point function: the linear density power spectrum $P_L$.
However, all kernels $F_n$ can be written in terms of the building blocks $R_L$ and
$K_s$, as in Fig.~\ref{figdiagtKs} and Eqs.(\ref{psiseries})-(\ref{tKsdef}) below,
which are also the basic blocks of the expansion in Fig.~\ref{figC21loop}.

\begin{figure}[htb]
\begin{center}
\epsfxsize=8.5 cm \epsfysize=9.9 cm {\epsfbox{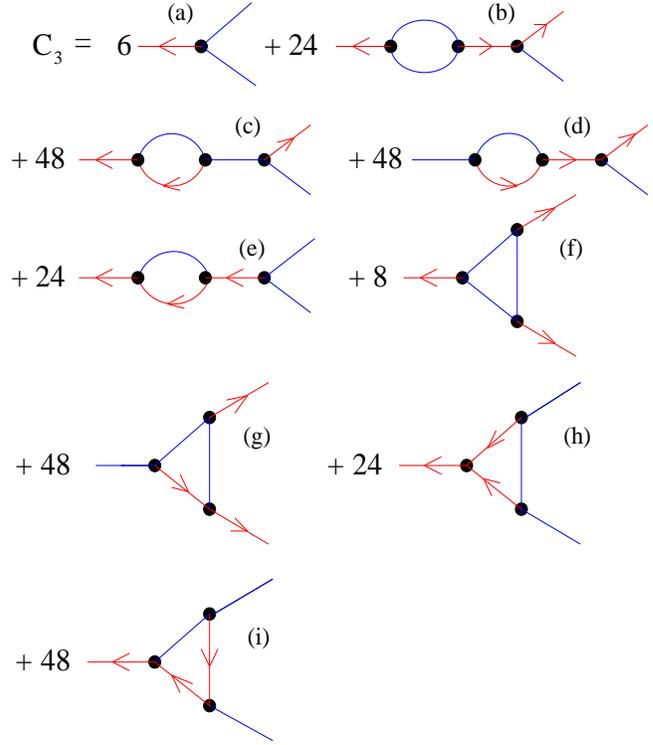}}
\end{center}
\caption{The diagrams associated with the expansion (\ref{C3Z}) over powers 
of $K_s$ for the three-point correlation $C_3$. The symbols are as in 
Fig.~\ref{figC21loop}. Again we only display the diagrams obtained up to 
one-loop order (i.e. $P_L^3$ for $C_3$).}
\label{figC31loop}
\end{figure}

Higher-order correlation functions can be obtained as in Eq.(\ref{C2Z}) 
by expanding over $K_s$. This gives for the three-point correlation:
\beq
C_3 \! = \lag \psi(x_1)\psi(x_2) \psi(x_2)\left( \! \lambda K_s \psi^2 + 
\frac{(\lambda K_s \psi^2)^3}{6} + .. \! \right) \rag_0 .
\label{C3Z}
\eeq
Thus, we obtain up to 1-loop order
\beq
C_3^{\rm tree} = R_L K_s C_L C_L + 5 \, {\rm perm.} ,
\label{C3tree}
\eeq
and
\beqa
\lefteqn{ C_3^{\rm 1loop} = (24+48+48) (K_s^2 R_L^2 C_L^2) K_s R_L C_L }
\nonumber \\
&& \!\!\! + 24 (K_s^2 R_L^3 C_L) K_s C_L^2 + 8 R_L^3 (K_s^3 C_L^3) \nonumber \\
&& \!\!\! + 48 R_L^2 C_L (K_s^3 R_L C_L^2) 
+ (24+48) R_L C_L^2 (K_s^3 R_L^2 C_L) .
\label{C3loop}
\eeqa
We show the diagrams associated with Eqs.(\ref{C3tree})-(\ref{C3loop})
in Fig.~\ref{figC31loop}. We can note that the first four one-loop diagrams, 
(b), (c), (d), and (e),
are 1-particle reducible: they can be disconnected by cutting the two-point
function which connects the ``bubble'' to the rightmost vertex $K_s$.
The comparison with Fig.~\ref{figC21loop} shows that they correspond
to the perturbative corrections of the two-point functions. Thus, the first 
three one-loop diagrams can be obtained from the tree diagram (a) 
by inserting the one-loop corrections (b) and (c) of Fig.~\ref{figC21loop} 
into the two-point correlations $C_L$ whereas the
fourth one can be obtained by inserting the one-loop correction of the
response $R$. The last four one-loop diagrams (f), (g), (h), and (i),  
of Fig.~\ref{figC31loop},
correspond to perturbative corrections to the vertex $K_s$ itself.
Note that whereas the last two diagrams respect the structure of the
bare vertex $K_s$ (attached to one response field $\lambda$ and two fields
$\psi$) the other two diagrams bring up vertices attached to three fields
$\lambda$ or to two fields $\lambda$ and one field $\psi$. Thus all possible
configurations are generated beyond tree-level for the ``renormalized'' vertex.
As for the two-point function diagrams shown in Fig.~\ref{figC21loop}, 
we have only displayed in Fig.~\ref{figC31loop} the non-vanishing diagrams.
For instance, the renormalized vertex which would arise from
a closed triangular loop over three responses $R_L$ vanishes because of the
Heaviside factors associated with the response $R_L$ (that express causality).

Again, one can check that Eqs.(\ref{C3tree})-(\ref{C3loop}) give back the
results obtained from standard perturbation theory in Eq.(\ref{C3standard})
and Fig.~\ref{figC3standard}. Thus, from the expression (\ref{Ksdef}) of the 
bare vertex $K_s$, the tree-order diagram (\ref{C3tree}) reads as
\beqa
\lefteqn{ \! C_3^{\rm tree}(x_1,x_2,x_3) \! = \! \delta_D(\bk_1+\bk_2+\bk_3) 
\!\! \int_{-\infty}^{\eta_1} \!\!\!\!\! \d\eta_1' 
R_{Li_1 i_1'}(k_1;\eta_1,\eta_1') } \nonumber \\
&& \times C_{Li_2 i_2'}(k_2;\eta_2,\eta_1') C_{Li_3 i_3'}(k_3;\eta_3,\eta_1')
\gamma^s_{i_1';i_2',i_3'}(\bk_2,\bk_3)
\nonumber \\
&& + \; 5 \; {\rm perm.}
\label{C3tree1}
\eeqa
Substituting the expressions of the linear two-point functions in the
right hand side of Eq.(\ref{C3tree1}) and performing the integral over the time
$\eta_1'$ we recover Eq.(\ref{Btreestandard}) for $i_1=i_2=i_3=1$.

Higher-order correlations $C_p$ are computed in the same fashion, by inserting 
the factor $(\psi_1 .. \psi_p)$ in front of the exponential in 
Eq.(\ref{Zjpsilambda}) and expanding over the cubic part of the action.
The same technique also provides higher-order response functions, such as
\beqa
R_{\psi\psi\lambda} & \equiv & \lag \psi(x_1) \psi(x_2) \lambda(x_3) \rag 
= \left. \frac{\delta}{\delta \zeta(x_3)} \right|_{\zeta=0} 
\lag \psi(x_1) \psi(x_2) \rag \nonumber \\
& = & \left. \frac{\delta C(x_1,x_2)}{\delta \zeta(x_3)} \right|_{\zeta=0} .
\label{R21}
\eeqa
Thus, $R_{\psi\psi\lambda}$ measures the response of the two-point correlation
$C$ to an infinitesimal external perturbation $\zeta$ added to the right hand 
side of the equation of motion (\ref{OKsdef}). It is also related to the 
cross-correlation of the nonlinear field $\psi$ with the initial condition
$\psi_I$ at a finite time $\eta_I$ as
\beq
\lag \psi(x_1)\psi(x_2) \psi_I(\overline{x}_3) \rag = 
R_{\psi\psi\lambda}(x_1,x_2;\overline{x},\eta_I) \times 
G_I(\overline{x},\overline{x}_3) ,
\label{R21GI}
\eeq
where the product does not involve any integration over time and we note
$\overline{x}=(\bk,i)$, see Eq.(49) of Valageas (2007b) for details in the 
case of the two-point response function.

Note that throughout this article we take the limit $\eta_I\rightarrow -\infty$
(i.e. $D_I\rightarrow 0$). This applies both to the standard perturbation theory
of Sect.~\ref{Standard-perturbation-theory} (and the expansion 
(\ref{psiseries}) below), which allows us to write the solution of the equations
of motion as an expansion over the linear growing mode, and to the methods 
derived from the path integral (\ref{Zjpsilambda}). It is possible to
apply the path-integral formalism while keeping $\eta_I$ finite, 
see Valageas (2007b), but computations simplify in the limit 
$\eta_I\rightarrow -\infty$. In a similar fashion, keeping $\eta_I$ finite
would add decaying terms to the standard expansion (\ref{Fn}).

\section{Large-$N$ methods}
\label{LargeN}

We now describe in this section how to obtain the three-point density
and velocity correlations from the large-$N$ methods
introduced in Sect.~\ref{Path-integral}.
From the path-integral formalism (\ref{Zjpsilambda})-(\ref{Spsilambda}) the 
three-point correlation $\cC_3$ can be written as (Zinn-Justin 1989; 
Valageas 2007a)
\beqa
\lefteqn{ \cC_3(x_1,x_2,x_3) = - \cC(x_1,x_1') \cC(x_2,x_2') \cC(x_3,x_3') 
\Gamma_3(x_1',x_2',x_3') } \nonumber \\
\label{Gamma3}
\eeqa
where $\Gamma_3$ is the symmetric dressed three-point vertex (the so-called
three-point proper vertex of the 1PI effective action $\Gamma[\psi]$). 
Here we put on 
the same footing the fields $\psi$ and $\lambda$ so that two-point functions 
$\cC$ correspond to all three possibilities $\lag\psi\psi\rag=C$, 
$\lag\psi\lambda\rag=R$, and $\lag\lambda\lambda\rag=0$, and similarly for the
three-point function $\cC_3$. Then, the proper vertex $\Gamma_3$ can be
obtained from a series of three-leg diagrams, up to the required order over
$1/N$.

In this article we restrict ourselves to the three-point correlation 
$C_3(x_1,x_2,x_3)$ defined as
\beq
C_3(x_1,x_2,x_3)= \lag\psi(x_1)\psi(x_2)\psi(x_3)\rag .
\label{C3def}
\eeq
It describes both density and velocity correlations, as well as their
cross-correlations. We do not study in this paper three-point functions such as
$\lag\psi\psi\lambda\rag$ of Eq.(\ref{R21}), associated with the functional 
derivative of the two-point correlation with respect to an external noise
(and in particular with respect to the initial conditions).
Of course, higher-order correlation and response functions can be obtained in
a similar manner, from the two-point function $C$ and $R$ and the higher-order
proper vertices $\Gamma_p$. However, since for practical purposes higher-order
statistics are increasingly noisy we do not go beyond three-point functions
in this paper.

Note that the structure of the large-$N$ expansions is somewhat 
different from the expansions described in 
Sects.~\ref{Standard-perturbation-theory}-\ref{Expansion-over-cubic}.
Indeed, contrary to the latter cases, the contribution associated with a given
diagram is not fixed but depends on the order up to which the expansion is
performed. This arises from the fact that each diagram depends on the nonlinear
two-point functions $R$ and $C$ (more precisely the approximated nonlinear 
two-point functions computed at this order). Thus, the diagrammatic series
actually gives an implicit equation for $R$ and $C$, as in 
Eqs.(\ref{Ceq})-(\ref{Rforward}) where the right hand side depends on $R$ and $C$.

Then, the order of these expansions is set by the order of the diagrams
kept in the self-energies $\Sigma$ and $\Pi$. At tree-order, we merely have
$\Sigma=0$ and $\Pi=0$, at one-loop order we have either Eqs.(\ref{S0eq})-(\ref{P0eq})
or Eqs.(\ref{Seq})-(\ref{Peq}), and at higher orders we would need to include 
higher-order diagrams (that would no longer look identical for the steepest-descent
and 2PI methods).

\subsection{Direct steepest-descent method}
\label{Direct-steepest-descent}

We first consider in this section the direct steepest-descent method of 
Sect.~\ref{def-Direct-steepest-descent-method}.
At tree-order, the two-point functions are equal to the linear two-point
functions (Valageas 2004, 2007a) whereas the proper vertex $\Gamma_3$ is equal to 
the bare vertex of the action $S[\psi,\lambda]$ which yields from 
Eq.(\ref{Spsilambda})
\beq
\Gamma_3^{\rm tree}(x_1,x_2,x_3) = - K_s(x_1;x_2,x_3) + 5 \; {\rm perm.} ,
\label{Gamma3tree}
\eeq
with a multiplicity factor $6=3!$ which corresponds to all permutations of
the triplet $(x_1,x_2,x_3)$. Therefore, at tree-order we recover exactly the
tree-diagrams (a) of Figs.~\ref{figC21loop}-\ref{figC31loop}, obtained by
expanding over the nonlinear coupling $K_s$, hence the lowest-order results of
standard perturbation theory.

\begin{figure}[htb]
\begin{center}
\epsfxsize=6 cm \epsfysize=1. cm {\epsfbox{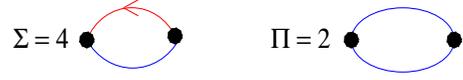}}
\end{center}
\caption{The one-loop diagrams for the self-energies $\Sigma$ and $\Pi$ obtained
from the direct steepest-descent expansion.}
\label{figSigPi1loop}
\end{figure}

\begin{figure}[htb]
\begin{center}
\epsfxsize=6.1 cm \epsfysize=2.4 cm {\epsfbox{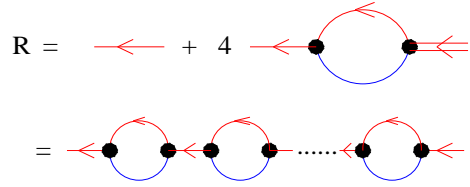}}
\end{center}
\caption{The diagrams obtained at one-loop order for the response $R$ from the 
direct steepest-descent expansion. The first equality corresponds to the implicit
equation (\ref{RSRLint}), where we note by double lines the nonlinear response
$R$. The second equality shows the infinite series of bubble diagrams resummed
by this large-$N$ method, as in Eq.(\ref{RSRL}).}
\label{figRsd}
\end{figure}

At one-loop order, the two-point functions are obtained from the system 
(\ref{Ceq})-(\ref{Rforward}) together with Eqs.(\ref{S0eq})-(\ref{P0eq})
for the self-energy. The diagrammatic expression of the self-energy is shown
at one-loop order in Fig.~\ref{figSigPi1loop}. Then, the solution of 
Eq.(\ref{Rforward}) can be written as
\beqa
R & = & R_L + R_L.\Sigma.R \label{RSRLint} \\
& = & R_L + R_L.\Sigma.R_L + R_L.\Sigma.R_L.\Sigma.R_L + ...
\label{RSRL}
\eeqa
In the first line we used the second Eq.(\ref{CLRL}) to obtain the integral 
form of Eq.(\ref{Rforward}) and in the second
line we wrote the solution as a series over powers of $\Sigma$. 
Equation (\ref{RSRL}) explicitly shows that at ``one-loop order'' the 
steepest-descent method has performed the resummation of the infinite series 
of bubble diagrams displayed in Fig.~\ref{figRsd}.

\begin{figure}[htb]
\begin{center}
\epsfxsize=8.5 cm \epsfysize=4.1 cm {\epsfbox{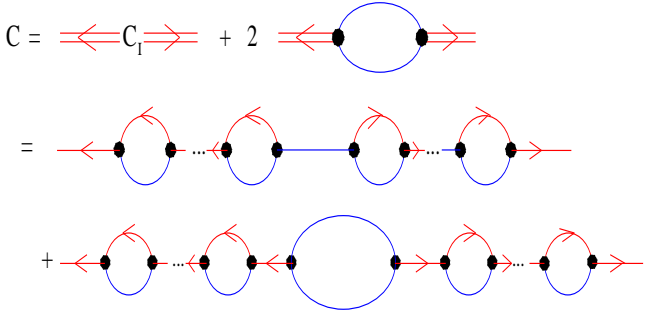}}
\end{center}
\caption{The diagrams obtained at one-loop order for the two-point correlation 
$C$ from the direct steepest-descent expansion. The first equality corresponds 
to Eq.(\ref{Cint}), where we note by double lines the nonlinear response
$R$. The second equality shows the two infinite series of bubble diagrams 
resummed by this large-$N$ method, in terms of the linear functions $C_L$ 
and $R_L$.}
\label{figCsd}
\end{figure}

Let us stress here that for the large-$N$ methods, the loop order of the 
approximation refers to the truncation order of the infinite series of diagrams 
obtained for the self-energies, and not to the truncation order of the correlation 
and response functions. Indeed, any finite order in terms of the self-energy
diagrams automatically leads to an infinite partial resummation over all orders
for the diagrams associated with the correlation and response functions.
However, these resummations are only complete up to the same order as for the
self-energy (we miss some of the higher-order diagrams). 
Thus, at one-loop order, taking into account the only two diagrams of 
Fig.~\ref{figSigPi1loop} for the self-energies actually gives rise to the
infinite series shown in Figs.~\ref{figRsd}-\ref{figCsd} for the physical
two-point functions $R$ and $C$, which include contributions that contain
an arbitrary number of loops. However, these partial resummations are complete
only up to one-loop, that is they miss some two-loop and higher-order diagrams,
such as the one of Fig.~\ref{figChighk} for $C$.

Next, the solution of Eq.(\ref{Ceq}) can be written as (Valageas 2004, 2007a)
\beq
C = R \times C_L(\eta_I) \times R^T + R . \Pi . R^T ,
\label{Cint}
\eeq
where the first product does not contain any integration over time,
\beqa
\lefteqn{ R \times C_L(\eta_I) \times R^T = \delta_D(\bk_1+\bk_2) 
\sum_{i_1'i_2'} R_{i_1i_1'}(k_1;\eta_1,\eta_I) } \nonumber \\
&& \times C_{Li_1'i_2'}(k_1;\eta_I,\eta_I) R_{i_2i_2'}(k_1;\eta_2,\eta_I) ,
\label{crossdef}
\eeqa
and we let the initial time go to the infinite past $\eta_I\rightarrow -\infty$.
Moreover, we can note that in the linear regime we have
\beq
C_L = R_L \times C_L(\eta_I) \times R_L^T ,
\label{CLRLcross}
\eeq
as can be checked from the explicit expressions of $C_L$ and $R_L$.
From Eqs.(\ref{RSRL})-(\ref{CLRLcross}) and Fig.~\ref{figRsd} we see that the 
infinite series of diagrams resummed by the steepest-descent method at one-loop
order are the two series of bubble diagrams shown in Fig.~\ref{figCsd}.
We can check that we recover the nonzero diagrams obtained at one-loop
order in Fig.~\ref{figC21loop} within the perturbative expansion over the cubic
interaction $K_s$. In addition, the implicit equations 
(\ref{Ceq})-(\ref{Rforward})
have allowed us to resum two infinite series of higher-order diagrams.
Thus, we recover the fact that the large-$N$ expansions and the standard 
perturbative expansions agree up to the truncation order (here one-loop) and 
only differ by higher-order terms (Valageas 2004, 2007a). 

Note that we do not recover the diagram (d) of Fig.~\ref{figC21loop} that 
was equal to zero. In fact, the property $K_s(0;\bq,-\bq)=0$ that sets this 
diagram to zero has already been used twice to derive the Schwinger-Dyson
equations (\ref{Ceq})-(\ref{P0eq}): i) to obtain the path integral 
(\ref{Zjpsilambda}) where we used that the Jacobian is an irrelevant constant,
and ii) to obtain Eqs.(\ref{Ceq})-(\ref{P0eq}) by expanding over $1/N$
where we used that $\lag\psi\rag=0$ (which also comes from $K_s(0;\bq,-\bq)=0$).
Then, the explicit use of these simplifications has already removed from the
large-$N$ expansion diagrams such as the diagram (d) of Fig.~\ref{figC21loop}.

\begin{figure}[htb]
\begin{center}
\epsfxsize=8.1 cm \epsfysize=8.8 cm {\epsfbox{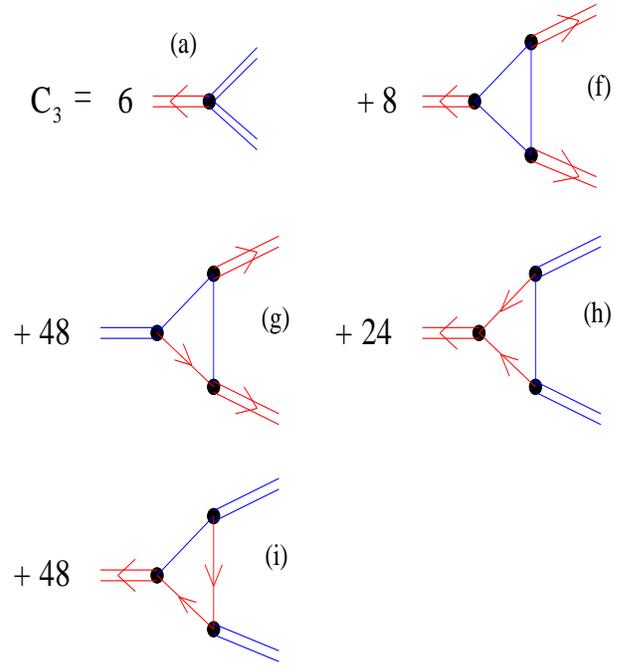}}
\end{center}
\caption{The diagrams obtained at one-loop order for the three-point correlation 
$C_3$ from the direct steepest-descent expansion. The double lines are again the
nonlinear response $R$ and the nonlinear correlation $C$. We use the same
labels as in Fig.~\ref{figC31loop}. The four diagrams (b) to (e) of 
Fig.~\ref{figC31loop} do not appear in this expansion scheme since their 
contribution is included in diagram (a) through the use of nonlinear two-point
functions instead of linear ones in the external legs.}
\label{figC3sd}
\end{figure}

\begin{figure}[htb]
\begin{center}
\epsfxsize=6.2 cm \epsfysize=4.7 cm {\epsfbox{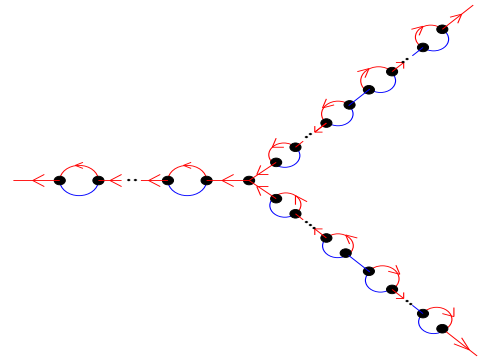}}
\end{center}
\caption{An infinite series of diagrams resummed by the ``tree-diagram'' of
Fig.~\ref{figC3sd} of the steepest-descent expansion at one-loop order. 
It is obtained from the diagrammatic expansions of the nonlinear $R$ and $C$ 
shown in Figs.~\ref{figRsd}-\ref{figCsd}. There are two more series (not shown) 
obtained by taking into account the second diagram of Fig.~\ref{figCsd}.}
\label{figC3treesd}
\end{figure}

For the three-point correlation $C_3$ we need the one-loop contribution to the 
three-point vertex $\Gamma_3$ which reads as (e.g. Zinn-Justin 1989; Valageas 2004)
\beq
\Gamma_3^{\rm 1loop}= - 6^3 K_s \cC_0 K_s \cC_0 K_s \cC_0 ,
\label{Gamma3loop_sd}
\eeq
where the two-point functions $\cC_0$ form a loop which joins the three bare
vertices $K_s$. Here $\cC_0$ is the auxiliary two-point function
introduced within the direct steepest-descent scheme as in 
Sect.~\ref{def-Direct-steepest-descent-method}, which happens to be equal to the
linear two-point function $\cC_L$. Then, from Eq.(\ref{Gamma3}) the 
``tree-diagram'' contribution reads as (using Eq.(\ref{Gamma3tree}))
\beq
\cC_3^{\rm tree}= 6 \, \cC\cC\cC . K_s ,
\label{C3treesd}
\eeq
whereas the ``one-loop diagram'' reads as (using Eq.(\ref{Gamma3loop_sd}))
\beq
\cC_3^{\rm 1loop}= 216 \, \cC\cC\cC . (K_s \cC_L K_s \cC_L K_s \cC_L) .
\label{C31loopsd}
\eeq
We show in Fig.~\ref{figC3sd} the diagrams associated with 
Eqs.(\ref{C3treesd})-(\ref{C31loopsd}) that give the three-point correlation
at one-loop order within the steepest-descent expansion.
In both equations (\ref{C3treesd})-(\ref{C31loopsd}) we must use the nonlinear 
two-point functions $\cC$ computed at the same order within the steepest-descent
method (here one-loop). That is, from the one-loop diagrams of 
Fig.~\ref{figSigPi1loop}
for the self-energy, which lead to the infinite loop resummations for two-point
functions shown in Figs.~\ref{figRsd}-\ref{figCsd}. (Let us recall that the 
``loop-order'' refers to the diagrams kept for the self-energy, but the two-point
functions $R$ and $C$ contain resummations over diagrams that contain an arbitrary
number of loops over the linear propagators.)

The three nonlinear two-point functions $\cC$ that appear in 
Eqs.(\ref{C3treesd})-(\ref{C31loopsd}) are 
shown in Fig.~\ref{figC3sd} by the double lines associated with the three 
external legs, as in Eq.(\ref{Gamma3}). Thus, the contribution (\ref{C3treesd}), 
shown by diagram (a) in Fig.~\ref{figC3sd}, is no longer equal to the standard 
tree-diagram (a) of Fig.~\ref{figC31loop} associated with Eqs.(\ref{C3tree}) 
and (\ref{C3tree1}). It includes three infinite series of additional diagrams,
such as the series shown in Fig.~\ref{figC3treesd}, which are obtained by 
replacing the nonlinear two-point functions $R$ and $C$ by their 
diagrammatic expressions shown in Figs.~\ref{figRsd}-\ref{figCsd}.
In particular, these series contain the first four one-loop diagrams (b), (c), 
(d), and (e), of the expansion over $K_s$ shown in Fig.~\ref{figC31loop}.
The other diagrams of Fig.~\ref{figC3sd} also contain such infinite series 
once written in terms of the linear two-point functions $R_L$ and $C_L$. 
Thus, the steepest-descent method has allowed us to resum in an automatic manner
some partial infinite series of diagrams for all many-body correlation functions.

\subsection{2PI effective action method}
\label{2PI-effective-action}

\begin{figure}[htb]
\begin{center}
\epsfxsize=7.9 cm \epsfysize=2.9 cm {\epsfbox{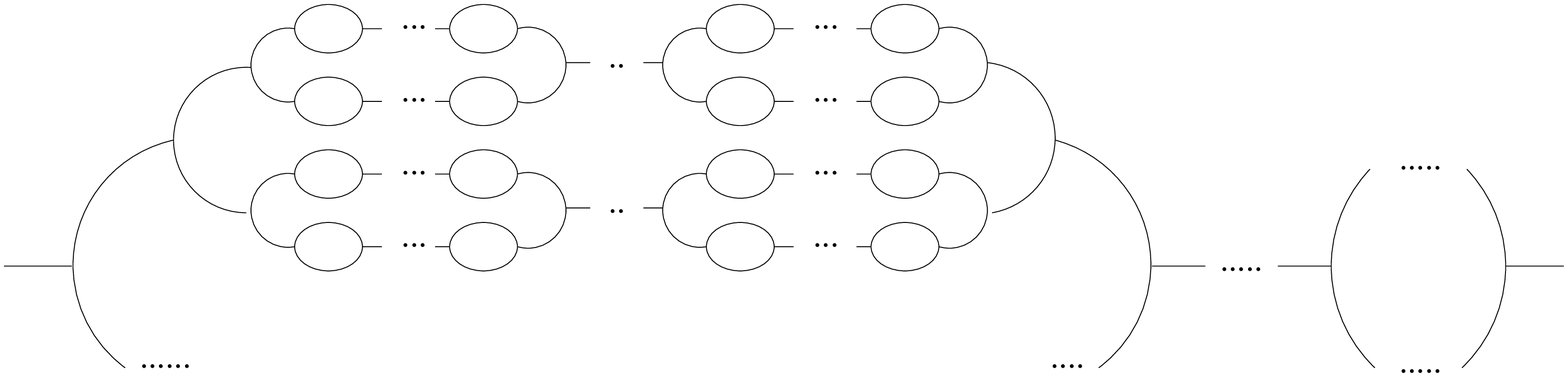}}
\end{center}
\caption{An infinite series of diagrams included in the two-point functions
given by the 2PI effective action expansion at one-loop order. There are 
additional infinite series of diagrams as compared with the steepest-descent 
method thanks to the ``renormalization'' of the self-energy terms.}
\label{figdiag2PI}
\end{figure}

As recalled in Sect.~\ref{def-2PI-effective-action-method}, the 2PI effective 
action method also gives the Schwinger-Dyson equations 
(\ref{Ceq})-(\ref{Rforward}).
In particular, Eqs.(\ref{RSRLint})-(\ref{Cint}) still apply but they are now
nonlinear as the self-energy terms now depend on the nonlinear two-point 
functions as in Eqs.(\ref{Seq})-(\ref{Peq}). At tree-order we again recover the 
results of the standard perturbative expansion. At one loop-order the 
self-energy 
is given by the same diagrams as those of Fig.~\ref{figSigPi1loop} except that 
the internal two-point functions are the nonlinear ones (i.e. single lines must 
be replaced by double lines). Then, it is clear that the two-point functions 
obtained from Eqs.(\ref{RSRL})-(\ref{Cint}) contain the same infinite series of 
diagrams as those derived in the steepest-descent method, shown in 
Figs.~\ref{figRsd}-\ref{figCsd}, but also additional series due to the 
``renormalization'' of the self-energy terms. For illustration we display
in Fig.~\ref{figdiag2PI} a typical diagram included in the 2PI effective action 
resummation for the two-point functions $R$ and $C$. 

Next, the three-point proper vertex $\Gamma_3$ at one loop-order can also be 
written as Eq.(\ref{Gamma3loop_sd}), where the linear two-point functions $\cC_0$
must be replaced by the nonlinear ones $\cC$ (i.e. all single lines including 
the internal ones must be replaced by double lines). Expanding back over the 
linear two-point functions $R_L$ and $C_L$ we recover again all diagrams 
obtained within the steepest-descent method as well as additional series which 
include terms such as those associated with Fig.~\ref{figdiag2PI}.

\section{Random advection in the large-$k$ limit}
\label{large-k-limit}

\begin{figure}[htb]
\begin{center}
\epsfxsize=7 cm \epsfysize=4.8 cm {\epsfbox{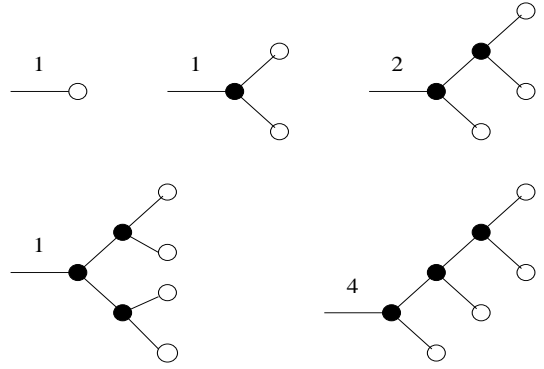}}
\end{center}
\caption{The expansion of the nonlinear field $\psi$ over the linear growing mode
$\psi_L$ from Eq.(\ref{psiseries}), up to order $\psi_L^4$. The filled circles 
are the vertex $\tKs$ of Eq.(\ref{tKsdef}), whereas the white circles are the 
linear growing mode $\psi_L$. The numbers are the multiplicity factor 
associated with each diagram.}
\label{figdiagtKs}
\end{figure}

As recalled in Sect.~\ref{Standard-perturbation-theory}, the standard 
perturbation theory associated with Eq.(\ref{Fn}) writes the nonlinear density
field $\delta(\bk)$ as a series over powers of the linear density contrast
$\delta_L$. The kernels $F_n$ of Eq.(\ref{Fn}) are computed by recursion from
the equation of motion (\ref{OKsdef}) so that each order involves a new vertex
$F_n$. However, it is clear that the latter can also be written in terms
of the vertex $K_s$ that appears in Eq.(\ref{OKsdef}). Thus, the equation of 
motion (\ref{OKsdef}) for the two-component field $\psi$ can be solved through
the expansion over the linear growing mode $\psi_L$ of Eq.(\ref{psiL}) as 
(Crocce \& Scoccimarro 2006a,b; Valageas 2007b)
\beq
\psi= \psi_L + \tKs \psi_L^2 + 2 \tKs^2 \psi_L^3 + 5 \tKs^3 \psi_L^4 + ... \, ,
\label{psiseries}
\eeq
where we introduced the integral vertex $\tKs$ defined from $K_s$ and the linear
response $R_L$ by
\beq
\tKs = R_L . K_s
\label{tKsdef}
\eeq
The diagrams associated with the series (\ref{psiseries}) are shown in 
Fig.~\ref{figdiagtKs}, see Crocce \& Scoccimarro (2006a).
Of course, Eq.(\ref{psiseries}) and Fig.~\ref{figdiagtKs} are equivalent to
the standard perturbation series of Eq.(\ref{Fn}) and 
Fig.~\ref{figdeltastandard}. Then, with the aim of computing the response 
function $R$, Crocce \& Scoccimarro (2006a) noticed that
it is possible to perform a partial resummation of the diagrams shown in
Fig.~\ref{figdiagtKs} in a high-$k$ limit. First, one only keeps the
diagrams such as the last one in Fig.~\ref{figdiagtKs}, where one can define
a ``principal path'' connecting a linear mode $\psi_L$ to the left root so
that all other linear modes $\psi_L$ are directly connected to this path.
(This is not the case for the other diagram of order $\psi_L^4$ in 
Fig.~\ref{figdiagtKs}.) Second, in a high-$k$ limit one may approximate the
vertex $K_s$ by
\beqa
k \! \gg \!k_1 : K_s(x;x_1,x_2) & \! \simeq \! & \delta_D(\bk_2-\bk) 
\delta_D(\eta_1-\eta) \delta_D(\eta_2-\eta) \nonumber \\ 
&& \times \frac{\bk.\bk_1}{2k_1^2} \; \delta_{i,i_2} \, \delta_{i_1,2} .
\label{Kshighk}
\eeqa
This simplifies the series (\ref{psiseries}) which can be resummed to give the
response function (Crocce \& Scoccimarro 2006a,b)
\beq
R(x_1,x_2) = R_L(x_1,x_2) e^{-(D_1-D_2)^2 k^2\sigma_v^2/2} ,
\label{RGauss}
\eeq
where $\sigma_v^2$ is the variance of the one-dimensional linear displacement 
field $s_{L0}$ (equal to the variance of the one-dimensional linear velocity
dispersion up to a normalization factor) given by
\beq
\sigma_v^2 = \frac{4\pi}{3} \int_0^{\infty} \d k P_{L0}(k) .
\label{sigmav}
\eeq
As seen in Valageas (2007b), keeping only the ``principal path'' diagrams 
amounts to approximate the equation of motion (\ref{OKsdef}) by the 
linearized equation
\beq
\cO.\psi= 2 K_s \psi_L \psi , \;\;\; \mbox{or} \;\;\; 
\psi= \psi_L+ 2 \tKs \psi_L \psi .
\label{Opsilin}
\eeq
Then, using the approximation (\ref{Kshighk}), the solution of Eq.(\ref{Opsilin})
reads as (Valageas 2007b)
\beq
\delta(\bk) = \delta_L(\bk) \; 
e^{\int\!\d\bq \,\frac{\bk.\bq}{q^2} \, \delta_L(\bq)} .
\label{deltahighk}
\eeq
The key simplification comes from the approximation 
$\delta_D(\bk_1+\bk_2-\bk) \simeq \delta_D(\bk_2-\bk)$ in Eq.(\ref{Kshighk}).
This makes all the intermediate legs in the simplified diagrammatic series
factorize as powers of $\int \d\bq \, \frac{\bk.\bq}{q^2} \, \delta_L(\bq,\eta')$
which allows the resummation (\ref{deltahighk}). Then, it is easy to derive
Eq.(\ref{RGauss}) from Eq.(\ref{deltahighk}), as well as
\beq
C(x_1,x_2) = C_L(x_1,x_2) e^{-(D_1-D_2)^2 k^2\sigma_v^2/2} .
\label{CGauss}
\eeq
As discussed in Valageas (2007b), this procedure has actually replaced the
gravitational dynamics by the simple effective dynamics
\beq
\delta(\bx,D)= \delta_L(\bx-\bs_L(\bx_{\rm Lag}=0,D),D) .
\label{advec}
\eeq
That is, the linear density field is uniformly advected by the linear displacement
$\bs_L(\bx_{\rm Lag}=0,D)$ associated with the Lagrangian position 
$\bx_{\rm Lag}=0$. 
Then, the average over this random Gaussian displacement leads to the 
apparent loss of memory seen in Eqs.(\ref{RGauss}), (\ref{CGauss}), through 
the Gaussian decay $e^{-D^2 k^2\sigma_v^2/2}$.
As discussed in Valageas (2007b), it happens that for the Zeldovich dynamics
Eq.(\ref{RGauss}) actually gives the exact nonlinear response whereas
Eq.(\ref{CGauss}) gives the correct Gaussian decay for different times 
$D_1\neq D_2$ but does not capture the equal-time behavior of the nonlinear 
correlation $C$.

\begin{figure}[htb]
\begin{center}
\epsfxsize=8 cm \epsfysize=1.4 cm {\epsfbox{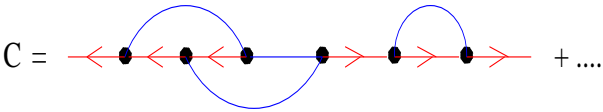}}
\end{center}
\caption{A diagram contained in the high-$k$ resummation associated with
Eq.(\ref{Opsilin}) for the two-point correlation $C$.}
\label{figChighk}
\end{figure}

As compared with the large-$N$ resummation of 
Sect.~\ref{Direct-steepest-descent}, which gives the bubble diagrams of 
Fig.~\ref{figCsd} for the two-point correlation $C$, the resummation associated
with Eq.(\ref{Opsilin}) (i.e. keeping the principal path diagrams for $\psi$)
contains additional diagrams such as the one shown in Fig.~\ref{figChighk}.
Indeed, when we join the relevant diagrams of Fig.~\ref{figdiagtKs} to compute
$C(x_1,x_2)=\lag\psi(x_1)\psi(x_2)\rag$, the Gaussian average draws all 
possible pairs from all secondary legs $\psi_L(\bq_j)$ that connect to the 
principal path (including those obtained by matching fields $\psi_L(\bq_j)$ 
that are associated with both nonlinear fields $\psi(x_1)$ and $\psi(x_2)$).
By contrast, the bubble diagrams of Fig.~\ref{figCsd} only involve the
Gaussian pairings performed in sequential order as we move along the principal
path (but this method does not require the high-$k$ approximation 
(\ref{Kshighk})).

Nevertheless, at one-loop order we can substitute the two-point functions 
(\ref{RGauss}), (\ref{CGauss}), obtained in this fashion into the diagrams
of Fig.~\ref{figC3sd} to derive an approximation for the three-point function
that is correct up to this order and contains infinite partial resummations.
We can obtain two approximations for $C_3$ in this manner. We can either use
Eqs.(\ref{RGauss}), (\ref{CGauss}), into the external lines only of the one-loop
diagrams of Fig.~\ref{figC3sd}, in the spirit of the steepest-descent
scheme of Sect.~\ref{Direct-steepest-descent}, or into both the external
and internal lines in the spirit of the 2PI effective action method of 
Sect.~\ref{2PI-effective-action}.
As for previous schemes, these two procedures may be applied 
to all higher-order correlation and response functions.

\section{Zeldovich dynamics}
\label{Zeldovich}

We investigate in this section the predictions of the various expansion 
schemes described in the previous sections for the simpler case of the Zeldovich 
dynamics (Zeldovich 1970), where the trajectories of particles follow the
linear displacement field as in Eq.(\ref{xq}). We focus on the three-point 
function $C_3$, and more specifically on the matter density bispectrum defined 
in Eq.(\ref{Bdef}), since the two-point functions $C$ and $R$ have already been 
studied in Valageas (2007b).

\subsection{Standard perturbation theory}
\label{ZelStandard}

As is well-known, from the exact solution (\ref{xq}) of the Zeldovich dynamics
the kernels $F_n$ introduced in the standard perturbative expansion (\ref{Fn}) 
can be directly derived at any order (Grinstein \& Wise 1987). For instance, 
expanding the exponential in Eq.(\ref{deltakq}) gives
\beq
\delta(\bk) = \sum_{n=1}^{\infty} \frac{1}{n!} 
\int\frac{\d\bq}{(2\pi)^3} \; e^{-i\bk.\bq} \; ( -i\bk.\bs_L )^n .
\label{deltakqs}
\eeq
Using Eq.(\ref{sLdeltaL}) we recover Eq.(\ref{Fn}) with the symmetric
kernels
\beq
F_n^s(\bq_1,..,\bq_n)= \frac{1}{n!} \; \frac{\bk.\bq_1}{q_1^2} ... 
\frac{\bk.\bq_n}{q_n^2} .
\label{ZelFns}
\eeq
Then, using Eqs.(\ref{Ptree})-(\ref{P1loopc}) the power spectrum reads up to
one-loop order as
\beqa
\lefteqn{ P(k) = P_L(k) - P_L(k) \int\!\d\bq \frac{(\bk.\bq)^2}{q^4} 
P_L(q) + \int\!\d\bq_1\d\bq_2 } \nonumber \\
&& \times \delta_D(\bq_1+\bq_2-\bk) P_L(q_1) P_L(q_2) 
\frac{(\bk.\bq_1)^2(\bk.\bq_2)^2}{2q_1^4q_2^4} .
\label{ZelPdel}
\eeqa
Eqs.(\ref{Btreestandard})-(\ref{Bestandard}) read for the bispectrum up to 
one-loop order as
\beq
\!\!B^{\rm tree}(k_1,k_2,k_3) \!\!= P_L(k_2) P_L(k_3)
\frac{\!(\bk_1.\bk_2)(\bk_1.\bk_3)\!}{k_2^2 k_3^2} + 2 \, {\rm perm.}
\label{ZelBtreedel}
\eeq
and,
\beqa
\!B^{(b)} \!\!& = &\!\! - P_L(k_2) P_L(k_3) \frac{(\bk_1.\bk_2) (\bk_1.\bk_3)}
{2k_2^2 k_3^2} \!\int\!\!\d\bq P_L(q) \frac{(\bk_1.\bq)^2\!}{q^4} 
\nonumber \\ 
&& + 2 \, \rm{perm.} ,
\label{ZelBb}
\eeqa
\beqa
\!B^{(c)} \!\!& = &\!\! - P_L(k_2) P_L(k_3) \frac{(\bk_1.\bk_2) (\bk_1.\bk_3)}
{2k_2^2 k_3^2} \!\int\!\!\d\bq P_L(q) \frac{(\bk_2.\bq)^2\!}{q^4} 
\nonumber \\ 
&& + 5 \, \rm{perm.} ,
\label{ZelBc}
\eeqa
\beqa
\lefteqn{ B^{(d)} = - P_L(k_1) \frac{\bk_1.\bk_3}{2k_1^2} \! 
\int\!\d\bq_1\d\bq_2 \; \delta_D(\bq_1+\bq_2-\bk_2) } \nonumber \\
&& \times P_L(q_1) P_L(q_2) 
\frac{(\bk_2.\bq_1)(\bk_2.\bq_2)(\bk_3.\bq_1)(\bk_3.\bq_2)}{q_1^4 q_2^4}
\nonumber \\
&& + 5 \, \rm{perm.} ,
\label{ZelBd}
\eeqa
\beqa
\lefteqn{B^{(e)} \! = \! - \!\! \int \!\! \d\bq_1\d\bq_2\d\bq_3 
\; \delta_D(\bq_1+\bq_2-\bk_1) } \nonumber \\
&& \times \; \delta_D(\bq_1+\bq_3+\bk_2) P_L(q_1) P_L(q_2) P_L(q_3) \nonumber \\
&& \times \frac{(\bk_1.\bq_1)(\bk_1.\bq_2)(\bk_2.\bq_1)(\bk_2.\bq_3)
(\bk_3.\bq_2)(\bk_3.\bq_3)}{q_1^4 q_2^4 q_3^4} .
\label{ZelBe}
\eeqa
Moreover, we can gather diagrams (b) and (c), which have the same form, as
\beqa
\lefteqn{\!B^{(b)}\!+\!B^{(c)} = - P_L(k_2) P_L(k_3) 
\frac{(\bk_1.\bk_2)(\bk_1.\bk_3)}{2k_2^2 k_3^2} \int\!\d\bq P_L(q) } 
\nonumber \\ 
&& \times \frac{(\bk_1.\bq)^2+(\bk_2.\bq)^2+(\bk_3.\bq)^2}{q^4} + 2 \, \rm{perm.}
\label{ZelBbc}
\eeqa
We can check that for a CDM-like linear power spectrum, with $P_L(k) \propto
k$ for $k\rightarrow 0$ and $P_L(k) \propto k^{-3}$ for $k\rightarrow \infty$,  
all contributions to the power spectrum and to the bispectrum converge.
More precisely, if we define the local spectral index $n$ as
\beq
n(k)= \frac{\d\ln P_{L0}(k)}{\d\ln k} ,
\label{ndef}
\eeq
we see that the one-loop contributions to the power spectrum and the bispectrum 
converge if $n(0)>-1$ and $n(\infty)<-1$. However, as is well-known 
(Vishniac 1983; Jain \& Bertschinger 1996), the infrared divergences at 
$q\rightarrow 0$ cancel out and the sum of all contributions is finite
for $n(0)>-3$. This can be easily checked from the explicit expressions 
(\ref{ZelPdel})-(\ref{ZelBbc}). On the other hand, the time-dependence of various
contributions is
\beq
P^{\rm tree} \propto D^2 \sigma_8^2, \;\;\; P^{\rm 1loop} \propto D^4 
\sigma_8^4 ,
\label{PDstandard}
\eeq
\beq
B^{\rm tree} \propto D^4 \sigma_8^4, \;\;\; B^{\rm 1loop} \propto D^6 
\sigma_8^6 ,
\label{BDstandard}
\eeq
where $\sigma_8$ is the normalization of the linear power spectrum today at $z=0$
(defined as usual from the rms linear density contrast within a sphere of radius
$8 h^{-1}$Mpc). Thus, we have the scaling $P_L \propto D^2 \sigma_8^2$ for a 
fixed shape of the linear power spectrum. Higher-order terms scale as higher 
powers of $P_L$ hence of $D^2\sigma_8^2$.

\subsection{Expansion over cubic interaction}
\label{Zelcubic}

We now turn to the method presented in Sect.~\ref{Expansion-over-cubic},
where we expand the path integral over the interaction vertex $K_s$.
As explained in Sect.~\ref{Expansion-over-cubic}, this method exactly recovers
the results of the standard perturbation theory of Sect.~\ref{ZelStandard}.

First, as discussed in Sect.~\ref{Expansion-over-cubic}, for the nonlinear 
power spectrum the diagrams (b) and 
(c) of Fig.~\ref{figC21loop} give back the contributions of diagram (b) and (c)
of Fig.~\ref{figC2standard} associated with the standard perturbative expansion.
Note that one can see at once that there is no mixing of diagrams as one goes 
from one expansion to the other one since it is clear that both diagrams (b) 
contain a fixed factor $P_L(k)$ whereas both diagrams (c) contain a double 
integral over $P_L(q_1)P_L(q_2)$ (hence there is a one-to-one correspondence). 
In any case, a simple explicit computation of the diagrams of 
Fig.~\ref{figC21loop} gives indeed Eq.(\ref{ZelPdel}) for the Zeldovich 
dynamics. Therefore, the conditions of convergence of various integrals
are the same as in Sect.~\ref{ZelStandard}.

We now consider the matter bispectrum. At tree-order, the diagram (a) of 
Fig.~\ref{figC31loop} gives back the diagram (a) of Fig.~\ref{figC3standard}.
Thus, the explicit computation of Eq.(\ref{C3tree1}) recovers 
Eq.(\ref{ZelBtreedel}) for the Zeldovich dynamics. At one-loop order, we no 
longer have such a one-to-one correspondence between both expansions since we 
have eight one-loop diagrams in Fig.~\ref{figC31loop} and only four one-loop 
diagrams in Fig.~\ref{figC3standard}.
First, let us consider the contributions which have the form of a triple 
integration over the linear power spectrum, as in Eq.(\ref{ZelBe}) associated 
with diagram (e) of Fig.~\ref{figC3standard}. One can see at once that only 
diagram (f) of Fig.~\ref{figC31loop} has this form and its explicit calculation 
gives back Eq.(\ref{ZelBe}). Next, contributions with a double integration, as 
in Eq.(\ref{ZelBd}), arise from diagrams (b) and (g) which read as
\beqa
\lefteqn{ B^{(b)} = - P_L(k_1) \int\!\d\bq_1\d\bq_2 \; 
\delta_D(\bq_1+\bq_2-\bk_2) \; \times} \nonumber \\
&& \!\!\!\!\!\!\!\! \frac{2 (\bk_1.\bk_3) (\bk_2.\bq_1)(\bk_2.\bq_2) \! + \! 
(2 k_1^2 (\bk_2.\bk_3)\!-\!k_3^2 (\bk_1.\bk_2)) (\bq_1.\bq_2)}
{12 k_1^2 q_1^2 q_2^2} \nonumber \\
&& \times \frac{(\bk_2.\bq_1)(\bk_2.\bq_2)}{q_1^2 q_2^2}  P_L(q_1) P_L(q_2)
+ 5 \, \rm{perm.} 
\label{ZelcubicBb}
\eeqa
and
\beqa
\lefteqn{ B^{(g)} \! = \! - P_L(k_1) \! \int\!\!\d\bq_1\d\bq_2 \, 
\delta_D(\bq_1+\bq_2-\bk_2) P_L(q_1) P_L(q_2)  } \nonumber \\
&& \times \left[ \frac{2 k_1^2 q_1^2 (\bk_3.\bq_2) + 2 (\bk_1.\bq_1) 
(\bk_3.\bq_2)^2} {6 k_1^2 q_1^2 q_2^2} \right. \nonumber \\
&& \left. + \frac{(\bk_1.\bq_1) [-k_3^2 q_2^2 + (k_1^2+2k_3^2+q_1^2-q_2^2) 
(\bk_3.\bq_2)]}{6 k_1^2 q_1^2 q_2^2} \right] \nonumber \\
&& \times \frac{(\bk_2.\bq_1)(\bk_2.\bq_2)}{q_1^2 q_2^2} 
+ 5 \, \rm{perm.}
\label{ZelcubicBg}
\eeqa
Then, summing $B^{(b)}+B^{(g)}$ and writing the result in a symmetric form over
$\{\bq_1,\bq_2\}$ we recover the contribution of Eq.(\ref{ZelBd}).
Finally, contributions with a single integration, as in Eq.(\ref{ZelBbc}),
arise from diagrams (c), (d), (e), (h) and (i), which read as
\beqa
\lefteqn{B^{(c)} = - P_L(k_2) P_L(k_3) \frac{(\bk_1.\bk_2)(\bk_1.\bk_3)}
{2k_2^2 k_3^2} \int\!\d\bq P_L(q) } \nonumber \\ 
&& \times \frac{(\bk_2.\bq)^2+(\bk_3.\bq)^2}{q^4} + 2 \, \rm{perm.} ,
\label{ZelcubicBc}
\eeqa
\beqa
\lefteqn{B^{(d)} \! = \! - P_L(k_2) P_L(k_3) \frac{3 k_2^2k_3^2\!
+\!2(k_2^2\!+\!k_3^2)(\bk_2.\bk_3)\!+\!(\bk_2.\bk_3)^2\!}{12k_2^2 k_3^2} } 
\nonumber \\ 
&& \times \int\!\d\bq P_L(q) \frac{(\bk_2.\bq)^2+(\bk_3.\bq)^2}{q^4} 
+ 2 \, \rm{perm.} ,
\label{ZelcubicBd}
\eeqa
\beqa
B^{(e)} & = & - P_L(k_2) P_L(k_3) \frac{2 k_2^2k_3^2 + (3 k_1^2+k_2^2+k_3^2)
(\bk_2.\bk_3)}{24k_2^2 k_3^2} \nonumber \\ 
&& \times \int\!\d\bq P_L(q) \frac{(\bk_1.\bq)^2}{q^4} + 2 \, \rm{perm.} ,
\label{ZelcubicBe}
\eeqa
\beqa
\lefteqn{\!B^{(h)} \! = \! P_L(k_2) P_L(k_3) \!\int\!\!\d\bq P_L(q) \biggl 
\lbrace \frac{(\bk_2.\bq) (\bk_3.\bq)}{12 k_2^2 k_3^2} 
\left[ 3 \frac{(\bk_1.\bq)^2\!} {q^4} \right. } \nonumber \\ 
&& \left. - 2 \frac{k_1^2 q^2 + 3 k_2^2 k_3^2 + 2 (k_2^2+k_3^2) (\bk_2.\bk_3) 
+ (\bk_2.\bk_3)^2}{q^4} \right] \nonumber \\ 
&& - \frac{3 q^2 (\bk_1.\bq) [k_3^2 (\bk_2.\bq)+k_2^2 (\bk_3.\bq)]}
{12 k_2^2 k_3^2 q^4} \biggl \rbrace \; + 2 \, \rm{perm.} ,
\label{ZelcubicBh}
\eeqa
and
\beqa
\lefteqn{\!\! B^{(i)}\!\!  = \! P_L(k_2) P_L(k_3) \!\int\!\!\d\bq P_L(q) \! 
\left[ \! \frac{\!(\bk_1.\bq) (\bk_2.\bq) k_3^2 (2k_2^2\!+\!3q^2)\!}
{12 k_2^2 k_3^2 q^4} \right. } \nonumber \\ 
&& \!\!\!\!\!\!\!\! + \frac{\!(\bk_1.\bq) (\bk_3.\bq) k_2^2 (2k_3^2\!+\!3q^2) 
\! - \!(\bk_1.\bq)^2 (\bk_2.\bk_3) (k_1^2\!\!+\!k_2^2\!\!+\!k_3^2)\!}
{12 k_2^2 k_3^2 q^4} \nonumber \\
&& \!\!\!\!\!\!\!\!  \left. + \frac{(\bk_2.\bq) (\bk_3.\bq) 
[2 k_1^2 q^2 - 3 (\bk_1.\bq)^2]}{12 k_2^2 k_3^2 q^4} \right] + 2 \, \rm{perm.} ,
\label{ZelcubicBi}
\eeqa
where we used the symmetries $\bq\leftrightarrow-\bq$ and 
$\bk_2\leftrightarrow\bk_3$,
and the property $\bk_1+\bk_2+\bk_3=0$, to simplify the expressions. Then, we can
check again that the sum $B^{(c)}+B^{(d)}+B^{(e)}+B^{(h)}+B^{(i)}$ of 
Eqs.(\ref{ZelcubicBc})-(\ref{ZelcubicBi}) gives back Eq.(\ref{ZelBbc}) obtained
from the standard perturbative expansion.
This consistency check allows us to check these expressions obtained 
through two different methods and to track possible mistakes in the 
computations. 

We again find that all one-loop contributions to the bispectrum 
converge if $n(0)>-1$ and $n(\infty)<-1$, except for $B^{(h)}$ and $B^{(i)}$ 
which require $n(\infty)<-3$.
As for the standard perturbative expansion (which we recover by summing all 
terms) infrared divergences compensate so that we only need $n(0)>-3$. 
On the other hand, the UV divergences of $B^{(h)}$ and $B^{(i)}$ also compensate 
so that we recover the standard constraint $n(\infty)<-1$. Thus, although the 
standard expansion of Sect.~\ref{ZelStandard} and the expansion over $K_s$ of 
this section are actually identical, the contributions obtained at each order 
are split in different ways between different diagrams so that the convergence 
properties of individual parts can be different for the two schemes. 
In particular, we have seen that for the Zeldovich dynamics one obtains an 
artificial UV divergence for $-3\leq n(\infty)<-1$ which disappears by summing 
all diagrams. For numerical purposes, since the contributions (h) and (i)
of Eqs.(\ref{ZelcubicBh})-(\ref{ZelcubicBi}) are not well-defined because of
this UV divergence, we introduce the regularized contributions $B^{(h)}_R$ and 
$B^{(i)}_R$. They are equal to $B^{(h)}$ and $B^{(i)}$ from which we subtract
the UV divergent part, that is, we remove the part that scales as $q^0$ from the
terms in the brackets in Eqs.(\ref{ZelcubicBh})-(\ref{ZelcubicBi}). This only
leaves terms that scale as $q^{-2}$ and gives
\beqa
\lefteqn{\!\!B^{(h)}_R \! = \! - P_L(k_2) P_L(k_3) \frac{\!3 k_2^2 k_3^2 
+ 2 (k_2^2\!+\!k_3^2) (\bk_2.\bk_3) + (\bk_2.\bk_3)^2\!}{6 k_2^2 k_3^2} } 
\nonumber \\
&& \times \!\int\!\!\d\bq P_L(q)  \frac{(\bk_2.\bq) (\bk_3.\bq)}{q^4}
 + 2 \, \rm{perm.} ,
\label{ZelcubicBhR}
\eeqa
and
\beqa
\lefteqn{B^{(i)}_R = - P_L(k_2) P_L(k_3) \frac{2 k_2^2 k_3^2 + (\bk_2.\bk_3) 
(k_1^2+k_2^2+k_3^2)}{12 k_2^2 k_3^2} } \nonumber \\
&& \times \!\int\!\!\d\bq P_L(q)  \frac{(\bk_1.\bq)^2}{q^4}
 + 2 \, \rm{perm.} 
\label{ZelcubicBiR}
\eeqa
Since the UV divergences of $B^{(h)}$ and $B^{(i)}$ compensate we have
$B^{(h)}+B^{(i)}=B^{(h)}_R+B^{(i)}_R$ and we can use $B^{(h)}_R$ and 
$B^{(i)}_R$ instead of $B^{(h)}$ and $B^{(i)}$.

\subsection{Direct steepest-descent method}
\label{Zelsteepest-descent}

We now consider the direct steepest-descent method presented in 
Sect.~\ref{Direct-steepest-descent}. First, the two-point functions $C$ and $R$
are obtained from the diagrams of Figs.~\ref{figRsd}-\ref{figCsd} that
correspond to Eqs.(\ref{RSRLint})-(\ref{Cint}). Thus, in terms of nonlinear
two-point functions we have only one one-loop diagram for either $C$ or $R$, 
which corresponds to an infinite series over the linear two-point functions 
$C_L$ and $R_L$. The results obtained at one-loop order have been described 
in detail in Valageas (2007b). Thus, the nonlinear response $R$ reads as
\beq
R(x_1,x_2)= R_L(x_1,x_2) \cos[\omega(k)(D_1-D_2)] ,
\label{Rcos}
\eeq
where $R_L$ is the linear response given in Eq.(\ref{RL}), $D=e^{\eta}$ is
the linear growth factor and $\omega(k)$ (where $k=k_1=k_2$ is the wavenumber) 
is given by
\beq
\omega(k)=  k \sigma_v  .
\label{omega}
\eeq
Here $\sigma_v^2$ is the variance of the one-dimensional linear displacement 
field $s_{L0}$ given in Eq.(\ref{sigmav}). Note that the nonlinear response
$R$ only depends on the linear power spectrum through this velocity variance. 
From the expression of the linear response $R_L$, Eq.(\ref{Rcos})
gives
\beqa
R(k;D_1,D_2) & = & \theta(D_1-D_2) \cos[\omega(k)(D_1-D_2)] \nonumber \\
&& \times \sum_{\ell=1}^2 \left(\frac{D_1}{D_2}\right)^{\alpha_{\ell}} 
R_{(\ell)} , 
\label{ZelRcos12}
\eeqa
with
\beq
\alpha_1=1 , \;\;\; \alpha_2=0 ,
\label{Zelalpha12}
\eeq
and the two constant matrices $R_{(1)}$ and $R_{(2)}$, associated with the 
linear growing and decaying (here constant) modes are
\beq
R_{(1)}= \left(\bea{cc} 0 & 1 \\ 0 & 1 \ea\right) \;\; \mbox{and} \;\; 
R_{(2)}= \left(\bea{cc} 1 & -1 \\ 0 & 0 \ea\right) .
\label{ZelR1R2}
\eeq

The expression of the two-point correlation $C$ is more intricate
and it depends on the details of the linear power spectrum. From Eq.(\ref{Rcos})
the first term of Eq.(\ref{Cint}) reads exactly as (Valageas 2007b)
\beq
R \times C_L(\eta_I) \times R^T = C_L(x_1,x_2) \cos(\omega D_1) 
\cos(\omega D_2) .
\label{Ccross_cos}
\eeq
The second term of Eq.(\ref{Cint}) has a more intricate structure but we shall
use the qualitative approximation
\beq
R . \Pi . R^T \simeq C_L(x_1,x_2) \sin(\omega D_1) \sin(\omega D_2) .
\label{Ccross_sin}
\eeq
As seen in Valageas (2007b), the exact one-loop result contains an additional
term but it has no qualitative effect on $C$ hence we shall use the simple
approximation obtained from the sum of 
Eqs.(\ref{Ccross_cos})-(\ref{Ccross_sin}),
\beq
C(x_1,x_2) \simeq C_L(x_1,x_2) \cos[\omega(k)(D_1-D_2)] ,
\label{Ccos}
\eeq
which gives:
\beq
\!\!C(k;D_1,D_2) \! = \! D_1 D_2 \cos[\omega(k)(D_1-D_2)] P_{L0}(k) \!
\left(\bea{cc} 1 & 1 \\ 1 & 1 \ea\right) \! .
\label{Ccosk}
\eeq
Equations (\ref{Rcos}),(\ref{Ccos}), also correspond to a simple effective 
dynamics
where the nonlinearity is reduced to a simple non-Gaussian random advection 
by a large-scale flow which gives rise to the cosine factors (Valageas 2007a,b). 
Then, these cosine factors simply describe the (weak) decorrelation, or 
loss of memory, due to this random advection (compare with Eqs.(\ref{RGauss}), 
(\ref{CGauss})). Note that the time dependence
obtained from the steepest-descent method is quite different from the one
obtained in Eq.(\ref{PDstandard}) since Eqs.(\ref{Ccross_cos})-(\ref{Ccross_sin})
give for the equal-time power spectrum
\beq
P^{\rm tree} \sim D^2 \sigma_8^2, \;\;\; P^{\rm 1loop} \sim D^2 \sigma_8^2 ,
\label{PDcos}
\eeq
where we label as the ``tree'' contribution the first diagram of 
Fig.~\ref{figCsd}, associated with Eq.(\ref{Ccross_cos}), and as the ``1loop'' 
contribution the second diagram of Fig.~\ref{figCsd}, associated with 
Eq.(\ref{Ccross_sin}).
Therefore, we now find that both contributions exhibit the same scaling in the
nonlinear regime and they only grow as $D^2\sigma_8^2$ instead of 
$D^4\sigma_8^4$ as in Eq.(\ref{PDstandard}).

The three-point correlation is now given up to one-loop
order by the five diagrams of Fig.~\ref{figC3sd}. As explained in 
Sect.~\ref{Direct-steepest-descent}, this simplification with respect to the
expansion over powers of $K_s$, studied in Sects.~\ref{Expansion-over-cubic} and
\ref{Zelcubic}, comes from the fact that one-loop diagrams associated with
a ``renormalization'' of the two-point functions $C$ and $R$ are included in the
tree-diagram (a) of Fig.~\ref{figC3sd} and only the four diagrams associated with
the ``renormalization'' of the three-point vertex $K_s$ are new structures
with respect to the previous order (tree-order).
Let us first consider the tree-diagram (a). Its explicit expression is of the
form of Eq.(\ref{C3tree1}) but using the nonlinear two-point functions $R$ 
and $C$. From Eqs.(\ref{ZelRcos12}), (\ref{Ccosk}), we see that the integral
over time $\eta_1'$ in Eq.(\ref{C3tree1}) gives rise to the two quantities
(for $\ell=1,2$):
\beqa
\lefteqn{ \int_0^D \frac{\d D_1'}{D_1'} 
\left(\frac{D}{D_1'}\right)^{\alpha_{\ell}} 
D^2 D_1'^2 \prod_{j=1}^3 \cos[\omega_j(D-D_1')] } \nonumber \\
&& \;\;\;\;\;\; = \frac{D^4}{2-\alpha_{\ell}} 
T_{(\ell)}^{(a)}(\omega_j D) .
\label{Ta}
\eeqa
This defines the two time-dependent functions $T_{(\ell)}^{(a)}$ of the 
three variables $\omega_j D$, associated with diagram (a). In Eq.(\ref{Ta}) 
we considered the case of equal-time statistics $D_1=D_2=D_3=D$ and we 
introduced:
\beq
\omega_j = \omega(k_j) = k_j \sigma_v .
\label{omegaj}
\eeq
We also normalized $T_{(\ell)}^{(a)}$ so that for $\omega_j=0$
we have $T_{(\ell)}^{(a)}(\omega_j=0)=1$. Thus, by putting 
$T_{(\ell)}^{(a)}=1$ we must recover the result (\ref{ZelBtreedel}) obtained 
by integrating Eq.(\ref{C3tree1}) over the linear two-point functions, as in the
expansion scheme of Sect.~\ref{Zelcubic} where we expanded over powers of $K_s$.
This provides a convenient consistency check. Note that all the time-dependence
is included in the standard prefactor $D^4$ and the function 
$T_{(\ell)}^{(a)}$, which only depends on the linear power spectrum and on the 
wavenumbers $k_j$ through the three combinations $k_j\sigma_v$.
The explicit expressions of $T_{(\ell)}^{(a)}$ are
\beqa
\lefteqn{ T_{(1)}^{(a)} \! = \! \frac{1}{4} \biggl \lbrace 
{\rm sinc}[D(-\omega_1+\omega_2+\omega_3)]
+{\rm sinc}[D(\omega_1-\omega_2+\omega_3)] } \nonumber \\
&& \!\!\! +{\rm sinc}[D(\omega_1+\omega_2-\omega_3)]
+{\rm sinc}[D(\omega_1+\omega_2+\omega_3)] \biggl \rbrace ,
\label{ZelTa1}
\eeqa
and
\beqa
\lefteqn{ T_{(2)}^{(a)} \! = \! \frac{1}{4} \biggl \lbrace 
{\rm cosc}[D(-\omega_1+\omega_2+\omega_3)]
+{\rm cosc}[D(\omega_1-\omega_2+\omega_3)] } \nonumber \\
&& \!\!\! +{\rm cosc}[D(\omega_1+\omega_2-\omega_3)]
+{\rm cosc}[D(\omega_1+\omega_2+\omega_3)] \biggl \rbrace ,
\label{ZelTa2}
\eeqa
where we introduced the functions
\beq
{\rm sinc}(x) \! = \! \frac{\sin(x)\!}{x} , \;\; {\rm cosc}(x) \! = \! 
\frac{2(1-\cos(x))\!}{x^2} \! = \!  {\rm sinc}^2(\frac{x}{2}).
\label{sinccosc}
\eeq
We can check that $T_{(\ell)}^{(a)}(0)=1$, and in the highly nonlinear regime
we have
\beq
D\omega_j \gg 1 : \;\;\; T_{(1)}^{(a)} \sim \frac{1}{D\omega} , \;\;\; 
T_{(2)}^{(a)} \sim \frac{1}{(D\omega)^2} ,
\label{ZelTa_asymp}
\eeq
(with $|\sum \pm \omega_j| \sim \omega$). Then, diagram (a) of 
Fig.~\ref{figC3sd} reads as
\beqa
\! B^{(a)} \!\!\! & = & \!\! P_L(k_2)P_L(k_3) \frac{T^{(a)}_{(1)} k_1^2 
(\bk_2.\bk_3) + T^{(a)}_{(2)} [ k_2^2 k_3^2 \! - \! (\bk_2.\bk_3)^2 ]}
{k_2^2 k_3^2} \nonumber \\
&& + 2 \, {\rm perm.}
\label{ZelsdBa}
\eeqa
We can check that by substituting $T^{(a)}_{(\ell)}=1$ (and using 
$\bk_1+\bk_2+\bk_3=0$) we recover the standard tree-order contribution 
(\ref{ZelBtreedel}). Moreover, in the nonlinear regime we have the scaling
\beq
D\omega_j \gg 1 : \;\;\; B^{(a)} \sim \frac{D^4\sigma_8^4}{D\omega} \propto D^3 
\sigma_8^3 ,
\label{ZelsdBascal}
\eeq
since $\omega \propto \sigma_8$.
Thus, the nonlinear corrections resummed in the large-$N$ steepest-descent method
have modified the scaling over $D$ of the tree diagram (a) as compared with
the standard perturbation theory result (\ref{ZelBtreedel}) which scales as the
first term in (\ref{BDstandard}).

Next, let us consider the first one-loop diagram (f) of Fig.~\ref{figC3sd}.
We can again factorize the time-dependent parts which read as
\beqa
\lefteqn{ \int_0^D \prod_{j=1}^3 \frac{\d D_j'}{D_j'} \left(\frac{D}{D_j'}
\right)^{\alpha_{\ell_j}} D_j'^2 \cos[\omega_j(D-D_j')] } \nonumber \\
&& \;\;\;\; = \frac{D^6}{(2-\alpha_{\ell_1})(2-\alpha_{\ell_2})
(2-\alpha_{\ell_3})} 
T_{(\ell_1,\ell_2,\ell_3)}^{(f)}(\omega_j D) .
\label{Tf}
\eeqa
This defines the functions $T_{(\ell_j)}^{(f)}$ of the three variables
$\omega_j D$. They are parameterized by the indices $\ell_j$ of the response
matrices $R_{(\ell_j)}$ of Eq.(\ref{ZelR1R2}). Since the diagram (f) involves 
three response functions, there are $2^3=8$ such functions 
$T_{(\ell_j)}^{(f)}$. However, thanks to the symmetry of diagram (f), 
$T_{(\ell_j)}^{(f)}$ can be split into a product of three terms as 
\beq
T_{(\ell_1,\ell_2,\ell_3)}^{(f)}(\omega_j D) \! = \! 
T_{(\ell_1)}^{(f)}(\omega_1 D) T_{(\ell_2)}^{(f)}(\omega_2 D) 
T_{(\ell_3)}^{(f)}(\omega_3 D) ,
\label{Tfprod}
\eeq
with
\beq
T_{(\ell)}^{(f)}(\omega D) = (2-\alpha_{\ell}) \int_0^1 \frac{\d x}{x} 
x^{2-\alpha_{\ell}} \cos[\omega D(1-x)] .
\label{Tf1}
\eeq
The integration over $x$ yields
\beq
T_{(1)}^{(f)}(\omega D) = {\rm sinc}(\omega D) , \;\;\; 
T_{(1)}^{(f)}(\omega D) = {\rm cosc}(\omega D) ,
\label{ZelTf1}
\eeq
which gives the nonlinear scaling
\beq
D\omega_j \gg 1 : \;\;\; T_{(1,1,1)}^{(f)}(\omega_j D) \sim
\frac{1}{(D\omega)^3} ,
\label{ZelTf111}
\eeq
and higher powers of $1/(D\omega)$ arise if some indices $\ell_j$ are equal 
to 2. Then, diagram (f) reads as
\beqa
\lefteqn{B^{(f)} \!\! = \!\! \int \!\! \d\bq_1\d\bq_2\d\bq_3 
\; \delta_D(\bq_1+\bq_2-\bk_1)  \delta_D(\bq_1+\bq_3+\bk_2) } \nonumber \\
&& \!\!\! \times \frac{T_{(1)}^{(f)}(\omega_1 D) k_1^2 (\bq_1.\bq_2) 
+ T_{(2)}^{(f)}(\omega_1 D) [q_1^2q_2^2-(\bq_1.\bq_2)^2]}{q_1^2q_2^2} \nonumber \\
&& \!\!\! \times \frac{T_{(1)}^{(f)}(\omega_2 D) k_2^2 (\bq_1.\bq_3) 
+ T_{(2)}^{(f)}(\omega_2 D) [q_1^2q_3^2-(\bq_1.\bq_3)^2]}{q_1^2q_3^2} \nonumber \\
&& \!\!\! \times \frac{-T_{(1)}^{(f)}(\omega_3 D) k_3^2 (\bq_2.\bq_3) 
+ T_{(2)}^{(f)}(\omega_3 D) [q_2^2q_3^2-(\bq_1.\bq_3)^2]}{q_2^2q_3^2} \nonumber \\
&& \!\!\! \times \; P_L(q_1) P_L(q_2) P_L(q_3) .
\label{ZelsdBf}
\eeqa
We can check that by substituting $T_{(\ell)}^{(f)}=1$ we again recover the 
corresponding one-loop diagram (f) of the expansion over $K_s$ of 
Sect.~\ref{Zelcubic}, which is also given by Eq.(\ref{ZelBe}) obtained from 
the standard perturbative expansion of Sect.~\ref{ZelStandard}. Then, we obtain 
for this one-loop diagram (f) the scaling
\beq
D\omega_j \gg 1 : \;\;\; B^{(f)} \sim \frac{D^6\sigma_8^6}{(D\omega)^3} \propto 
D^3\sigma_8^3 .
\label{ZelsdBfscal}
\eeq
Thus, the scaling of this one-loop diagram is again modified as compared with 
the corresponding one-loop diagram obtained in the standard
perturbation theory, which was equal to Eq.(\ref{ZelBe}) and which scaled as the
second term of (\ref{BDstandard}). Moreover, contrary to the standard expansion
scheme, the one-loop diagram shows the same scaling as the tree diagram
(compare with Eq.(\ref{ZelsdBascal})). 

The three other one-loop diagrams of Fig.~\ref{figC3sd} can be computed in the 
same fashion. In particular, because of the triple integration over inner 
times $D_j'$, weighted by the factors $\cos[\omega_j(D-D_j')]$ as in 
Eq.(\ref{Tf}), we can check that all time-dependent factors $T$ obey the same 
scaling as in Eq.(\ref{ZelTf111}),
\beq
D\omega_j \gg 1 : \;\;\; T_{(\ell_j)}^{(f,g,h,i)}(\omega_j D) \sim
\frac{1}{(D\omega)^3} ,
\label{ZelTfghi}
\eeq
with possible higher powers if some $\ell_j$ are equal to $2$.
Therefore, all one-loop diagrams scale as Eq.(\ref{ZelsdBfscal}) and as the
tree diagram of Eq.(\ref{ZelsdBascal}) computed at this same one-loop order.
Thus, we recover the nice improvement over the standard perturbation theory
already obtained for the two-point correlation: the one-loop correction does not
grow with an additional power $D^2$ as compared with the tree-order result and 
seems better behaved. Unfortunately, as shown in Valageas (2007b), at higher 
orders exponentially growing terms appear within this steepest-descent expansion 
scheme. Therefore, at two-loop order and beyond, this large-$N$ method becomes 
even more ill-behaved than the standard expansion in the sense of exhibiting 
strongly growing high-order terms.

On the other hand, we can note that the growth as $D^2\sigma_8^2$ of the 
two-point correlation (see Eq.(\ref{PDcos})) and as $D^3\sigma_8^3$ of the 
three-point correlation breaks physical constraints, that is, it cannot be 
achieved by any matter density distribution. Indeed, the positivity of the 
matter density $\rho(\bx)$ implies that $C_{\delta 3}$ should grow at least as 
$(C_{\delta 2})^2$ in the highly nonlinear regime, see Eq.(6) in 
Valageas (1999). 
This implies that the strong constraint $\rho(\bx)\geq 0$ is not satisfied by 
the large-$N$ steepest-descent expansion (nor by the usual perturbative 
expansion). The growth factors $D^2$ and $D^3$ are rather reminiscent of generic
unconstrained fields $\psi$ with a time-dependence which would approximately
factorize as $\psi(\bx,D) \sim D$.  

Finally, we must take care of the UV divergences of diagrams (h) and (i) already
encountered in Sect.~\ref{Zelcubic}. In the expansion over $K_s$, described in
Sect.~\ref{Zelcubic}, the UV divergences of diagrams (h) and (i) exactly 
compensate and we could as well use the regularized contributions $B^{(h)}_R$
and $B^{(h)}_R$ of Eqs.(\ref{ZelcubicBhR})-(\ref{ZelcubicBiR}). For the
steepest-descent scheme this is no longer true because the divergent parts
are multiplied by the independent functions $T^{(h)}$ and $T^{(i)}$.
On the other hand, at order $P_L^3$ we could substitute $T^{(h)}=T^{(i)}=1$
into diagrams (h) and (i), which already involve a product over three
terms $P_L$, so that the divergences again compensate.
In fact, at order $P_L^3$ we must recover the results
of the standard expansion scheme hence we know that such a UV divergence can
only appear at higher orders (if at all). Therefore, in a fashion similar
to the procedure of Sect.~\ref{Zelcubic}, we define regularized contributions
$B^{(h)}_R$ and $B^{(i)}_R$ by subtracting from $B^{(h)}$ and $B^{(i)}$
their UV divergent part. This makes no difference for the sum of contributions
(h) and (i) at order $P_L^3$ (but we have removed higher-order terms which 
happened to diverge and did not cancel in the sum).
This gives:
\beqa
\lefteqn{\!\!B^{(h)}_R \! = \! - \frac{T^{(h)}_{(111)} 2 (\bk_2.\bk_3) k_1^2 
+ T^{(h)}_{(211)} 3  [k_2^2 k_3^2 - (\bk_2.\bk_3)^2]}{6 k_2^2 k_3^2} } 
\nonumber \\
&& \!\!\! \times P_L(k_2) P_L(k_3) \! \int\!\!\d\bq P_L(q) 
\frac{\!(\bk_2.\bq) (\bk_3.\bq)\!}{q^4} + 2 \, \rm{perm.} 
\label{ZelsdBhR}
\eeqa
and
\beqa
\lefteqn{\!\!B^{(i)}_R \! = \! - \frac{T^{(i)}_{(111)} (\bk_2.\bk_3) k_1^2 
+ T^{(h)}_{(212)} [k_2^2 k_3^2 - (\bk_2.\bk_3)^2]}{6 k_2^2 k_3^2} } 
\nonumber \\
&& \!\!\! \times P_L(k_2) P_L(k_3) \! \int\!\!\d\bq P_L(q) 
\frac{(\bk_1.\bq)^2}{q^4} + 2 \, \rm{perm.} 
\label{ZelsdBiR}
\eeqa
Of course, we can check that by substituting $T^{(h)}=T^{(i)}=1$ into 
Eqs.(\ref{ZelsdBhR})-(\ref{ZelsdBiR}), and using $\bk_1+\bk_2+\bk_3=0$,
we recover the regularized contributions (\ref{ZelcubicBhR})-(\ref{ZelcubicBiR})
introduced in Sect.~\ref{Zelcubic} for the expansion over $K_s$.
On the other hand, this analysis shows that the partial resummations involved
in the large-$N$ methods have introduced spurious UV divergences.
For the Zeldovich dynamics, we know that such UV divergences are not seen in
the  exact nonlinear results that can be obtained from the solution (\ref{xq})
of the dynamics. Therefore, they must cancel out by including further diagrams
or by expanding back over powers of $P_L$, as in the standard perturbation 
theory. The fact that they do not cancel out automatically may be seen as a 
shortcoming of resummations schemes such as these large-$N$ methods. 
Indeed, the use of partial resummations is intended to improve the behavior
of the predicted nonlinear correlations as compared with the results of the 
standard perturbative expansions. For instance, one would like to recover
the damping of the response function in the nonlinear regime and a better control
of higher-order terms. At one-loop order this is partly achieved for the 
bispectrum as all contributions show the same scaling (\ref{ZelsdBfscal}) 
instead of the hierarchy (\ref{BDstandard}). However, this comes at the 
price of an increased
sensitivity to the smallest scales that even leads to a divergence for 
$n(\infty)\geq -3$. Nevertheless, since this divergence disappears in the 
expansion over $P_L$ it is always possible to remove it by subtraction as
in Eqs.(\ref{ZelsdBhR})-(\ref{ZelsdBiR}).

\section{Gravitational dynamics}
\label{Gravitational-dynamics}

As seen in Sects.~\ref{Equations-of-motion}-\ref{LargeN}, we can apply to the
gravitational dynamics the expansion schemes described in Sect.~\ref{Zeldovich}
for the Zeldovich dynamics. We do not give the explicit expressions of various
diagrams here as they are rather long, except for the expansion over $K_s$
in the Appendix.

\subsection{Standard perturbation theory}
\label{gravStandard}

For the standard perturbation theory recalled in 
Sect.~\ref{Standard-perturbation-theory}, the kernels $F_n$ are computed from
recursion relations derived from the equation of motion (\ref{OKsdef}).
Following the procedure described in Sect.~\ref{Standard-perturbation-theory},  
this gives the well-known results for the matter power spectrum and bispectrum
that we do not repeat here (Bernardeau et al. 2002 and references therein; 
Scoccimarro \& Frieman 1996a,b; Scoccimarro 1997).
The conditions of convergence of individual diagrams are the same as for the 
Zeldovich dynamics of Sect.~\ref{ZelStandard}, that is, $n(0)>-1$ and 
$n(\infty)<-1$. Again, since the Galilean invariance applies to both dynamics,
the infrared divergences cancel out as we sum all diagrams
so that we only require $n(0)>-3$ (Vishniac 1983; Jain \& Bertschinger 1996).
Finally, the dependence on the linear growth factor $D(\eta)$ and on the 
amplitude $\sigma_8$ of 
the linear power spectrum are as in Eqs.(\ref{PDstandard})-(\ref{BDstandard}).
We do not plot the results here since they are identical to the results obtained
from the expansion over $K_s$ displayed in the next section.

\subsection{Expansion over cubic interaction}
\label{gravcubic}

\begin{figure}[htb]
\begin{center}
\epsfxsize=8 cm \epsfysize=7 cm {\epsfbox{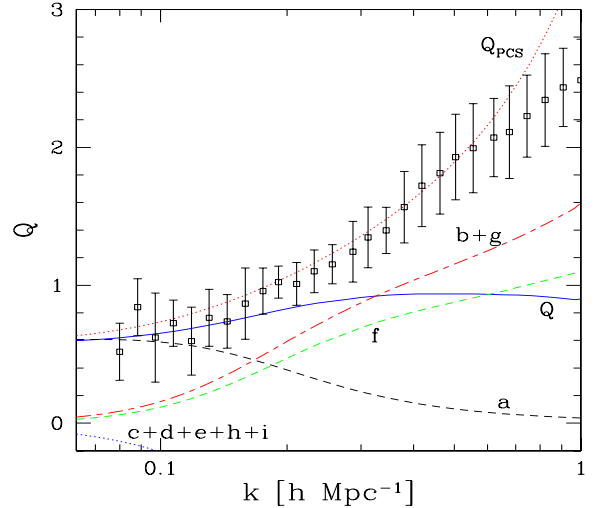}}
\end{center}
\caption{The reduced bispectrum $Q$ defined at one-loop order by the rational
function (\ref{Qfrac}). We show $Q$ as a function of wavenumber $k$ for 
equilateral triangles $k_1=k_2=k_3=k$. The data points are taken from Fosalba 
et al. (2005), for a $\Lambda$CDM simulation at $z=0$. 
The letters correspond to the various contributions,
such as $B^{(a)}/(S^{\rm tree}+S^{\rm 1loop})$, associated with tree and one-loop
diagrams obtained for the bispectrum $B$ within the expansion over $K_s$.
The solid line $Q$ is the full reduced bispectrum $Q$ (i.e. the sum of all these
contributions). It is also equal to the prediction of the standard 
perturbation theory. The rising dotted line $Q_{\rm PCS}$ is the 
phenomenological model of Pan et al. (2007). }
\label{figBkfrac_z0}
\end{figure}

\begin{figure}[htb]
\begin{center}
\epsfxsize=8 cm \epsfysize=7 cm {\epsfbox{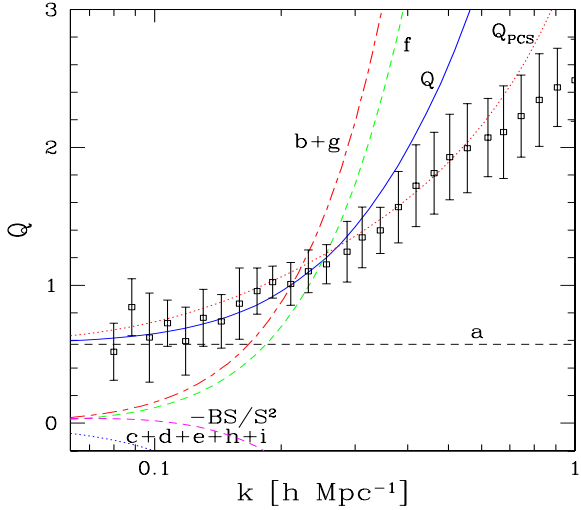}}
\end{center}
\caption{The reduced bispectrum $Q$ at one-loop order for equilateral triangles
as in Fig.~\ref{figBkfrac_z0}, but computed from the expansion (\ref{Qexp})
instead of the ratio (\ref{Qfrac}). The symbols are as in Fig.~\ref{figBkfrac_z0}
but the contributions to the bispectrum $B$ associated with tree and one-loop
diagrams are now divided by $S^{\rm tree}$, as in the first two terms of
Eq.(\ref{Qexp}), instead of $(S^{\rm tree}+S^{\rm 1loop})$, and the new curve
labeled $-BS/S^2$ corresponds to the last term of Eq.(\ref{Qexp}).}
\label{figBkexp_z0}
\end{figure}

\begin{figure}
\begin{center}
\epsfxsize=4.3 cm \epsfysize=4.6 cm {\epsfbox{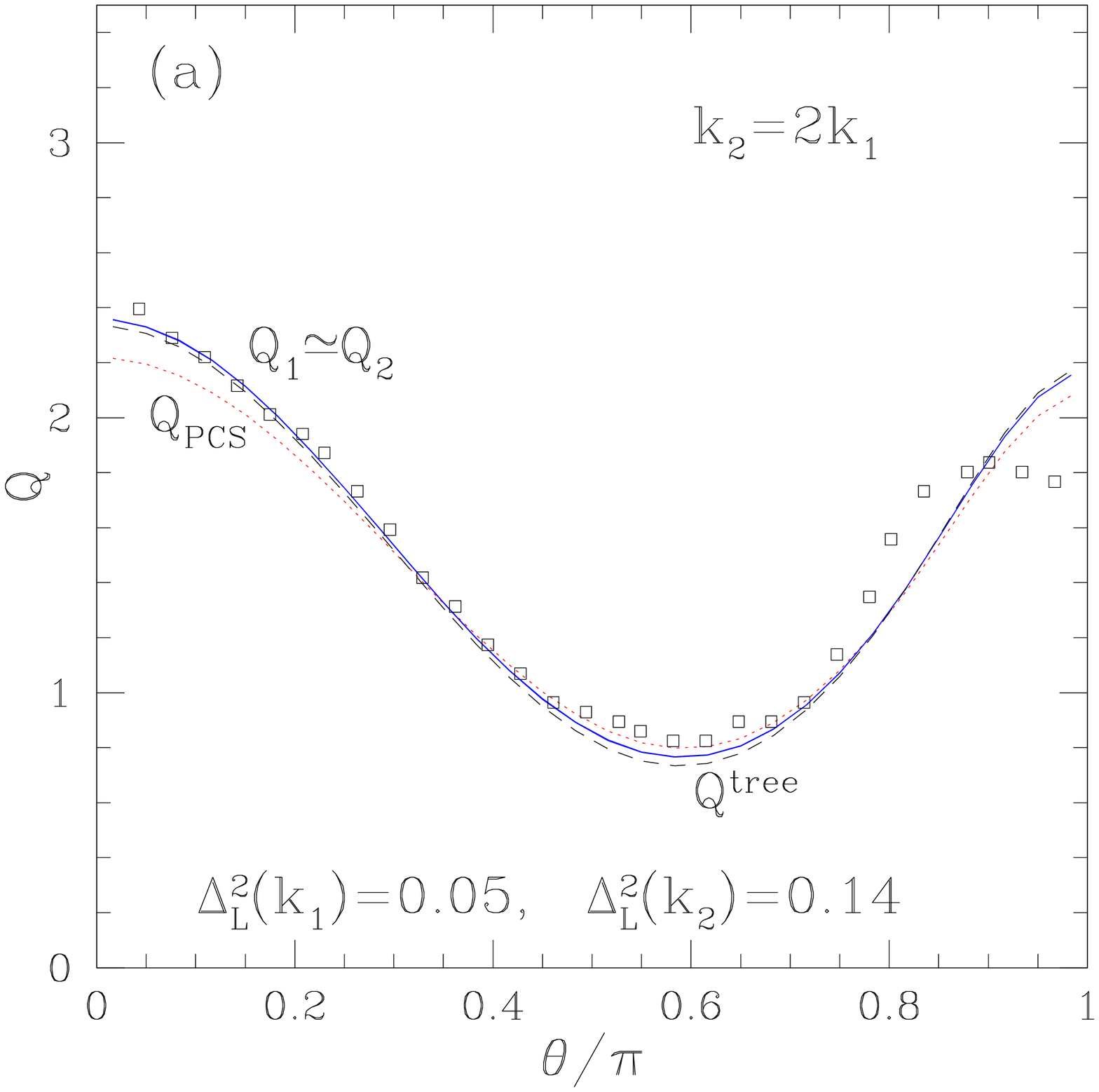}} 
\epsfxsize=4.3 cm \epsfysize=4.6 cm {\epsfbox{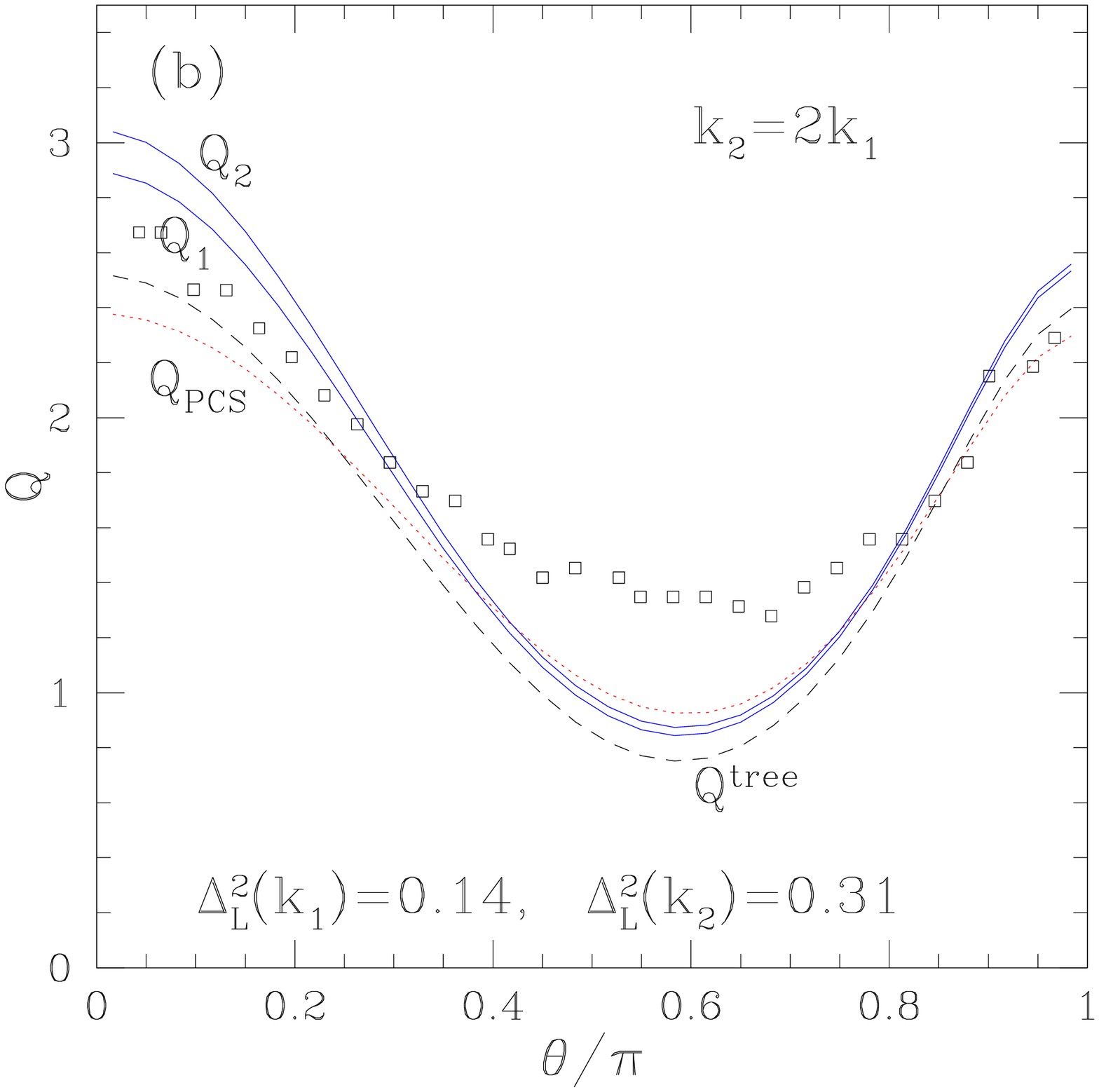}}\\
\epsfxsize=4.3 cm \epsfysize=4.6 cm {\epsfbox{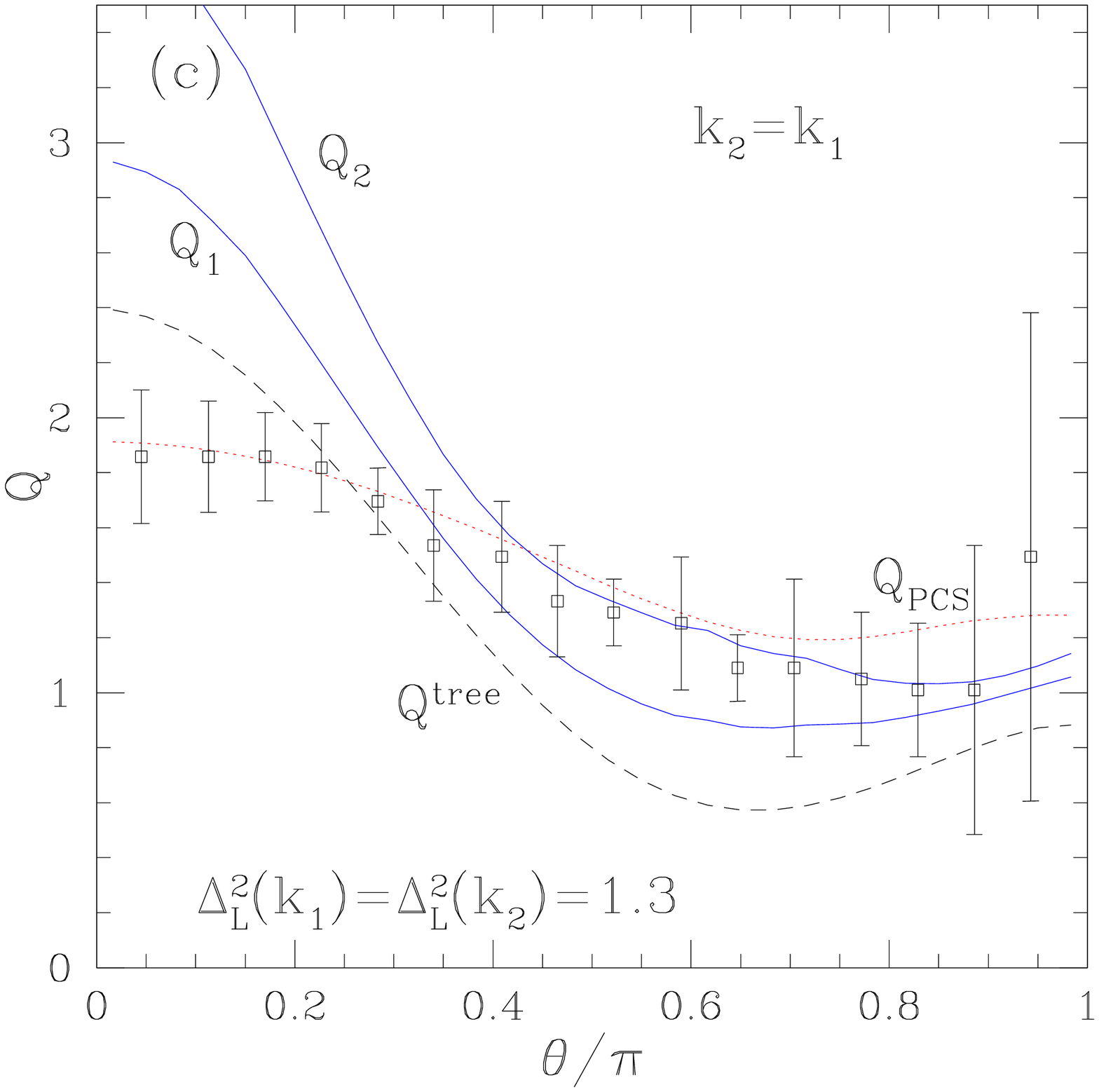}} 
\epsfxsize=4.3 cm \epsfysize=4.6 cm {\epsfbox{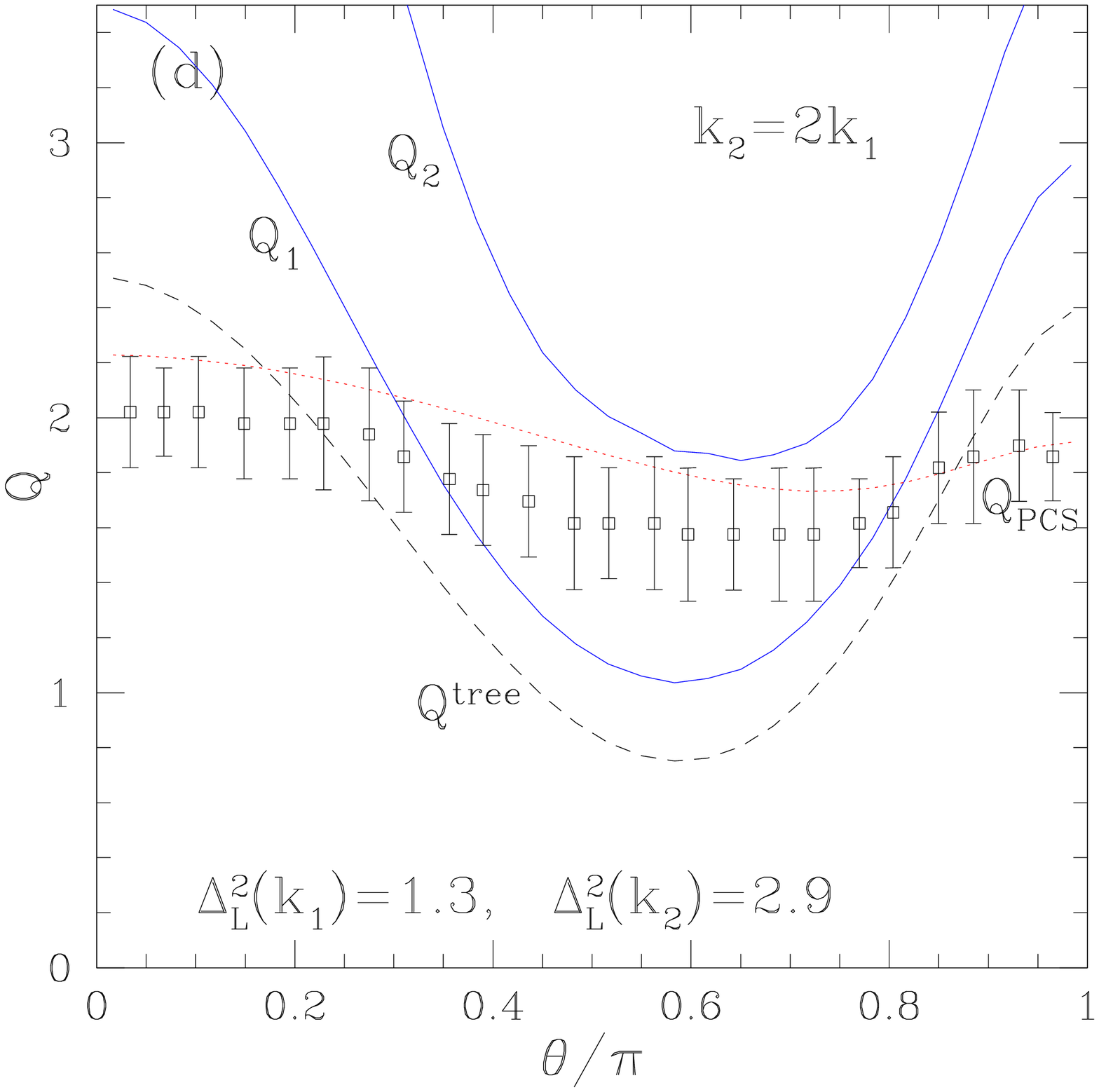}}
\end{center}
\caption{The reduced bispectrum $Q$ as a function of the angle $\theta$ between
wavenumbers $\bk_1$ and $\bk_2$. Panels (a), (b), and (d) have $k_2=2 k_1$ 
whereas panel (c) has $k_2=k_1$. We probe deeper into the nonlinear regime by
going from panel (a) to panel (d). The solid line $Q_1$ corresponds to the
ratio (\ref{Qfrac}) whereas the solid line $Q_2$ corresponds to its expansion
(\ref{Qexp}). They are almost identical in panel (a). The dashed line is
the tree-order result (\ref{Qtree}). The dotted line $Q_{\rm PCS}$ is the 
phenomenological model of Pan et al. (2007). Data points are taken from
Scoccimarro \& Couchman (2001), for panels (a) and (b), and
from Fosalba et al. (2005), for panels (c) and (d). They correspond to two 
different $\Lambda$CDM cosmologies.}
\label{figBt}
\end{figure}

The expansion over powers of $K_s$ proceeds exactly as for the Zeldovich dynamics
studied in Sect.~\ref{Zelcubic}. For the power spectrum there is again a 
one-to-one correspondence between the diagrams (b) and (c) of 
Fig.~\ref{figC21loop} and Fig.~\ref{figC2standard}. For the bispectrum, the
correspondence is again between the groups of diagrams $\{(a),(b+c),(d),(e)\}$
of Fig.~\ref{figC3standard} and $\{(a),(c+d+e+h+i),(b+g),(f)\}$ of 
Fig.~\ref{figC31loop}. For completeness, we give the expressions of
these diagrams in the Appendix.

These results are similar to those obtained for the Zeldovich dynamics in
Sect.~\ref{Zelcubic}, except that the rational functions that appear
in the integrals are more intricate. In particular, the denominators
do not factorize as products of $k_j$ and $q_j$ so that angular integrations 
require more care. The conditions of convergence are the same as for the 
Zeldovich dynamics, that is, $n(0)>-1$ and $n(\infty)<-1$, except for 
contributions $B^{(h)}$ and $B^{(i)}$ which require $n(\infty)<-3$.
Again, the infrared and ultraviolet divergences compensate so that the sum
is well-defined for $n(0)>-3$ and $n(\infty)<-1$. As in 
Eqs.(\ref{ZelcubicBhR})-(\ref{ZelcubicBiR}), it is convenient to introduce
the regularized contributions $B^{(h)}_R$ and $B^{(i)}_R$ by subtracting
from $B^{(h)}$ and $B^{(i)}$ their UV divergent part.

From the bispectrum $B(k_1,k_2,k_3)$ and the power spectrum $P(k)$ we define
as usual the reduced bispectrum $Q(k_1,k_2,k_3)$ by
\beq
Q(k_1,k_2,k_3) = \frac{B(k_1,k_2,k_3)}{P(k_2) P(k_3)+ 2 \, {\rm perm.}} .
\label{Qdef}
\eeq
From Eq.(\ref{Btreestandard}), in the quasi-linear regime it is independent of 
time and of the normalization of the linear power spectrum. Let us note 
$S(k_1,k_2,k_3)$ the denominator of Eq.(\ref{Qdef}):
\beq
S(k_1,k_2,k_3) = P(k_2) P(k_3)+ 2 \, {\rm perm.}
\label{Sdef}
\eeq
Within the standard perturbative expansion over powers of $P_L$, or the equivalent
expansion over $K_s$, the ratio (\ref{Qdef}) reads up to one-loop order as
\beqa
Q &  = & \frac{B^{\rm tree}+B^{\rm 1loop}}{S^{\rm tree}+S^{\rm 1loop}} 
\label{Qfrac} \\
& = & \frac{B^{\rm tree}}{S^{\rm tree}} + \frac{B^{\rm 1loop}}{S^{\rm tree}} 
-  \frac{B^{\rm tree}}{S^{\rm tree}} \frac{S^{\rm 1loop}}{S^{\rm tree}} , 
\label{Qexp} 
\eeqa
where in the second line we have expanded the ratio $B/S$ of the first line.
The various contributions to the bispectrum have been described in the previous
sections, whereas the denominator $S$ is given up to one-loop order by
\beqa
S^{\rm tree} & = & P_L(k_2) P_L(k_3)+ 2 \, {\rm perm.} 
\label{Stree} \\
S^{\rm 1loop} & = & P_L(k_2) P^{\rm 1loop}(k_3) + 5 \, {\rm perm.}
\label{Sloop} 
\eeqa

We show in Fig.~\ref{figBkfrac_z0} the results obtained for the reduced 
bispectrum $Q$ as defined by Eq.(\ref{Qfrac}), that is, we keep the ratio
(\ref{Qfrac}) without expanding again as in Eq.(\ref{Qexp}). We plot $Q$ as
a function of wavenumber $k$ for equilateral triangles $k_1=k_2=k_3=k$.
The data points are taken from Fosalba et al. (2005). They correspond to 
a $\Lambda$CDM simulation from the Virgo archive at redshift $z=0$. 
The four dashed curves labelled $\{a,c+d+e+h+i,b+g,f\}$ are the contributions
of the various diagrams obtained for $B$ in the expansion over $K_s$. Diagrams
of the same form (i.e., that involve the same number of integrations over the
linear power spectrum) have been gathered and divided by the denominator
$S^{\rm tree}+S^{\rm 1loop}$. Their sum is the full reduced bispectrum $Q$
shown by the solid line. The rising dotted line $Q_{\rm PCS}$ is the 
phenomenological model of Pan et al. (2007). It is based on the scale 
transformation introduced by Hamilton et al. (1991) and it uses as input the
nonlinear power spectrum, obtained from the fit to numerical simulations
given by Smith et al.(2003). 
We show in Fig.~\ref{figBkexp_z0} the results obtained for the same case when
we use the expansion (\ref{Qexp}) instead of the ratio (\ref{Qfrac}). The new
curve labeled $-BS/S^2$ is the last term of Eq.(\ref{Qexp}).

Let us recall here that the predictions for the bispectrum $B$, hence for the
reduced bispectrum $Q$, obtained from this expansion over $K_s$, are equal
to the results of the standard perturbation theory.
Note that the tree-order result, given by
\beq
Q^{\rm tree} = \frac{B^{\rm tree}}{S^{\rm tree}} ,
\label{Qtree}
\eeq
also corresponds to the horizontal line $(a)$ in Fig.~\ref{figBkexp_z0}. 
It is constant since
both $B^{\rm tree}$ and $S^{\rm tree}$ scale as $P_L(k)^2$. We can check that
the reduced bispectra $Q$ obtained from the two Eqs.~(\ref{Qfrac})-(\ref{Qexp})
agree on quasi-linear scales, up to $0.2 h$ Mpc$^{-1}$, where 
$\Delta^2_L(k) \simeq 1$ (here $\Delta^2_L(k)=4\pi k^3 P(k)$ is the power per
logarithmic wavenumber). At smaller scales the ratio (\ref{Qfrac}) goes to
a constant (that is, a value that is independent of the normalization of $P_L$
but shows a weak dependence on the shape of the linear power spectrum, since
both $B^{\rm 1loop}$ and $S^{\rm 1loop}$ scale as $P_L^3$), whereas the expanded
version (\ref{Qexp}) grows as $\Delta^2_L$. This leads to a better agreement of 
Eq.(\ref{Qexp}) with the numerical results up to $0.4 h$ Mpc$^{-1}$, where 
$\Delta^2_L(k) \simeq 4$. However, it is clear that this better match has no
strong foundations. We can also note that in the nonlinear regime there are some
cancellations between various contributions, for both 
Eqs.~(\ref{Qfrac})-(\ref{Qexp}).
On the other hand, it appears that the phenomenological model
of Pan et al. (2007) works best, up to  $0.8 h$ Mpc$^{-1}$, where 
$\Delta^2_L(k) \simeq 17$. However, at $k \sim 0.1-0.2 h$ Mpc$^{-1}$ it may 
slightly overestimate $Q$.

We show in Fig.~\ref{figBt} the reduced bispectrum $Q$ as a function
of the angle $\theta$ between the two wavenumbers $\bk_1$ and $\bk_2$:
\beq
\cos(\theta)=\frac{\bk_1.\bk_2}{k_1 k_2} , \;\;\;
k_3 = \sqrt{k_1^2+k_2^2+2 k_1 k_2 \cos(\theta)} .
\label{theta}
\eeq
In the lower left panel (c) we have $k_2=k_1$ whereas in other panels 
$k_2=2 k_1$. We probe smaller scales farther into the nonlinear regime going
from panel (a) to panel (d). On large scales, all curves match the tree-order
result and numerical data, as seen in panel (a), and show a strong dependence
on the angle $\theta$. On smaller scales, the dependence on $\theta$ decreases, 
as shown by the simulations. This behavior is captured by the phenomenological
model of Pan et al. (2007) but not by the one-loop predictions of 
Eqs.~(\ref{Qfrac})-(\ref{Qexp}). In agreement with Figs.~\ref{figBkfrac_z0}-
\ref{figBkexp_z0}, the ratio (\ref{Qfrac}) yields a smaller value for $Q$
than the expansion (\ref{Qexp}), but none of these two predictions matches the
data for $\Delta^2_L(k) > 0.3$.
The results displayed in Figs.~\ref{figBkfrac_z0}-\ref{figBt} agree with
previous studies of the standard perturbation theory, discussed for instance
in great details in the review of Bernardeau et al. (2002).

\subsection{Direct steepest-descent method}
\label{gravsteepest-descent}

For the gravitational dynamics, the nonlinear response function is no longer
given by Eq.(\ref{Rcos}) at one-loop order in the steepest-descent expansion.
However, Eqs.(\ref{Rcos}) and (\ref{Ccosk}) still provide a reasonable
approximation of the exact one-loop result, as seen in Valageas (2007a).
Thus, in this article we use the approximations (\ref{Rcos}) and (\ref{Ccosk})
to investigate the behavior of the  steepest-descent method for the
gravitational dynamics as well. Then, from Eq.(\ref{RL}) the powers 
$\alpha_{\ell}$ and the matrices $R_{(\ell)}$ of 
Eqs.(\ref{Zelalpha12})-(\ref{ZelR1R2}) must be replaced by
\beq
\alpha_1=1,  \;\;\; \alpha_2=-\frac{3}{2} ,
\label{gravalpha12}
\eeq
\beq
R_{(1)}= \frac{1}{5} \left(\bea{cc} 3 & 2 \\ 3 & 2 \ea\right) , \;\;\;
R_{(2)}= \frac{1}{5} \left(\bea{cc} 2 & -2 \\ -3 & 3 \ea\right) .
\label{gravR1R2}
\eeq
Next, the computation of various diagrams proceeds as in 
Sect.~\ref{Zelsteepest-descent}. In particular, since the linear growing modes
are identical for both dynamics (exponent $\alpha_1=1$),
the time-dependent functions $T_{(\ell_j)}^{(.)}$ are identical if all
$\ell_j$ are equal to $1$. Then, we recover the scalings (\ref{PDcos}) for
the power spectrum and (\ref{ZelsdBascal}), (\ref{ZelsdBfscal}), for the
bispectrum.

Again, in order to handle the UV divergences associated with diagrams (h) and
(i) it is necessary to introduce the regularized contributions $B^{(h)}_R$ 
and $B^{(i)}_R$, as in Eqs.(\ref{ZelsdBhR})-(\ref{ZelsdBiR}). 
As discussed in Sect.~\ref{Zelsteepest-descent}, this makes no difference
at order $P_L^3$ as the UV divergence is associated with higher order terms.
On the other hand, for the gravitational dynamics this UV divergence may have
a physical meaning. That is, it could be a signature of the breakdown of the
equations of motion beyond shell-crossing and may not compensate at higher orders.
Indeed, as noticed in Valageas (2002), within the
standard perturbative expansion one also encounters UV-divergent diagrams
beyond a finite order if $n(\infty)>-3$. Since the large-$N$ expansion has
performed a partial resummation of an infinite series of diagrams it is not
too surprising to encounter such UV-divergent terms. 
For the Zeldovich dynamics this is not the case, since
we know from the exact nonlinear results that these divergences must eventually
compensate once we include all diagrams of a given order, even though the Eulerian
description also breaks down beyond shell-crossing. However, for the
gravitational dynamics, where there is no such explicit nonlinear solution,
it is not known whether these divergences must also fully cancel out 
(nevertheless they must compensate at least up to the last regular order 
predicted by the standard expansion scheme).
We do not give the explicit expressions of the diagrams obtained in this fashion
here as they are too lengthy. They are similar to those obtained within the
expansion over $K_s$, given in the Appendix, with the addition of time-dependent
functions $T_{(\ell_j)}^{(a,f,g,h,i)}(\omega_j D)$, as described in details
in Sect.~\ref{Zelsteepest-descent} for the simpler case of the Zeldovich 
dynamics.

Since we use Eqs.(\ref{Rcos}) and (\ref{Ccosk}) for the nonlinear two-point
functions the reduced bispectrum $Q$ defined in Eq.(\ref{Qdef}) reads as
\beq
Q_{\cos}(k_1,k_2,k_3) = \frac{B_{\cos}(k_1,k_2,k_3)}
{P_L(k_2) P_L(k_3) + 2 \, {\rm perm.}} ,
\label{Qcos}
\eeq
where the subscript ``cos'' stands for the steepest-descent method
with the original cosine cutoffs of Eqs.(\ref{Rcos}), (\ref{Ccosk}).
At this one-loop order, $B_{\cos}$ is given by the diagrams of 
Fig.~\ref{figC3sd}, as discussed above.

\begin{figure}[htb]
\begin{center}
\epsfxsize=8 cm \epsfysize=7 cm {\epsfbox{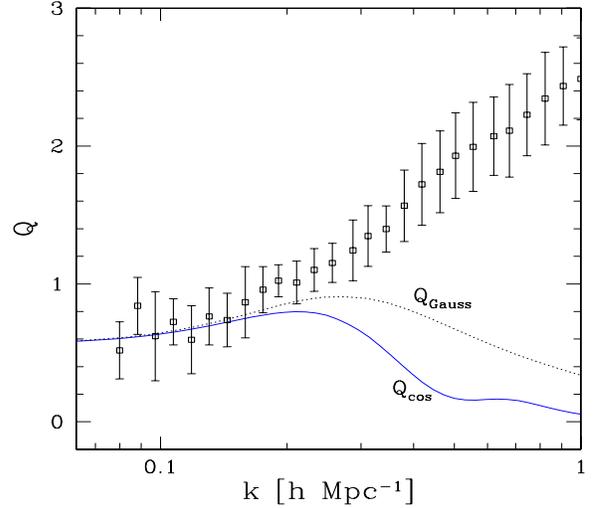}}
\end{center}
\caption{The reduced bispectrum $Q$ as a function of wavenumber $k$ for 
equilateral triangles $k_1=k_2=k_3=k$, as in Fig.~\ref{figBkfrac_z0}.
We show the prediction of the one-loop steepest-descent method from 
Eq.(\ref{Qcos}) (solid line $Q_{\cos}$) and the results obtained by using 
a Gaussian cutoff as in Eqs.(\ref{RGauss}), (\ref{CGauss}) 
(dotted line $Q_{\rm Gauss}$).}
\label{figBksd_z0}
\end{figure}

\begin{figure}
\begin{center}
\epsfxsize=4.3 cm \epsfysize=4.6 cm {\epsfbox{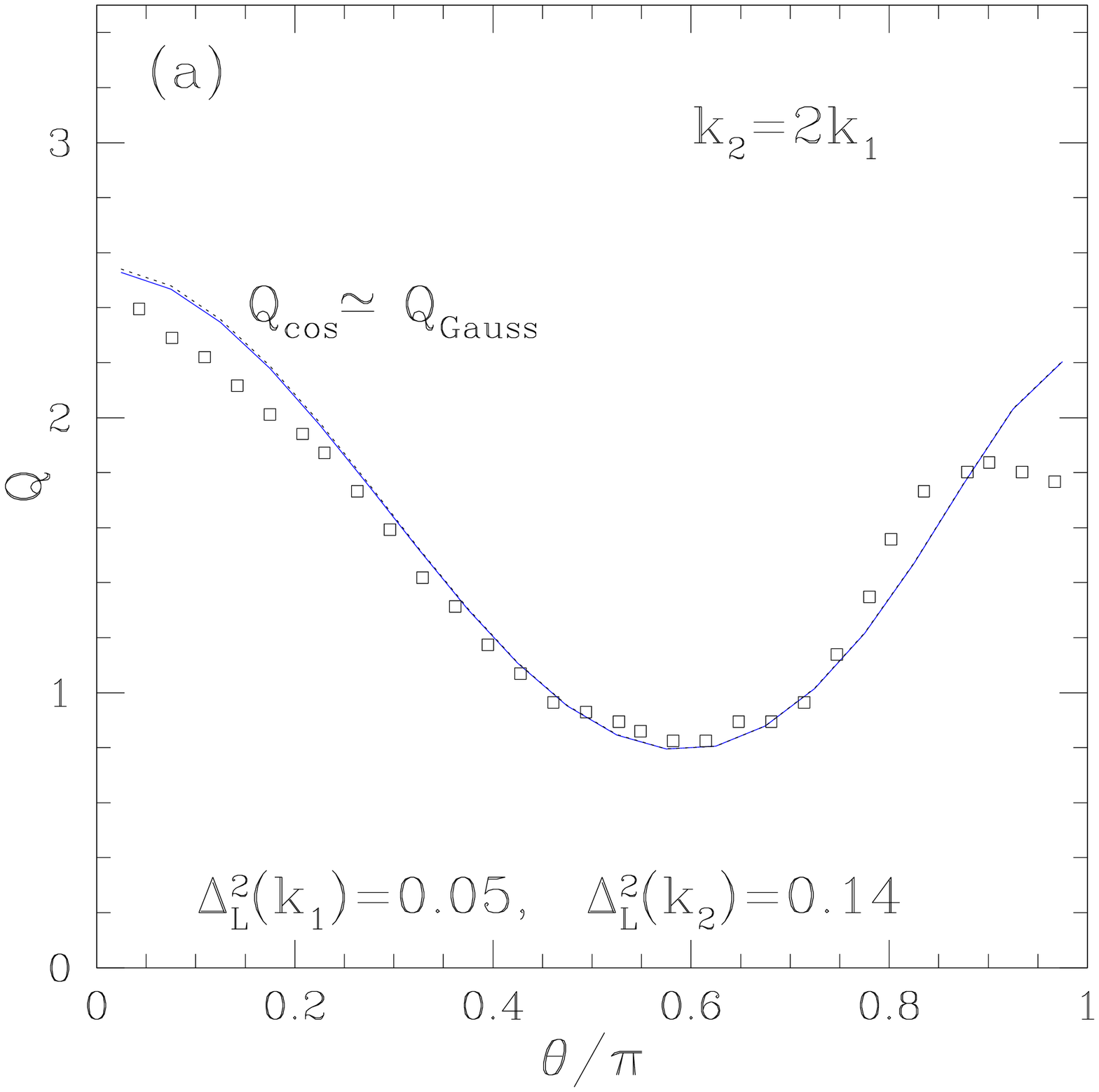}} 
\epsfxsize=4.3 cm \epsfysize=4.6 cm {\epsfbox{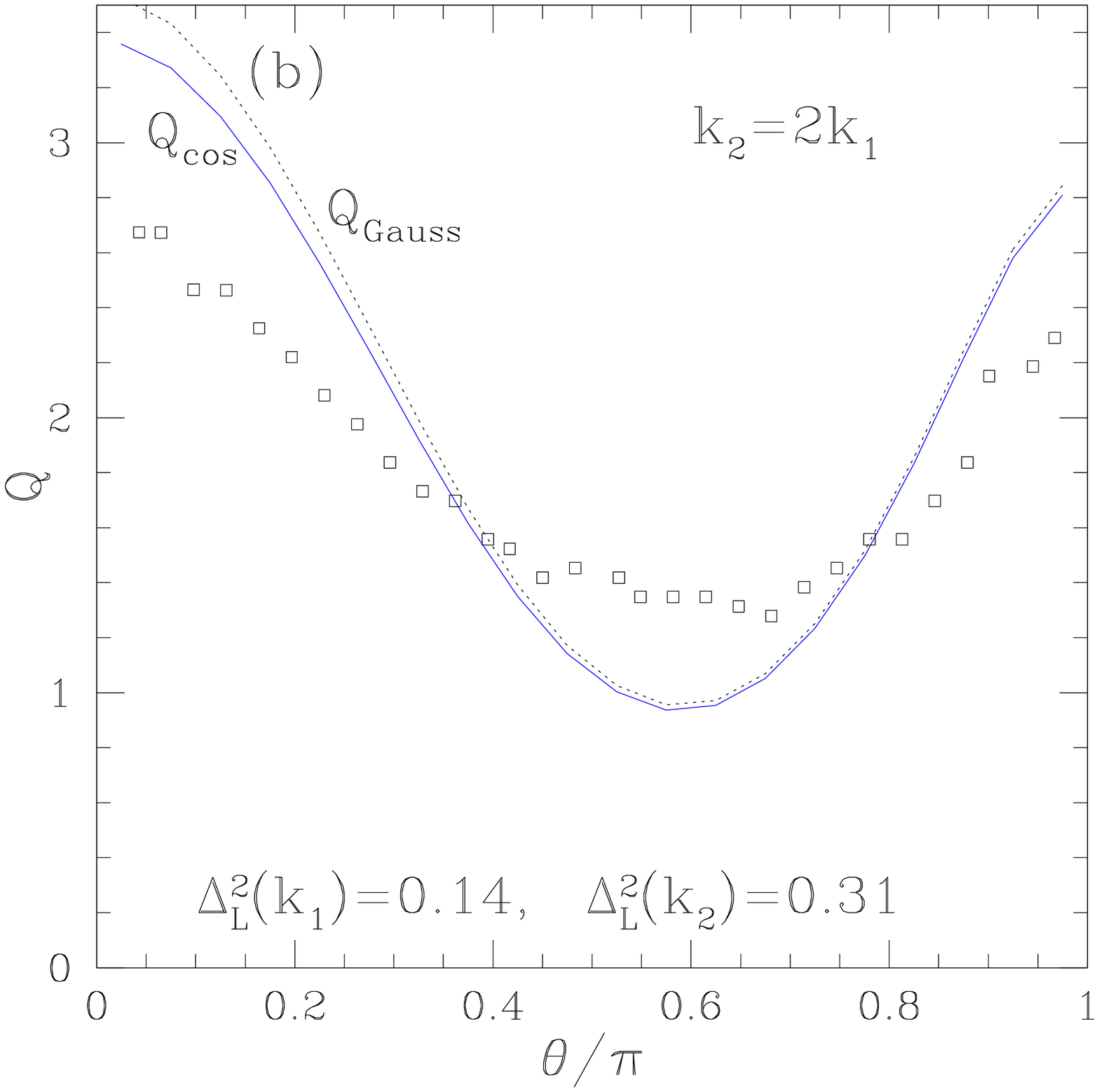}}\\
\epsfxsize=4.3 cm \epsfysize=4.6 cm {\epsfbox{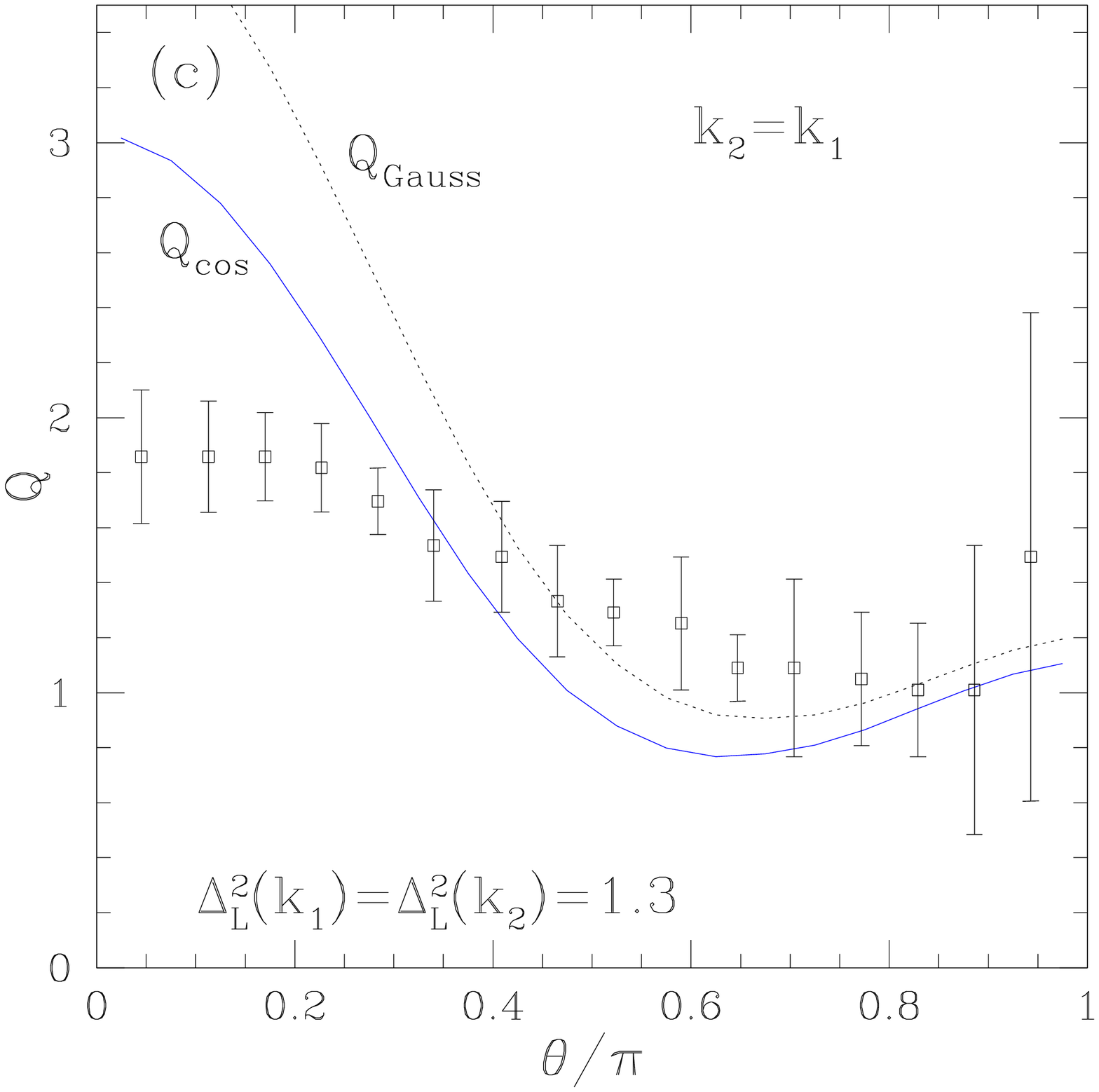}} 
\epsfxsize=4.3 cm \epsfysize=4.6 cm {\epsfbox{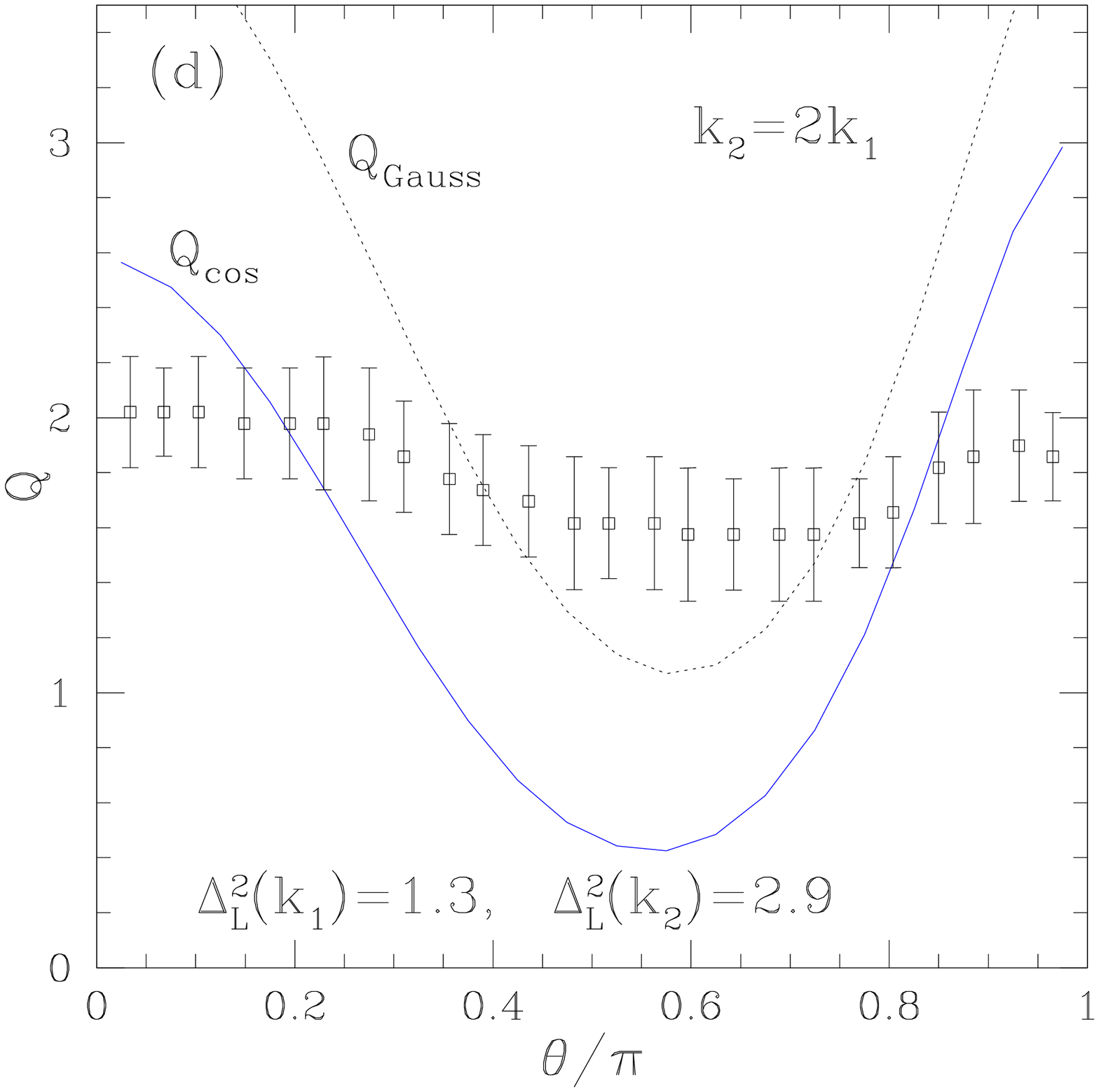}}
\end{center}
\caption{The reduced bispectrum $Q$ as a function of the angle $\theta$ between
wavenumbers $\bk_1$ and $\bk_2$, as in Fig.~\ref{figBtsd}. The curves
$Q_{\cos}$ (solid line) and $Q_{\rm Gauss}$ (dotted line)
are the predictions obtained from the one-loop steepest-descent 
method by using, i) the original nonlinear two-point functions with a cosine
cutoff, ii) the two-point functions with a Gaussian cutoff.}
\label{figBtsd}
\end{figure}

As noticed in Sect.~\ref{large-k-limit}, it is possible to consider
an alternative resummation scheme for the bispectrum, based on the diagrams
of Fig.~\ref{figC3sd} associated with the large-$N$ steepest-descent method,
but using the two-point functions (\ref{RGauss}), (\ref{CGauss}), given by the
large-$k$ partial resummation associated with the random uniform advection
of large-scale structures. This amounts to replace the cosine cutoffs of
the one-loop steepest-descent predictions (\ref{Rcos}), (\ref{Ccosk}),
by the Gaussian cutoffs of Eqs.(\ref{RGauss}), (\ref{CGauss}). Therefore,
the structure of the various contributions to the bispectrum $B$ is the same
as for the genuine steepest-descent method discussed above. The only change is
that the time-dependent functions $T_{(\ell_j)}^{(a,f,g,h,i)}(\omega_j D)$
are different as they involve integrals over Gaussian factors instead of
cosines. Nevertheless, the scalings (\ref{ZelTa_asymp}) and (\ref{ZelTfghi})
remain valid,
\beq
D\omega_j \gg 1 : \;\;\; T^{({\rm tree})} \sim \frac{1}{D\omega} , 
T^{({\rm 1loop})} \sim \frac{1}{(D\omega)^3} ,
\label{Tscal}
\eeq
where we noted by $T^{({\rm tree})}$ the time-dependent functions associated with
the tree diagram of Fig.~\ref{figC3sd} and by $T^{({\rm 1loop})}$ the 
time-dependent functions associated with the one-loop diagrams of 
Fig.~\ref{figC3sd}. Indeed, the scalings (\ref{Tscal}) simply come from the
single and triple integrations over past times associated with the tree
and one-loop diagrams. For large $D\omega$, only recent times $D'$, with
$(D-D') \la 1/\omega$, contribute because of the Gaussian or cosine cutoffs.
This leads to Eq.(\ref{Tscal}). Therefore, we again recover the same scalings 
(\ref{ZelsdBascal}) and (\ref{ZelsdBfscal}) for the tree and one-loop
contributions to the bispectrum $B$. Then, the reduced bispectrum $Q$ is again
given by Eq.(\ref{Qcos}), but using a subscript ``Gauss'' to denote the
Gaussian cutoffs associated with the resummation of Sect.~\ref{large-k-limit}.

Thus, for the equal-time bispectrum the
form of the different-time decay of two-point functions does not really matter
since it merely yields the scalings (\ref{Tscal}). 
Of course, for the different-time 
three-point correlation we recover the decay of the two-point functions.
For instance, we obtain up to one-loop order from the diagrams
of Fig.~\ref{figC3sd}
\beq
\! D_j \omega_j \! \gg 1 \! :  B  \sim 
e^{-\frac{1}{2}[(D_2-D_1)^2\omega_2^2+(D_3-D_1)^2\omega_3^2]} + 2 {\rm perm.} 
\label{BdiffDscal}
\eeq
if we use the Gaussian decay.

We show in Fig.~\ref{figBksd_z0} the results obtained from the two methods
described above for the reduced bispectrum as a function of wavenumber $k$
for equilateral triangles. We can see that the the curves agree on large
scales with each other and with the numerical simulations, up to $0.2 h$ 
Mpc$^{-1}$. The comparison with Figs.~\ref{figBkfrac_z0}, \ref{figBkexp_z0},
shows that the approximations (\ref{Rcos}), (\ref{Ccosk}), or
(\ref{RGauss}), (\ref{CGauss}), for the nonlinear two-point functions, are
accurate enough to recover the deviations from the constant tree-order result
associated with the one-loop order corrections. On small scales, the 
reduced bispectra $Q_{\cos}$ and 
$Q_{\rm Gauss}$ show a slow decay, in agreement with the scalings (\ref{PDcos}) 
and (\ref{ZelsdBascal}), (\ref{ZelsdBfscal}), which give
\beq
D \omega_j \gg 1 : \;\; Q \sim \frac{1}{D\sigma_8} .
\label{Qscal}
\eeq
The curve $Q_{\cos}$ displays some oscillations due to the cosine factors
whereas the curve $Q_{\rm Gauss}$ shows a monotonous decay due to the Gaussian
factors.

We show in Fig.~\ref{figBtsd} our results for the bispectrum as a function
of the angle $\theta$ between wavenumbers $\bk_1,\bk_2$.
We can see that the results are similar to those obtained in Fig.~\ref{figBt}
from the expansion over $K_s$ (equivalent to the standard perturbation theory).
In particular, they
do not manage to reproduce the flattening of the reduced bispectrum as a function
of $\theta$ on smaller scales. Therefore, the predictions obtained in this manner
from the steepest-descent formalism do not bring a significant improvement
over the standard perturbation theory, as shown by the comparison with
Figs.~\ref{figBkfrac_z0}-\ref{figBt}.

\subsection{Using the nonlinear power spectrum}
\label{nonlinear-power-spectrum}

\begin{figure}[htb]
\begin{center}
\epsfxsize=8 cm \epsfysize=7 cm {\epsfbox{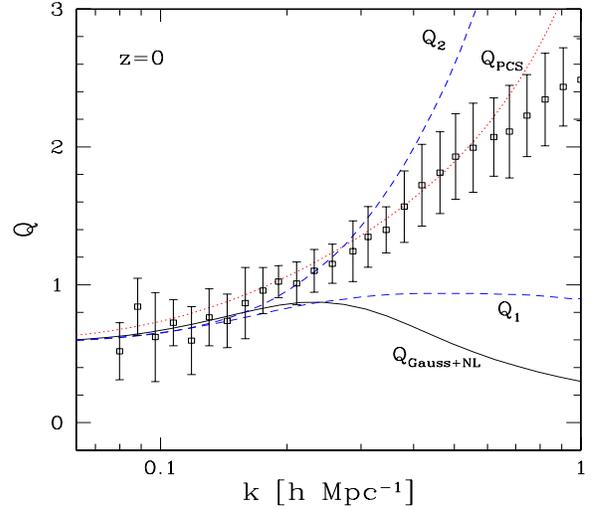}}
\end{center}
\caption{The reduced bispectrum $Q$ as a function of wavenumber $k$ for 
equilateral triangles $k_1=k_2=k_3=k$, as in Fig.~\ref{figBksd_z0}, but using
the nonlinear power spectrum from Smith et al. (2003) for the two-point
correlations within the one-loop steepest-descent diagrams, together with
the Gaussian decay ansatz (solid line $Q_{\rm Gauss+NL}$). We also plot for 
comparison the standard one-loop results, from the ratio (\ref{Qfrac}) 
(lower dashed line $Q_1$), and from its expansion (\ref{Qexp}) (upper dashed line
$Q_2$), as well as the phenomenological model of Pan et al. (2007) (dotted line
$Q_{PCS}$).}
\label{figBksdNL_z0}
\end{figure}

\begin{figure}
\begin{center}
\epsfxsize=4.3 cm \epsfysize=4.6 cm {\epsfbox{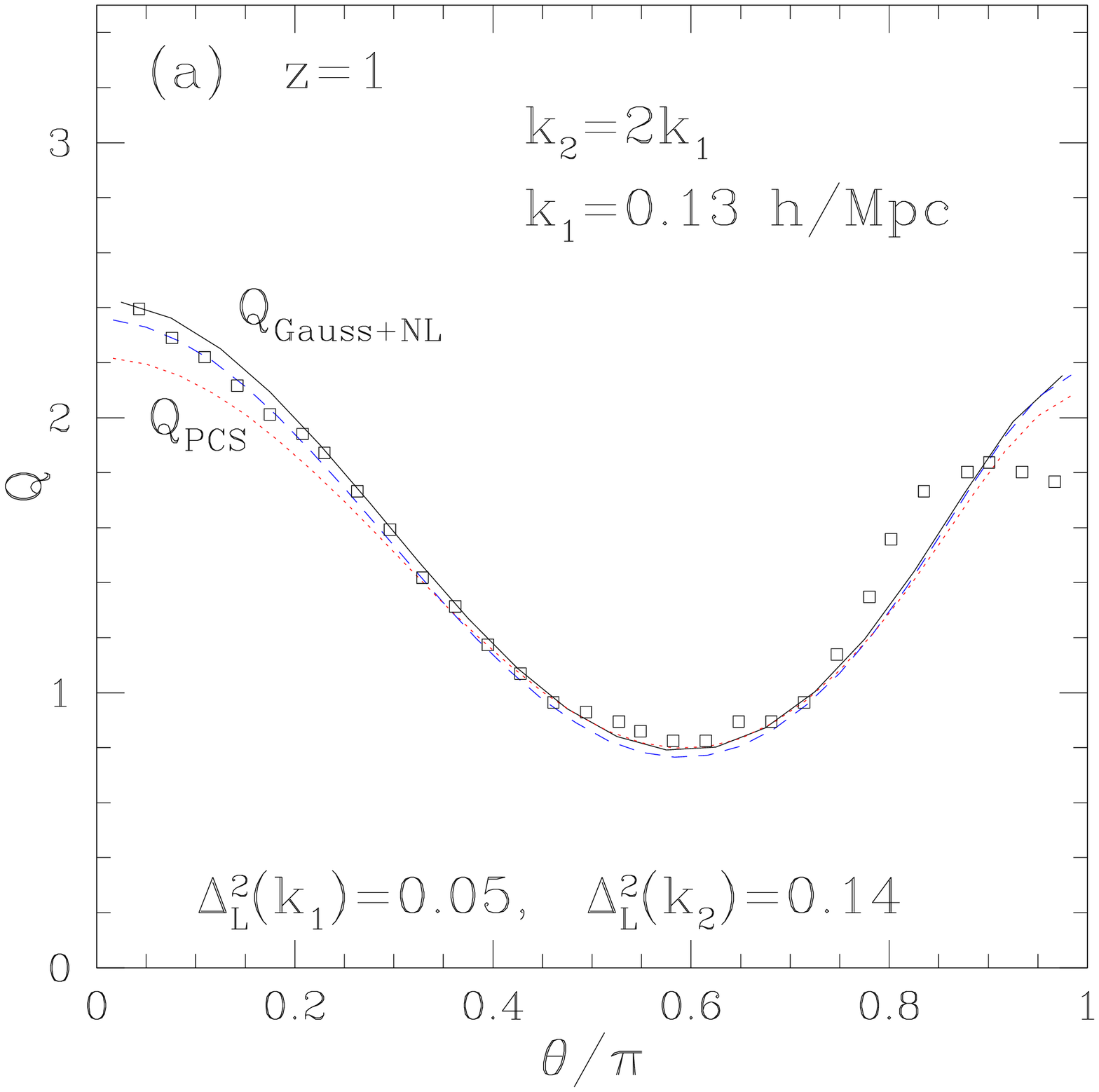}} 
\epsfxsize=4.3 cm \epsfysize=4.6 cm {\epsfbox{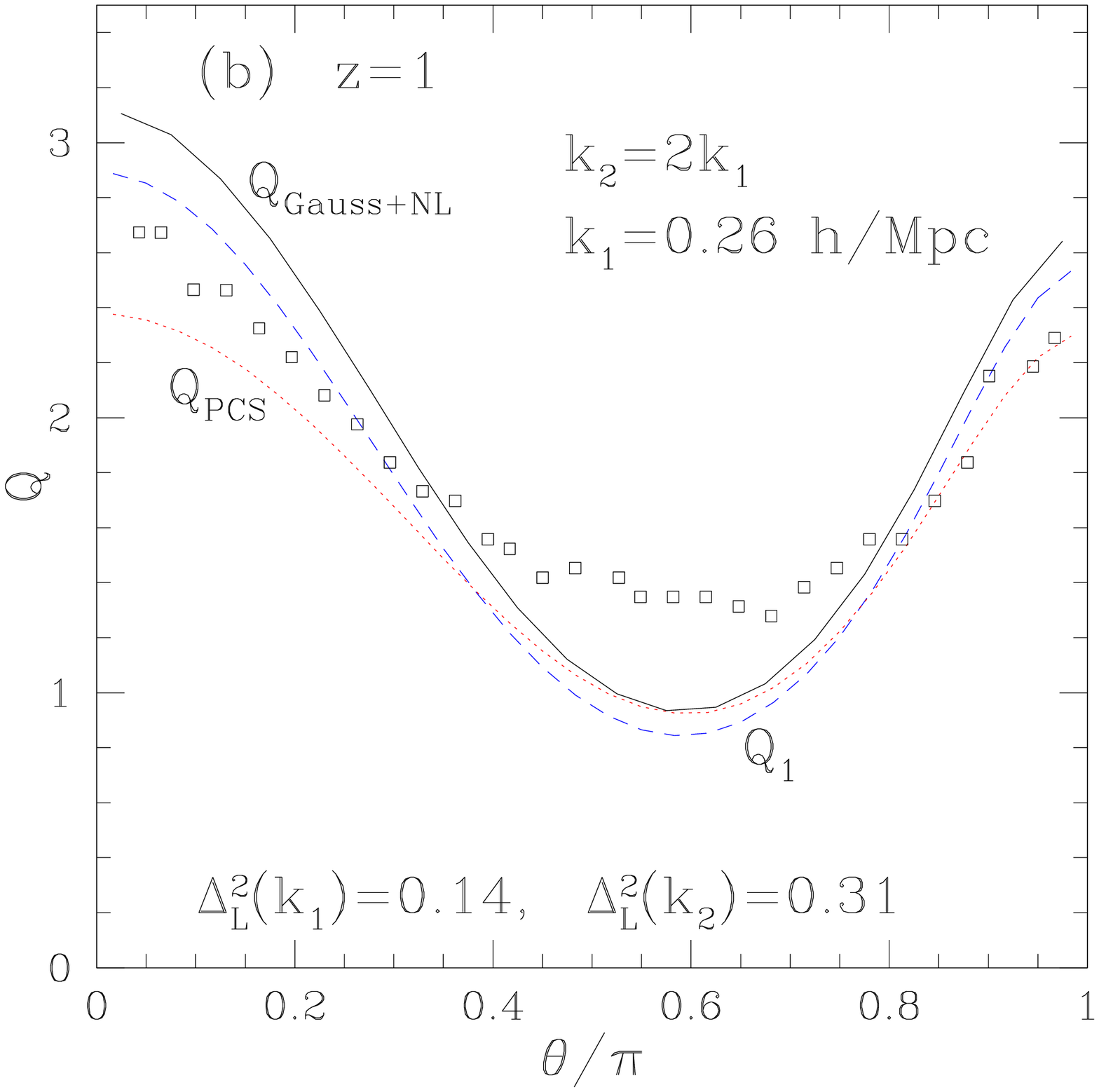}}\\
\epsfxsize=4.3 cm \epsfysize=4.6 cm {\epsfbox{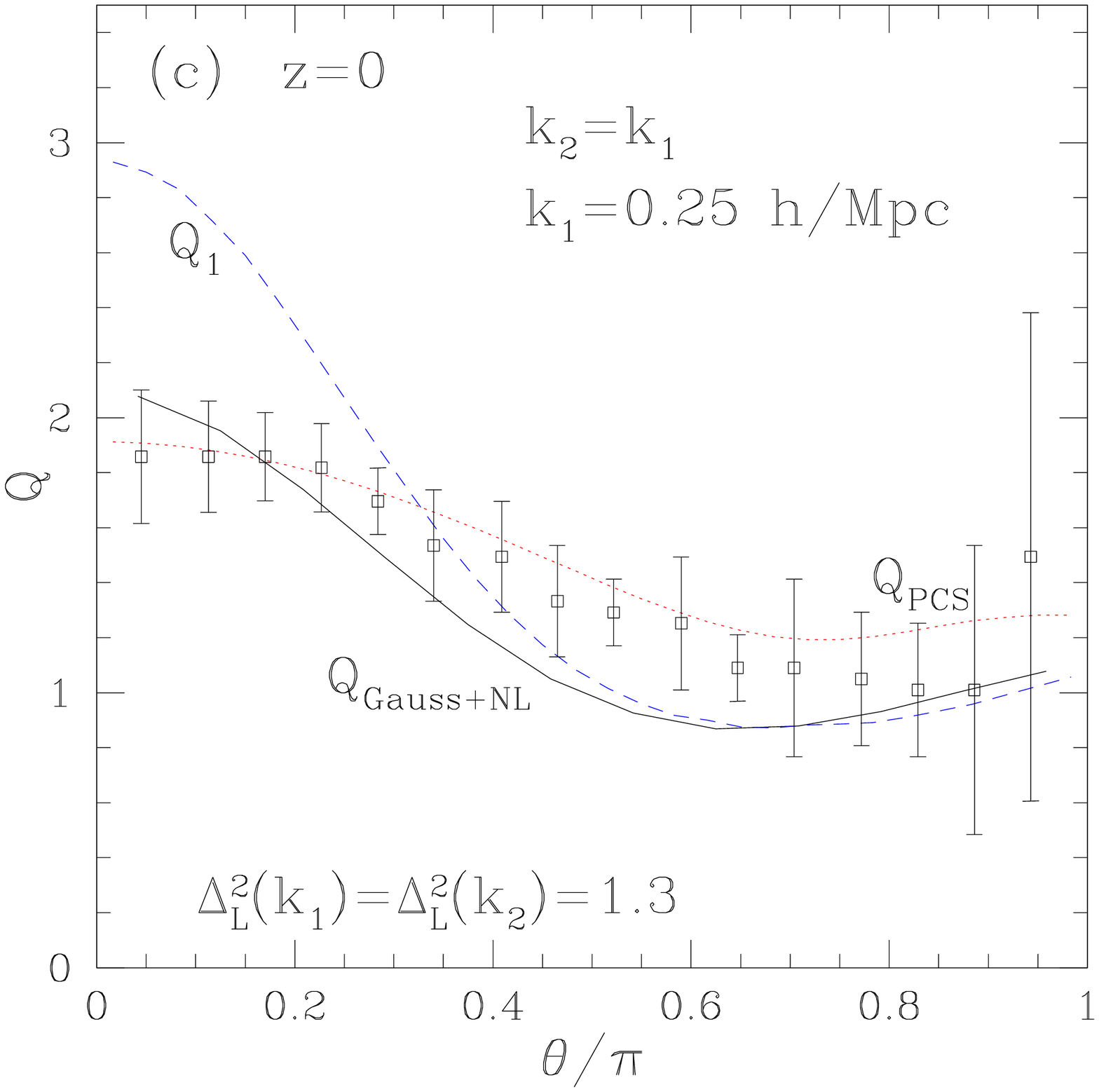}} 
\epsfxsize=4.3 cm \epsfysize=4.6 cm {\epsfbox{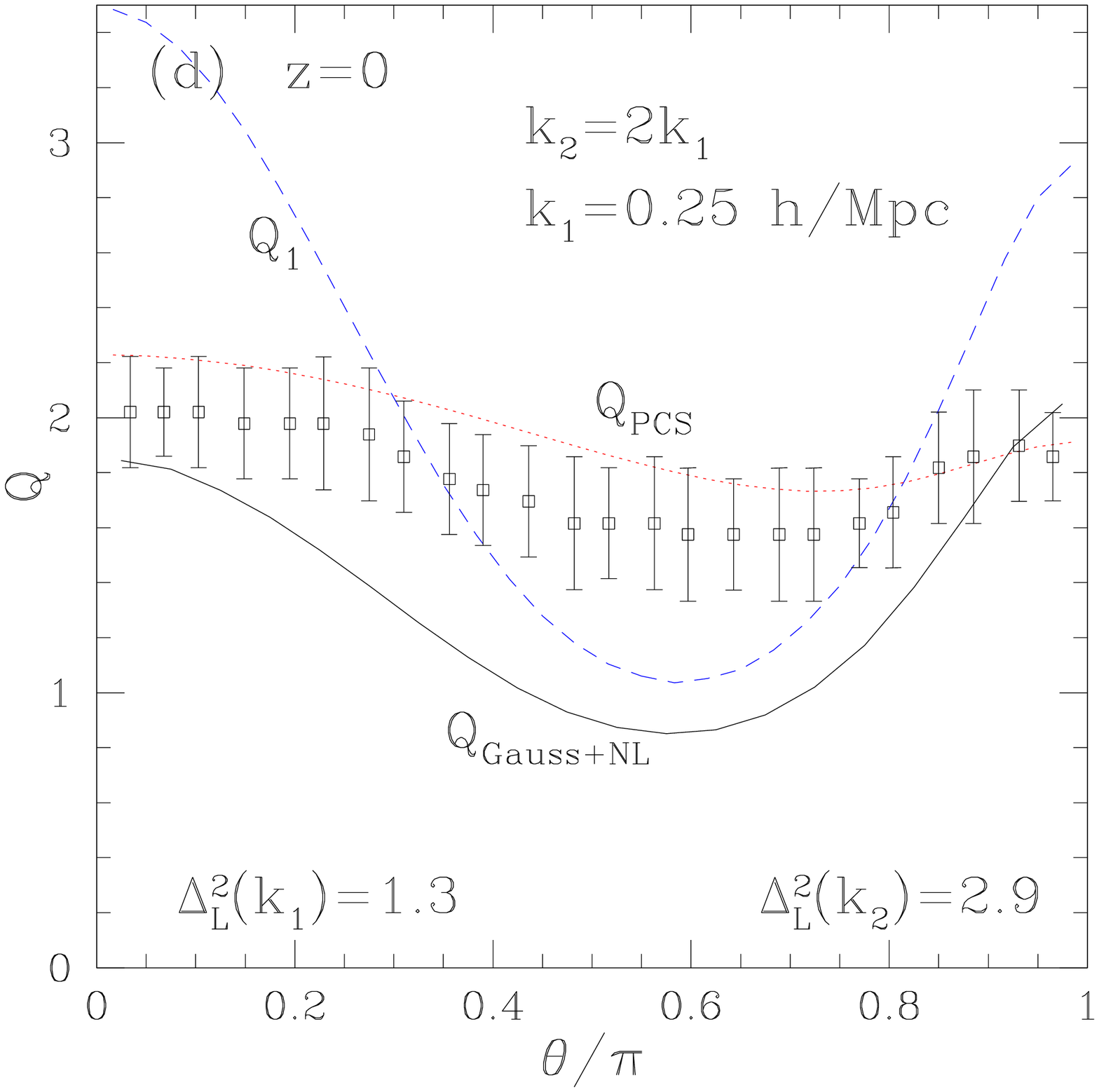}}
\end{center}
\caption{The reduced bispectrum $Q$ as a function of the angle $\theta$ between
wavenumbers $\bk_1$ and $\bk_2$, as in Fig.~\ref{figBtsd}, but using
the nonlinear power spectrum from Smith et al. (2003) for the two-point
correlations within the one-loop steepest-descent diagrams, together with
the Gaussian decay ansatz (solid line $Q_{\rm Gauss+NL}$).}
\label{figBtsdNL}
\end{figure}

We do not compute numerically in this paper the predictions associated with
the 2PI effective action method of Sect.~\ref{2PI-effective-action}
because there is no simple approximation such as Eqs.(\ref{Rcos}), (\ref{Ccosk}),
or Eqs.(\ref{RGauss}), (\ref{CGauss}), for the two-point functions. 
Moreover, the integrals over time no longer 
factorize through functions $T_{(\ell_j)}^{(a,f,g,h,i)}(\omega_j D)$, since
the internal lines of the one-loop diagrams are nonlinear two-point functions.
Thus, even if the latter are approximated by Eqs.(\ref{Rcos}), (\ref{Ccosk}), or
Eqs.(\ref{RGauss}), (\ref{CGauss}), one encounters integrals over internal
times $D_j'$ and wavenumbers $q_j$ of terms such as 
$e^{-(D_j-D_j') q_j^2 \sigma_v^2/2}$ instead of power-law prefactors 
$(D_j'/D_j)^{\alpha}$. All this would make the computation of the 2PI effective
action predictions much more complex. 

However, to investigate the effect
of using more accurate nonlinear two-point functions in all lines of the diagrams
of Fig.~\ref{figC3sd}, we modify the approximation $Q_{\rm Gauss}$ discussed above
by substituting the fit $P_{\rm NL}$ of the nonlinear power spectrum, obtained 
from numerical simulations by Smith et al. (2003), into the diagrams.
In other words, we use the exact equal-time nonlinear two-point functions in all
lines, and the dependence on the time-difference on external lines is given 
by the Gaussian factor of Eqs.(\ref{RGauss}), (\ref{CGauss}). 
We keep the power-law time factors in the internal lines so as to keep the 
factorization through the functions $T_{(\ell_j)}^{(a,f,g,h,i)}(\omega_j D)$.
In this fashion, we only need the nonlinear matter power-spectrum given by
numerical simulations at the redshift of interest, where we wish to compute
the equal-time bispectrum. Then, the equal-time
reduced bispectrum $Q$ is computed by using also the fit $P_{\rm NL}$ of 
the nonlinear power spectrum (from Smith et al. 2003) in the denominator of  
Eq.(\ref{Qdef}). We could apply the same procedure using the cosine time-dependent
cutoff of Eqs.(\ref{Rcos}), (\ref{Ccosk}), however since it does not fare as well
as the Gaussian cutoff, we only show in Figs.~\ref{figBksdNL_z0}-\ref{figBtsdNL}
our results obtained with the Gaussian decay.

In principle, we could improve this procedure in two
ways. First, we could use in the external lines the actual two-time density
correlation measured in simulations, instead of the Gaussian ansatz. Second,
we could also use this in the internal lines. However, we leave this for
future works, since the different-times density correlation has not been
studied in details in current numerical simulations, and this would imply
heavier computations for the relevant diagrams as explained above.    
Nevertheless, note that the different-time Gaussian decay agrees reasonably well 
with numerical 
simulations for the response function (Crocce \& Scoccimarro 2007b) and it 
should also provide a reasonable guess for the nonlinear correlation function
as discussed in Valageas (2007b), since it describes the random advection
of small-scale structures by the larger-scale coherent flow which affects
both two-point functions.

We show our results for equilateral triangles as a function of wavenumber $k$
in Fig.~\ref{figBksdNL_z0} (curve $Q_{\rm Gauss+NL}$). For comparison, we also 
display the predictions of the standard one-loop perturbation theory 
(curves $Q_1$ and $Q_2$ associated with Eqs.(\ref{Qfrac})-(\ref{Qexp})) 
and the phenomenological model of Pan et al. (2007).
We can see that using the ``exact'' nonlinear power spectrum $P_{\rm NL}$
does not change much the results as compared with Fig.~\ref{figBksd_z0}.
The failure to match the numerical simulations beyond
$0.2 h$ Mpc$^{-1}$ shows that in order to improve this prediction
it is not sufficient to improve the two-point functions alone. One needs to
include additional diagrams to those displayed in Fig.~\ref{figC3sd},
such as those obtained by adding new lines that connect the external to the
internal lines and which cannot be factorized in the form of Fig.~\ref{figC3sd}
(this requires new two-loop or higher-order diagrams).

The dependence on the angle $\theta$ between wavenumbers $\bk_1$ and $\bk_2$
is displayed in Fig.~\ref{figBtsdNL}. We only show the standard one-loop
prediction $Q_1$ for comparison, because it fares better than the alternative
$Q_2$, see Fig.~\ref{figBt}. Here we can see a significant
improvement over the previous predictions since we partly recover the weakening 
of the dependence on $\theta$ in the nonlinear regime. 
The improvement is most significant in panel (c) (where $\Delta^2_L(k)=1.3$).
At smaller scales the prediction $Q_{\rm Gauss}$ is too low as compared with
the simulations, since it decays as in Eq.(\ref{Qscal}), as seen in 
Fig.~\ref{figBksdNL_z0} too.
Thus, although it did not make much change for equilateral
configurations (Fig.~\ref{figBksdNL_z0}), using accurate two-point functions
within such frameworks can improve the detailed angular behavior of the 
bispectrum in the weakly nonlinear regime. 

This means that resummation
schemes can indeed improve over the standard perturbation theory (as shown
by the comparison with the standard one-loop prediction $Q_1$ in 
Fig.~\ref{figBtsdNL}). However, phenomenological models (such as Pan et al. 2007)
can still offer a better match to numerical simulations. This is not completely
surprising since the resummation schemes discussed in this paper are still expected
to break down at small scales, would it be only for the effect of shell-crossing
that invalidates the equations of motion from which they are derived 
(although one could apply similar methods to the Vlasov equation, see 
Valageas 2004). On the other hand, being derived in a systematic fashion from the
equations of motion, perturbative methods (that can involve various resummation
procedures) are better controlled. Therefore, they should be useful and more 
reliable than simple models in their regime of validity.

\section{Conclusion}
\label{Conclusion}

In this article, we have described several methods to obtain systematic 
expansions for two-point and three-point correlation functions for the
cosmological large-scale density field. These methods also extend to higher-order
correlations and response functions (e.g., the response of the two-point and
three-point correlations to external perturbations or to the initial conditions).
They also apply to related systems, defined by similar hydrodynamical equations,
such as the Zeldovich dynamics. We showed how the simplest perturbative
expansion derived from the the path-integral formulation of the system (i.e.
expanding over the cubic part of the action) corresponds to the equivalent 
standard expansion directly obtained from the hydrodynamical equations of motion.
Next, we described two resummations, associated with ``large-$N$'' limits,
that can be conveniently derived from the path integral (Valageas 2007a,b), 
and a high-$k$ resummation that is most easily derived from the equation of 
motion (Crocce \& Scoccimarro 2006a,b; Valageas 2007b). Thus, we showed how to
derive expansions for high-order correlations (or responses) and we compared
the resummations involved in these various schemes.

We first applied these methods up to one-loop order to the simpler case 
of the Zeldovich dynamics, where explicit expressions are simple enough to be 
given here. We explicitly 
checked that the perturbative expansion of the path integral recovers the
standard perturbative expansion over powers of the linear density field.
However, since these two expansions are organized in different manners, we 
found that two diagrams, obtained within the former method for the 
three-point correlation, exhibit an additional UV divergence (for 
$n(\infty)>-3$). This divergence cancels out as we sum over all diagrams
but it shows explicitly that some care must be taken in using these expansions.
For numerical purposes, it is best to first regularize each diagram by 
subtracting the divergent parts that cancel out.
Next, we applied the large-$N$ steepest-descent method to this Zeldovich 
dynamics. We showed how the time-dependence can be factorized, through
functions $T(\omega_j D)$, that measure the decay of different-time correlation
and response functions in the nonlinear regime (within simple approximations).
We found that, contrary to the standard expansions, tree diagrams and
one-loop diagrams obey the same scaling laws over the linear amplitude
$D \sigma_8$. This is due to the cutoff $T(\omega_j D)$ associated with the
partial resummations. We also explained how it is again possible to
regularize the expansion, as for the previous expansion.

Second, we applied these expansion schemes to the gravitational dynamics.
All properties obtained for the Zeldovich dynamics remain valid and we have
compared our results with numerical simulations. We checked that on
quasi-linear scales all expansion schemes agree with one another and with the
numerical simulations, up to $0.2 h$ Mpc$^{-1}$ (where $\Delta^2_L(k)\simeq 1$ 
for a $\Lambda$CDM power spectrum with $\sigma_8=0.9$). At smaller scales
they deviate from one another and from the numerical simulations.
We recovered the known failure of the standard perturbative expansion
to reproduce the observed flattening in the nonlinear regime of the reduced 
bispectrum $Q(k_1,k_2,k_3)$ as a function of the angle $\theta$ between 
wavenumbers $\bk_1$ and $\bk_2$. We found that the steepest-descent
resummation does not fare much better. Even substituting the Gaussian decay
obtained in a high-$k$ limit for the different-time two-point functions
does not help (it actually increases somewhat further the variation with 
$\theta$). Nevertheless, the agreement with numerical simulations significantly
improves if we use the ``exact'' nonlinear power spectrum (obtained from a
fit to simulations) for the two-point correlation within this formalism
(together with the Gaussian decay). This does not extend the range of
validity for equilateral configurations towards higher $k$, but it manages
to partly recover the observed flattening of the reduced bispectrum as a
function of $\theta$.

We have also compared the predictions obtained from these expansion schemes
with the simple phenomenological model of Pan et al. (2007). It appears
that the validity of the latter extends to smaller scales, up to
$0.8 h$ Mpc$^{-1}$ where $\Delta^2_L(k) \simeq 17$ in the case studied here.
Since this model is also much simpler to use, it seems that the scale
transformations devised by Hamilton et al. (1991) may indeed provide remarkably 
efficient tools to investigate various statistics associated with gravitational
clustering. At quasi-linear scales $k \sim 0.1 h$ Mpc$^{-1}$, the systematic
expansion schemes studied in this article may be somewhat more accurate
(since they arise from rigorous systematic expansions) whereas the 
phenomenological model of Pan et al. (2007) may lead to a slight overestimate
of the bispectrum (which would not be surprising since its expansion is not
expected to agree with perturbation theory).

The results obtained in this article suggest that, in order to make further
progress along these lines, one needs to include higher-order diagrams 
(two-loop diagrams or higher) to obtain reliable predictions for the bispectrum
that accurately reproduce the weakening of the dependence on the angle $\theta$
in the nonlinear regime. Moreover, using accurate two-point functions provides
a significant improvement. This requires both the use of a Gaussian decay
for different-time quantities (or a similar cutoff that agrees with numerical
simulations) and of a reasonable equal-time correlation (e.g. using the
nonlinear power spectrum measured in simulations). On the other hand, the
path-integral formulation might serve as a basis for other approximation methods
than the expansion schemes discussed here. 

Alternative methods, based on 
phenomenological models, such as the halo model (Cooray \& Sheth 2002) or the
scaling transformation of Hamilton et al. (1991), may provide other means to
describe the statistical properties of gravitational clustering. Although
simpler and more efficient in the nonlinear regime, they do not have the
rigorous and systematic character of expansion schemes. Moreover, they often 
involve several free parameters that must be fitted to simulations and they
are not derived from first principles. It would be interesting to develop
a connection between such phenomenological approaches and more systematic
methods as those developed in this article. However, it is not clear whether
such a connection can be achieved at a detailed level. 
This issue is left for future works.

\begin{appendix}

\section{Bispectrum for the gravitational dynamics: tree and one-loop diagrams
for the expansion over $K_s$}

We give in this Appendix the expressions of the diagrams of 
Fig.~\ref{figC31loop}, associated with the expansion over powers of $K_s$ of 
the path integral defining the bispectrum, for the case of the gravitational 
dynamics. The corresponding results for 
the Zeldovich dynamics are given in Sect.~\ref{Zelcubic}. The expressions
below are similar to the former but the rational functions which appear in the
integrands are more intricate.
First, the tree-diagram (a) is:
\beqa
B^{(a)}(k_1,k_2,k_3) & \! = \! & P_L(k_2) P_L(k_3)
\left[\frac{10}{7}+\left(\frac{k_2}{k_3}\!+\!\frac{k_3}{k_2}\right) 
\frac{\bk_2.\bk_3}{k_2 k_3} \right.  \nonumber \\
&& \left.  + \frac{4(\bk_2.\bk_3)^2}{7 k_2^2 k_3^2} \right] + 2 \, {\rm perm.}
\label{gravBtreedel}
\eeqa
The diagram (f) which involves a triple integration over the linear power 
spectrum reads as
\beqa
B^{(f)} & = & \int \d\bq_1\d\bq_2\d\bq_3 \; \delta_D(\bq_1+\bq_2-\bk_1) 
\nonumber \\
&& \times \; \delta_D(\bq_1+\bq_3+\bk_2) \, P_L(q_1) P_L(q_2) P_L(q_3) 
\nonumber \\
&& \times  
\frac{10q_1^2q_2^2+7(q_1^2+q_2^2)(\bq_1.\bq_2)+4(\bq_1.\bq_2)^2}{7q_1^2q_2^2}
\nonumber \\
&& \times  
\frac{10q_1^2q_3^2+7(q_1^2+q_3^2)(\bq_1.\bq_3)+4(\bq_1.\bq_3)^2}{7q_1^2q_3^2}
\nonumber \\
&& \times  
\frac{10q_2^2q_3^2-7(q_2^2+q_3^2)(\bq_2.\bq_3)+4(\bq_2.\bq_3)^2}{7q_2^2q_3^2} .
\label{gravBf}
\eeqa
Next, the two diagrams (b) and (g) which involve a double integration over $P_L$
are:
\beqa
\lefteqn{B^{(b)} = - P_L(k_1) \int \d\bq_1\d\bq_2 \;
\delta_D(\bq_1+\bq_2-\bk_2) } \nonumber \\
&& \times \frac{10 q_1^2 q_2^2 + 7 (q_1^2+q_2^2) (\bq_1.\bq_2) 
+ 4(\bq_1.\bq_2)^2}{7q_1^2q_2^2} \nonumber \\
&& \times \biggl \lbrace \frac{7 k_2^2 (\bk_1.\bk_3) 
[10q_1^2q_2^2 + 7(q_1^2\!+\!q_2^2)(\bq_1.\bq_2) + 4(\bq_1.\bq_2)^2]}
{252 k_1^2 k_2^2 q_1^2 q_2^2}\nonumber \\
&& + \frac{7 k_1^2 (\bk_2.\bk_3) -2 k_3^2 (\bk_1.\bk_2)}{252 k_1^2 k_2^2}
\left[ 6 + \frac{7(q_1^2+q_2^2)(\bq_1.\bq_2)}{q_1^2 q_2^2} \right. \nonumber \\
&& \left. + \frac{8(\bq_1.\bq_2)^2}{q_1^2 q_2^2} 
\right] \biggl \rbrace P_L(q_1) P_L(q_2)
+ 5 \, \rm{perm.} ,
\label{gravcubicBb}
\eeqa
\beqa
\lefteqn{B^{(g)} = P_L(k_1) \int \d\bq_1\d\bq_2 \;
\delta_D(\bq_1+\bq_2-\bk_2) } \nonumber \\
&& \times P_L(q_1) P_L(q_2) \frac{10(\bk_2.\bq_1)(\bk_2.\bq_2) 
- 3 k_2^2 (\bq_1.\bq_2)}{q_1^2q_2^2} \nonumber \\
&& \times \biggl \lbrace \! \frac{2 (\bk_1.\bq1) 
[10 k_3^2 q_2^2+(7 q_2^2-11 k_3^2 -14 (\bk_3.\bq_2)) (\bk_3.\bq_2)]}
{882 k_1^2 q_1^2 q_2^2} \nonumber \\
&& + \frac{(\bk_1.\bk_3)[(14 q_2^2+41 k_3^2+70(\bk_3.\bq_2))(\bk_3.\bq_2) 
-15 k_3^2 q_2^2]}{882 k_1^2 q_2^2 |\bk_3+\bq_2|^2} \nonumber \\
&& - \frac{q_1^2 (\bk_1.\bq_2)+ k_1^2 (\bq_1.(\bk_3+\bq_2))}
{882 k_1^2 q_1^2} \left[ \frac{15 k_3^2}{|\bk_3+\bq_2|^2} \right. \nonumber \\
&& \left. - \frac{(\bk_3.\bq_2) (14 q_2^2+41 k_3^2+70 (\bk_3.\bq_2))}
{q_2^2 |\bk_3+\bq_2|^2} \right] \biggl \rbrace + 5 \, \rm{perm.} 
\label{gravcubicBg}
\eeqa
Finally, the five diagrams (c), (d), (e), (h), and (i), which involve a single 
integration over $P_L$, are given by:
\beqa
\lefteqn{\!B^{(c)} \!\! = \! P_L(k_2) P_L(k_3) \frac{10k_2^2k_3^2
\!+\!7(k_2^2+k_3^2)(\bk_2.\bk_3)\!+\!4(\bk_2.\bk_3)^2 \!}{7k_2^2 k_3^2} } 
\nonumber \\ 
&& \times \int\!\d\bq P_L(q) \left[ \frac{30 k_2^2 q^2 + (47k_2^2-7q^2) 
(\bk_2.\bq)}{126 q^2 |\bk_2-\bq|^2} \right. \nonumber \\ 
&& + \frac{-(63k_2^4+163k_2^2 q^2) (\bk_2.\bq)^2+2(71 k_2^2+35q^2) 
(\bk_2.\bq)^3}{126 k_2^2 q^4 |\bk_2-\bq|^2} \nonumber \\ 
&& \left. - \frac{56(\bk_2.\bq)^4}{126 k_2^2 q^4 |\bk_2-\bq|^2} \right]
 + 5 \, \rm{perm.} ,
\label{gravcubicBc}
\eeqa
\beqa
\lefteqn{B^{(d)} \! = \! P_L(k_2) P_L(k_3) \!\int\!\d\bq P_L(q) \biggl \lbrace
\frac{9 k_3^2 (\bk_1.\bk_2)-2k_1^2 (\bk_2.\bk_3)}{k_2^2 k_3^2} } \nonumber \\
&& \!\!\!\!\!\!\! \times \!\!\left[\! \frac{2(\bk_2.\bq)^3 
(4 (\bk_2.\bq)\!\!-\!\!5 q^2)\!+\!k_2^4 (6 q^4\!\!-\!\!29 q^2 (\bk_2.\bq) 
\!+\! 21 (\bk_2.\bq)^2)\!}{1386 k_2^2 q^4 |\bk_2-\bq|^2} \right. \nonumber \\
&& \left. + \frac{k_2^2 (\bk_2.\bq) 
(-11 q^4+49 q^2 (\bk_2.\bq) -34 (\bk_2.\bq)^2)}{1386 k_2^2 q^4 |\bk_2-\bq|^2}  
\right] \nonumber \\
&& \!\!\!\!\!\!\! + \frac{3 (\bk_1.\bk_3)}{k_3^2} \left[ - \frac{ k_2^2 
(30 q^4 + 47 q^2 (\bk_2.\bq) - 63 (\bk_2.\bq)^2)}{1386 q^4 |\bk_2-\bq|^2}
\right. \nonumber \\
&& + \frac{(\bk_2.\bq) (7 q^4 +163 q^2 (\bk_2.\bq) - 142 (\bk_2.\bq)^2)}
{1386 q^4 |\bk_2-\bq|^2} \nonumber \\
&& \left. + \frac{14 (\bk_2.\bq)^3 (4 (\bk_2.\bq)
- 5 q^2)}{1386 k_2^2 q^4 |\bk_2-\bq|^2} \right] \biggl \rbrace  
+ 5 \, \rm{perm.} ,
\label{gravcubicBd}
\eeqa
\beqa
\lefteqn{B^{(e)} \! = \! - P_L(k_2) P_L(k_3) \!\int\!\d\bq P_L(q) \biggl \lbrace
\frac{k_3^2 (\bk_1.\bk_2)+ k_2^2 (\bk_1.\bk_3)}{k_2^2 k_3^2} } \nonumber \\
&& \!\!\!\!\! \times \!\left[ \frac{56 k_1^2 q^4 \! + \! (107 k_1^2 q^2\!\!
+\!33 q^4) (\bk_1.\bq) \!-\! 5(29 k_1^2\!\!+\!75 q^2)(\bk_1.\bq)^2\!}
{1386 q^4 |\bk_1-\bq|^2} \right. \nonumber \\
&& \left. + \frac{18 (17 k_1^2+5 q^2) (\bk_1.\bq)^3 
-  72 (\bk_1.\bq)^4}{1386 k_1^2 q^4 |\bk_1-\bq|^2} \right] 
- \frac{2(\bk_2.\bk_3)}{k_2 k_3} \nonumber \\
&& \!\!\!\!\! \times \!\left[ \! \frac{21 k_1^4 q^4 \! + \! (34 k_1^4 q^2\!\!
-\!\!6 k_1^2 q^4) (\bk_1.\bq) \!-\! (43 k_1^4\!\!+\!\!124 k_1^2 q^2)
(\bk_1.\bq)^2\!}{1386 k_2 k_3 q^4 |\bk_1-\bq|^2} \right. \nonumber \\
&& \left. + \frac{2 (53 k_1^2+30 q^2) (\bk_1.\bq)^3 
-  48 (\bk_1.\bq)^4}{1386 k_2 k_3 q^4 |\bk_1-\bq|^2} \right] \biggl \rbrace  
+ 2\, \rm{perm.} ,
\label{gravcubicBe}
\eeqa
\beqa
\lefteqn{B^{(h)} =  - P_L(k_2) P_L(k_3) \int \d\bq P_L(q) \biggl \lbrace
\frac{3 \bk_1.(\bk_3+\bq)}{539|\bk_3+\bq|^2} } \nonumber \\
&& \times \frac{10 k_2^2 q^2 -7 (k_2^2+q^2) (\bk_2.\bq) +4 (\bk_2.\bq)^2}
{k_2^2 q^2} \nonumber \\
&& \times \frac{6 k_3^2 q^2 +7 (k_3^2+q^2) (\bk_3.\bq) +8 (\bk_3.\bq)^2}
{k_3^2 q^2} \nonumber \\
&& \!\!\!\! + \frac{k_1^2 [ \bq.(-\bk_2+\bk_3+\bq)-(\bk_2.\bk_3)]}
{1617 k_2^2 k_3^2} \nonumber \\
&& \times \frac{6 k_2^2 q^2 -7 (k_2^2+q^2) (\bk_2.\bq) +8 (\bk_2.\bq)^2}
{q^2 |\bk_2-\bq|^2} \nonumber \\
&& \times \frac{6 k_3^2 q^2 +7 (k_3^2+q^2) (\bk_3.\bq) +8 (\bk_3.\bq)^2}
{q^2 |\bk_3+\bq|^2} \biggl \rbrace + 5\, \rm{perm.} ,
\label{gravcubicBh}
\eeqa
\beqa
\lefteqn{\!B^{(i)} \!\! = \! P_L\!(k_2) P_L\!(k_3)\! \!\int\!\!\d\bq P_L\!(q) 
\biggl \lbrace \!\frac{\bk_1.\bq}{q^2} \left[ \frac{2 k_2^2 q^2 \!\!
+\!\! (k_2^2\!+\!q^2) (\bk_2.\bq)\!}{k_2^2 q^2} \right. } \nonumber \\
&& \!\!\!\! \times \left( \! \frac{2 \bk_3.(\bk_2\!+\!\bq) |\bk_1\!-\!\bq|^2 
- 7 k_3^2 [\bk_1.\bk_2 + \bq.(\bk_1\!-\!\bk_2\!-\!\bq)]}
{154 k_3^2 |\bk_2+\bq|^2} \right. \nonumber \\
&& \!\!\!\! \left. - \frac{5 \bk_3.(\bk_1-\bq)}{66 k_3^2} \right) 
+ \frac{2\bk_2.\bq}{k_2^2} \left( - \frac{\bk_3.(\bk_1-\bq) |\bk_2+\bq|^2}
{66 k_3^2 q^2} \right.   \nonumber \\
&& \!\!\!\! \left. \left. + \frac{2 \bk_3.(\bk_2\!+\!\bq) |\bk_1\!-\!\bq|^2 
- 7 k_3^2 [\bk_1.\bk_2 + \bq.(\bk_1\!-\!\bk_2\!-\!\bq)]}{231 k_3^2 q^2} 
\!\right) \right] \nonumber \\
&& \!\!\!\! + \frac{7 k_1^2 q^2 + (2 k_1^2-9q^2) (\bk_1.\bq)}{q^2 |\bk_1-\bq|^2} 
\left[ \frac{2 k_2^2 q^2 + (k_2^2+q^2) (\bk_2.\bq)}{k_2^2 q^2} \right.  
\nonumber \\
&& \!\!\!\! \times \left( \! \frac{2 \bk_3.(\bk_2\!+\!\bq) |\bk_1\!-\!\bq|^2 
- k_3^2 [\bk_1.\bk_2 + \bq.(\bk_1\!-\!\bk_2\!-\!\bq)]}{462 k_3^2 |\bk_2+\bq|^2} 
\right. \nonumber \\
&& \!\!\!\! \left. - \frac{5 \bk_3.(\bk_1-\bq)}{1386k_3^2} \right) 
+ \frac{2\bk_2.\bq}{k_2^2} \left( - \frac{\bk_3.(\bk_1-\bq) |\bk_2+\bq|^2}
{1386 k_3^2 q^2} \right.   \nonumber \\
&& \!\!\!\! \left. \left. + \frac{2 \bk_3.(\bk_2\!+\!\bq) |\bk_1\!-\!\bq|^2 
- k_3^2 [\bk_1.\bk_2 + \bq.(\bk_1\!-\!\bk_2\!-\!\bq)]}{693 k_3^2 q^2} \!\right) 
\right] \! \biggl \rbrace \nonumber \\
&& \!\! + 5\, \rm{perm.}
\label{gravcubicBi}
\eeqa
Because of the property $\bk_1+\bk_2+\bk_3=0$, which has been used to simplify
somewhat these expressions, it is possible to write each contribution under
many different forms (for instance by replacing $\bk_1$ by $-(\bk_2+\bk_3)$).

We do not give in this paper the expressions obtained within the 
steepest-descent expansion, as they are longer than 
Eqs.(\ref{gravBtreedel})-(\ref{gravcubicBi}). However, as explained in the main
text, by putting all time-dependent functions $T^{(.)}_{(\ell_j)}$ equal to unity,
they reduce to the expressions above for diagrams (a), (f), (g), (h), 
and (i).

\end{appendix}

\end{document}